\documentclass[fleqn]{annalen}
\usepackage{psfrag,epsfig}
\usepackage{amssymb}
\usepackage{citesort}
\begin{document}
\renewcommand{\textfraction}{-0.1} 
\renewcommand{\floatpagefraction}{1.0001} 
\renewcommand{\topfraction}{1.0001} 
\renewcommand{\bottomfraction}{1.0001} 
\renewcommand{\footnoterule}{\rule{12cm}{0.5pt} \vspace{3pt}}
\renewcommand{\thefootnote}{\arabic{footnote}}
\setcounter{tocdepth}{2}
%
%
\newcommand{\volume}{1}              
\newcommand{\xyear}{2001}            
\newcommand{\issue}{1}               
\newcommand{\recdate}{dd.mm.yyyy}    
\newcommand{\revdate}{dd.mm.yyyy}    
\newcommand{\revnum}{0}              
\newcommand{\accdate}{dd.mm.yyyy}    
\newcommand{\coeditor}{ue}           
\newcommand{\firstpage}{1}           
\newcommand{\lastpage}{60}          
\setcounter{page}{\firstpage}        
\newcommand{\keywords}{Peierls chains, fluctuating gap model, pseudogap} 
\newcommand{\PACS}{71.23.-k, 02.50.Ey, 71.10.Pm}
\newcommand{\shorttitle}{L. Bartosch, Fluctuation effects in disordered 
Peierls systems} 
\title{Fluctuation effects in disordered Peierls systems}
\author{Lorenz Bartosch} 
\newcommand{\address}
  {Institut f\"{u}r Theoretische Physik, Universit\"{a}t Frankfurt,
Robert-Mayer-Strasse 8-10, 
60054 Frankfurt am Main, Germany}
\newcommand{\email}{\tt bartosch@th.physik.uni-frankfurt.de} 
\maketitle
\begin{abstract}
We review the density of states and related quantities of quasi one-dimensional disordered Peierls systems in which fluctuation effects of a backscattering potential play a crucial role. The low-energy behavior of non-interacting fermions which are subject to a static random backscattering potential will be described by the fluctuating gap model (FGM).
Recently, the FGM has also been used to explain the pseudogap phenomenon in high-$T_c$ superconductors.
After an elementary introduction to the FGM
in the context of commensurate and incommensurate Peierls chains,
we develop a non-perturbative method
which allows for a simultaneous calculation of the density of states
(DOS) and the inverse localization length. First, we recover all known
results in the limits of zero and infinite correlation lengths of the
random potential. Then, we attack the problem of finite correlation
lengths. While a complex order parameter, which describes
incommensurate Peierls chains, leads to a suppression of the DOS,
i.e. a pseudogap,
the DOS exhibits a singularity at the Fermi energy if the order parameter
is real and therefore refers to a commensurate system. We confirm
these results by calculating the DOS and the inverse localization
length for finite correlation lengths and Gaussian statistics of the
backscattering potential with unprecedented accuracy numerically. 
Finally, we consider the case of classical phase fluctuations which
apply to low temperatures where amplitude fluctuations are frozen
out. In this physically important regime, which is also characterized by finite
correlation lengths, we present analytic results
for the DOS, the inverse localization length, the specific heat, and the
Pauli susceptibility.
\end{abstract}

\tableofcontents

\section{Introduction}
\label{section:Introduction}

As the temperature is lowered,
some inorganic and organic conductors with a highly anisotropic
crystal and electronic structure become unstable and undergo a Peierls
transition, i.e.\ they develop a 
charge-density wave. 
This
instability is due to their quasi one-dimensional nature which results
in a (perfectly) nested Fermi surface. A qualitative understanding of
the Peierls instability can already be gained by treating the phonon
field of a quasi one-dimensional electron-phonon system 
in a mean-field picture
\cite{Froehlich54,Peierls55,Kuper55,Rice73a,Rice73b,Gruener85,Gruener88,Gruener94}.
However, because of reduced dimensionality, fluctuations of the phonon
field which can be identified as the order
parameter field $\Delta(x)$ are crucial and significant deviations are
to be expected.

In a seminal paper, Lee, Rice and Anderson \cite{Lee73}
introduced the one-dimensional so-called fluctuating gap model (FGM),
in which fluctuations of the phonon field are modeled by a static disorder
potential. Calculating the leading-order correction of the electronic
self energy 
of an incommensurate chain which is described by a complex order
parameter field with $\langle \Delta(x) \rangle = 0$ and
$\langle \Delta(x) \Delta^{\ast} (x') \rangle = \Delta_s^2 \, e^{-|x-x'|/\xi}$,
where $\xi$ is the temperature-dependent correlation length, Lee,
Rice, and Anderson obtained an 
approximate expression for the density of states (DOS), showing a
suppression of the DOS near the Fermi energy, which is called a
pseudogap.

A few years later, Sadovskii \cite{Sadovskii79} apparently obtained an
exact expression for the Green function of the FGM using Gaussian
statistics for the higher correlation functions of the order parameter
field 
which he could assume to be real or complex, referring to a band
filling being commensurate or incommensurate with the underlying lattice. 
Recently, the experimental observation of a pseudo-gap state in the overdoped
cuprates above the superconducting phase transition led to a
reincarnation of the FGM and Sadovskii's exact solution
in the field of high-temperature
superconductivity \cite{Schmalian98,Schmalian99,Sadovskii01}. 
However, the revived interest in Sadovskii's
solution also brought to light a subtle error in this solution 
\cite{Tchernyshyov99} which questions not only the
solution itself, but also the work based on it.

Besides the limit $\xi \rightarrow \infty$
where Sadovskii's solution is indeed 
exact \cite{Kuchinskii99,Tchernyshyov99},
Sa\-dov\-skii's solution can also be easily tested in the white-noise
limit $\xi \to 0$, keeping $D \equiv \Delta_s^2 \xi$ constant, such
that $\langle \Delta(x) \Delta^{\ast} (x') \rangle = 2 D \,
\delta(x-x')$. Solving a stationary Fokker-Planck equation, Ovchinnikov and
Erikhman \cite{Ovchinnikov77} obtained an exact expression for the DOS
for real $\Delta(x)$. They showed that for small $\omega$ and
$\langle \Delta (x) \rangle = 0$, the DOS diverges as 
$\langle \rho ( \omega ) \rangle \propto
| \omega   \ln^3  | \omega |  |^{-1}$.
Singularities of this type at the band
center of a random Hamiltonian
have been discovered
by Dyson \cite{Dyson53} in the fifties
and have recently also been found
in one-dimensional spin-gap systems \cite{McKenzie96a,Fabrizio97}.
It is important to note that in the FGM, the
singularity is a consequence of
phase resonance,
and is {\it{not}} related to concrete probability properties
of $\Delta (x)$ \cite{Lifshits88,Mostovoy98}. 
In particular, the singularity 
is {\it{not}} an artifact of the exactly solvable limit 
$\xi \rightarrow 0$ considered in Ref.\ \cite{Ovchinnikov77}. As
argued in Ref.\ \cite{Bartosch99a}, 
it is therefore reasonable to expect that
for any $\xi < \infty$ the average DOS of the FGM 
exhibits a singularity at $\omega = 0$.
This general argument is in disagreement
with Sa\-dov\-skii's solution \cite{Sadovskii79}
which for large but finite $\xi$ shows 
a pseudogap and no singularity. 
In this work we shall reexamine the DOS of the FGM which determines
the whole thermodynamics of the FGM and resolve the above
contradictions. 
This emerged from the PhD thesis of the author \cite{Bartosch00PhD}.
Some of the results presented here have been published in a series 
of recent research articles \cite{Bartosch99a,Bartosch99d,Bartosch00c}.

The organisation of this article is as follows: 
In Section \ref{chap:FGM}, we give an elementary
introduction to diordered Peierls systems whose Fermi wave vector
can be commensurate or incommensurate with the underlying lattice
structure.
Starting from a Fr\"ohlich
Hamiltonian which describes a one-dimensional electron-phonon system,
we will introduce the fluctuating gap model (FGM) as a low energy model
in which the phonon system is essentially replaced by a fluctuating static 
backscattering potential which will serve as the order parameter field.

Section \ref{section:formalism} focuses on the one-particle Green
function of the FGM. After calculating the Green function in the
leading-order Born approximation which reproduces the result obtained
by Lee, 
Rice and Anderson, we will develop a formally exact non-perturbative
expression of the Green function as a functional of the disorder
potentials based on a non-Abelian generalization of the 
Schwinger-ansatz.
To calculate the DOS and inverse localization length, the
introduction of phase variables will turn out to be very convenient. While
one phase variable is simply related to the integrated DOS and
satisfies a non-linear equation of motion which is equivalent to a
Riccati equation, the other phase variable is related to the inverse
localization length and can be expressed in terms of the first phase
variable. 
These equations of motions will serve as the starting point for
detailed calculations of the DOS and inverse localization length for
various probability distributions of the disorder potentials in the
next sections.

In Section \ref{chap:whitenoise}, we will review known exact results of
the DOS and inverse localization length in the limit of infinite
correlation lengths and in the white noise limit. 
Generalizing the phase formalism developed in Ref.\ \cite{Lifshits88}
such that $\Delta(x)$ is allowed to be complex,
we will derive a linear fourth-order Fokker-Planck equation
previously only obtained within the framework of the method of
supersymmetry \cite{Hayn96}. The solution of this stationary
Fokker-Planck 
equation encapsulates all known results for the DOS and inverse
localization length of the FGM in the
white noise limit including the above mentioned Dyson singularity in
the Ovchinnikov and Erikhman limit.
Results for the case of infinite correlation lengths will finally be
obtained by averaging the DOS and the inverse localization length
calculated for a constant disorder potential over
an appropriate probability distribution of the disorder potentials.

The case of finite correlation lengths of the order parameter field
will be attacked in Section \ref{section:finite_xi}. Considering the
equation of motion related to the integrated DOS, we will first 
argue that we expect for any finite $\xi$ a Dyson singularity in the
DOS. We will then set up an algorithm based on the equations of motion
derived in Section \ref{section:formalism} which will allow for a
simultaneous numerical calculation of the DOS and inverse localization
length for arbitrary disorder potentials with unprecedented
accuracy. For complex 
$\Delta(x)$, Sadovskii's solution is not too far off from our
numerical solution. In particular, for large correlation lengths
$\Delta_s \xi \gg 1$, the DOS at the Fermi energy vanishes as
$\rho(0) \propto (\Delta_s \xi)^{-0.64}$ instead of
$\rho(0) \propto (\Delta_s \xi)^{-1/2}$, as predicted by
Sadovskii. However, for real $\Delta$, we will find a 
pseudogap in the DOS for $\Delta_s \xi \gg 1$ which for any finite
$\xi$ is overshadowed by a Dyson singularity of the form 
$\rho ( \omega )  = A \,
| \omega   \ln^{\alpha + 1}  | \omega |  |^{-1}$, where $A$ and the exponent
$\alpha$ depend on the correlation length $\xi$. As the correlation
length $\xi$ increases, $\alpha$ assumes the finite value $\alpha =
0.41$, but the weight of the Dyson singularity vanishes with
increasing correlation length. 
At the end of Section \ref{section:finite_xi}, we shall also discuss
the case of only phase fluctuations of the order parameter which
applies to sufficiently low temperatures where the amplitude of the
order parameter is confined to a narrow region around $\Delta_s$ such
that Gaussian statistics do not apply any more.
We will find exact analytic results for the DOS and inverse
localization length and we will also calculate the low-temperature Pauli
paramagnetic susceptibility and the electronic low-temperature
specific heat.

\section{Peierls systems and the fluctuating gap model}
\label{chap:FGM}

{\em In this introductory section we will replace the dynamic phonon 
ensemble in Peierls systems by a static backscattering potential which 
we will identitfy as 
the order parameter. Due to reduced dimensionality, fluctuations of this 
order parameter field are very important and we will determine its statistics
in the context of a brief discussion of a generalized Ginzburg-Landau 
functional. Finally, we introduce the fluctuating gap model
  (FGM) as a low-energy model which takes into account these fluctuations.}

\subsection{Fr\"ohlich Hamiltonian and Peierls instability}


The formation of
periodic lattice distortions and  charge-density waves in Peierls
chains is due to the electron-phonon interaction in these quasi
one-dimensional
materials \cite{Froehlich54,Peierls55,Kuper55,Gruener85,Gruener88,Gruener94}.  
Since particle-hole excitations with
momentum\footnote{In this work we choose units such that 
$\hbar = k_B = 1$.} 
$2k_F$ are possible for very small excitation energies, the
Lindhard density-density response function exhibits a singularity at
$q=2k_F$. Kohn showed that this singularity should be conveyed into a
kink in the phonon spectrum \cite{Kohn59}. While these Kohn anomalies
are rather weak in isotropic materials, they can lead to a substantial
alteration to the phonon dispersion in quasi one-dimensional
materials with a topology of the Fermi surface which shows
perfect nesting. At low enough temperatures, the renormalized phonon
mode at $2k_F$ can scale all the way down to zero, i.e.\ become
gapless. This process is called softening of the phonon mode. Since
$\omega_{\rm ren} (2k_F) \to 0$, a static lattice distortion with
wave vector $2k_F$ may now arise. Simultaneously, there is a formation
of a charge density wave. As a consequence, the discrete translational
invariance 
is broken. The same physics can also be described by considering the
thermodynamics of a Peierls system. This approach will also allow to
go beyond a mean-field picture and will therefore be followed here.

A Hamiltonian to describe a one-dimensional electron-phonon system was
proposed in 
1954 by Fr\"ohlich \cite{Froehlich54}:
\begin{equation}
  \label{eq:Froehlich}
  \mathcal{H} = \sum_{k,\sigma} \epsilon_k\, c_{k,\sigma}^{\dagger}
  c_{k,\sigma} + 
  \sum_q \omega_q \, b_q^{\dagger} b_q +
  \sum_{q} \frac{g_q}{\sqrt{L}}\, \hat \varrho_{q}^{\dagger}\, (b_{q} +
  b_{-q}^{\dagger}) \;. 
\end{equation}
The system has length $L=Na$  where $a$ is the lattice
spacing, and periodic boundary conditions are assumed.
$c_k^{\dagger}$ and $c_k$ are fermionic creation and annihilation
operators with momentum $k$, spin $\sigma$, and energy
$\epsilon_k$.
While $\epsilon_k = k^2/2m$ for free electrons, in the
tight-binding approximation one has $\epsilon_k=- 2t \cos
ka$.
The second term in the Fr\"ohlich Hamiltonian (\ref{eq:Froehlich})
describes phonons with phonon dispersion $\omega_q$. 
$b_q^{\dagger}$ and $b_q$ are bosonic creation and annihilation
operators with momentum $q$ which 
is confined to the first Brillouin zone. For a chain with only
nearest-neighbor interactions we have (see for example Ref.\ 
\cite{Ashcroft76})  $\omega_q =
2\omega_0 \sin|q a/2|$. 
Finally, the last term in Eq.\ (\ref{eq:Froehlich}) models the
interaction of the phonon system with the fermions. The phonons are
linearly coupled via the electron-phonon coupling constant $g_q$ to
the Fourier components of the electron density
\begin{equation}
  \label{eq:density:fermions}
  \hat \varrho_q^{\dagger} \equiv \sum_{k,\sigma} c_{k+q,\sigma}^{\dagger}
  c_{k,\sigma}\;. 
\end{equation}
The phonon operators $b_q$ and $b_q^{\dagger}$ are directly related to
the operators of the normal coordinates $u_q$ of the lattice system by
\begin{equation}
  \label{eq:uq}
  u_q = \left(\frac{1}{2M\omega_q}\right)^{1/2} \left(b_q +
    b_{-q}^{\dagger} \right) \;.
\end{equation}
Here, $M$ is the ionic mass. The lattice displacement operators of the
ions at $x_n = n a$ are given by its Fourier transform,
\begin{equation}
  \label{eq:ux}
  u(x_n) = \sum_q e^{iqx_n} \left(\frac{1}{2NM\omega_q}\right)^{1/2}
  \left(b_q + 
    b_{-q}^{\dagger} \right) \;.
\end{equation}
In a mean-field picture, the Peierls transition
will lead to a non-vanishing expectation value $\left\langle u(x_n)
\right\rangle$ which implies that the system exhibits a static
lattice distortion \cite{Gruener94,Bartosch00PhD}. This static lattice
distortion is accompanied by a single-particle gap, a 
charge density wave and unusual 
electronic transport properties 
\cite{Gruener85,Gruener88,Gruener94,Monceau85,Eckern86}.

\subsection{Euclidean action}

In an Euclidean functional integral approach, the Fr\"ohlich Hamiltonian
is conveyed into the action (see, for example, Negele and Orland
\cite{Negele88}) 
\begin{equation}
  \label{eq:action}
  S\{\psi^{\ast},\psi;b^{\ast},b\} = S_{\rm el} \{\psi^{\ast},\psi\} +
  S_{\rm ph} \{b^{\ast},b\} + S_{\rm int}\{\psi^{\ast},\psi;b^{\ast},b\}\;,
\end{equation}
where
\begin{eqnarray}
  \label{eq:action:el}
  S_{\rm el} \{\psi^{\ast},\psi\} & \hspace{-2.3mm} = &
  \hspace{-2.3mm} \beta \sum_{k,\tilde \omega_n,\sigma} \psi_{k,\tilde
    \omega_n,\sigma}^{\ast} \left[i\tilde \omega_n - \tilde \epsilon_k
  \right] \psi_{k,\tilde \omega_n,\sigma}\;, \\
  S_{\rm ph} \{b^{\ast},b\} & \hspace{-2.3mm} = &
  \hspace{-2.3mm} - \beta \sum_{q,\omega_m}
  b_{q,\omega_m}^{\ast}\left[i\omega_m - \omega_q \right]
  b_{q,\omega_m} \;, \\
  S_{\rm int}\{\psi^{\ast},\psi;b^{\ast},b\} & \hspace{-2.3mm} = &
  \hspace{-2.3mm} \beta \sum_{q,\omega_m} \frac{g_q}{\sqrt{L}}
  \left(\sum_{k,\tilde \omega_n,\sigma}\! \psi_{k+q,\tilde \omega_n + \omega_m,\sigma}^{\ast}
    \psi_{k,\tilde \omega_n,\sigma}  \right) \nonumber \\
  & & \hspace{42mm} \times \,
  \left[b_{q,\omega_m} +
    b_{-q,-\omega_m}^{\ast} \right] \;.\qquad
\end{eqnarray}
Here, $\beta \equiv 1/ T$ is the inverse
temperature and $\tilde \epsilon_k \equiv \epsilon_k -\mu$ is the energy
dispersion measured with respect to the chemical potential $\mu$. While the
conjugated Grassmann variables $\psi_{k,\tilde \omega_n}$ and
$\psi_{k,\tilde \omega_n}^{\ast}$
describe fermions with momentum $k$ and fermionic Matsubara frequency
$\tilde \omega_n \equiv (2n+1)\pi/\beta$, $b_{q,\omega_m}$ and
$b_{q,\omega_m}^{\ast}$ are complex (bosonic) phonon fields with
momentum $q$ and bosonic Matsubara frequency $\omega_m \equiv 2\pi
m/\beta$. 
In terms of the above action, the partition function reads
\begin{equation}
  \label{eq:partition_function}
  \mathcal{Z} = \int \mathcal{D} \left\{\psi^{\ast},\psi\right\} \,
  \mathcal{D} \left\{b^{\ast},b \right\} \,
  \exp\left[-S\{\psi^{\ast},\psi;b^{\ast},b\}\right] \;,
\end{equation}
where $\mathcal{D} \left\{\psi^{\ast},\psi\right\}$ and $\mathcal{D}
\left\{b^{\ast},b \right\}$ are appropriately normalized fermionic and
bosonic integration measures \cite{Negele88}.
Using the variable transformation
$\phi_{q,\omega_m} \equiv \frac{g_q}{\sqrt{L}} \left( b_{q,\omega_m} + 
b_{-q,-\omega_m}^{\ast} \right)$, $\eta_{q,\omega_m} \equiv 
-i \frac{g_q}{\sqrt{L}} \left(b_{q,\omega_m} -
    b_{-q,-\omega_m}^{\ast} \right)$, such that 
$ \phi_{q,\omega_m}^{\ast} =
\phi_{-q,-\omega_m}$ and $ \eta_{q,\omega_m}^{\ast} =
\eta_{-q,-\omega_m}$, $\eta$ may easily be integrated out
resulting  in 
\begin{equation}
  \label{eq:action2}
  S\{\psi^{\ast},\psi;\phi\} = S_{\rm el} \{\psi^{\ast},\psi\} +
  S_{\rm ph} \{\phi\} + S_{\rm int}\{\psi^{\ast},\psi;\phi\}\;,
\end{equation}
where $S_{\rm el} \{\psi^{\ast},\psi\}$ is unchanged and
\begin{eqnarray}
  S_{\rm ph} \{\phi\} & \hspace{-2.3mm} = &
  \hspace{-2.3mm} \frac{1}{2} \beta L \sum_{q,\omega_m} \frac{1}{|g_q|^2} \,
  \phi_{q,\omega_m}^{\ast}\left[\frac{\omega_m^2 + \omega_q^2}{
      \omega_q} \right] \phi_{q,\omega_m} \;, \\
  S_{\rm int}\{\psi^{\ast},\psi;\phi\} & \hspace{-2.3mm} = &
  \hspace{-2.3mm} \beta \sum_{q,\omega_m}
  \left(\sum_{k,\tilde \omega_n} \psi_{k+q,\tilde \omega_n + \omega_m}^{\ast}
    \psi_{k,\tilde \omega_n}  \right) \phi_{q,\omega_m}
  \;.\qquad
\end{eqnarray}
So far, no approximation has been made. In the following, we will
restrict ourselves to the low-energy physics of the weak-coupling
limit, so that only fermions in the vicinity of
the Fermi energy are involved. In this case the Fermi energy may be
linearized 
around the two Fermi points, such that it assumes the form
\begin{equation}
  \label{eq:energy_dispersion:linearised}
  \tilde \epsilon_k = v_F (|k|-k_F)\;.
\end{equation}
To separate right- and left-moving Fermions, let us introduce
the spinor field  
\begin{equation}
  \label{eq:spinor}
  \bar \psi_{k,\tilde \omega_n,\sigma} \equiv \left(\begin{array}{c}
      \psi_{+,k,\tilde \omega_n,\sigma} \\ \psi_{-,k,\tilde
        \omega_n,\sigma} \end{array} \right) \equiv \left(\begin{array}{c}
      \psi_{k_F+k,\tilde \omega_n,\sigma} \\ \psi_{-k_F+k,\tilde
        \omega_n,\sigma} \end{array} \right)\;
\end{equation}
and its conjugated counterpart
\begin{equation}
  \label{eq:spinor:conjugated}
  \bar \psi_{k,\tilde \omega_n,\sigma}^{\dagger} \equiv \left(
      \psi_{+,k,\tilde \omega_n,\sigma}^{\ast}\;,\; \psi_{-,k,\tilde
        \omega_n,\sigma}^{\ast} \right) \equiv \left(
      \psi_{k_F+k,\tilde \omega_n,\sigma}^{\ast}\;,\;\psi_{-k_F+k,\tilde
        \omega_n,\sigma}^{\ast} \right)\;.
\end{equation}
The electronic part of the action may easily be rewritten in terms of
these spinor fields and the inverse non-interacting Matsubara Green
function 
\begin{equation}
  \label{eq:inverse_G:matrix}
  \mathbf{G}_{0}^{-1} (k,\tilde \omega_n) 
  \equiv
  \left(
    \begin{array}{cc}
      i\tilde \omega_n -v_F k  & 0 \\
      0 & i\tilde \omega_n + v_F k
    \end{array}
  \right) \;.
\end{equation}
Since the momentum transfer of the phonons is either small compared
with the Fermi momentum or approximately $2k_F$,
we decompose $\phi_{q,\omega_m}$ according to
\begin{equation}
  \label{decomposition:phi}
  \mathbf{V}_{q,\omega_m} \equiv \left (
    \begin{array}{cc}
        V_{q,\omega_m} & \Delta_{q,\omega_m} \\
        \Delta_{-q,-\omega_m}^{\ast} & V_{q,\omega_m}
    \end{array}
    \right)
    \equiv \left (
    \begin{array}{cc}
        \phi_{q,\omega_m} & \phi_{q+2k_F,\omega_m} \\
        \phi_{q-2k_F,\omega_m} & \phi_{q,\omega_m} \\
    \end{array}
    \right) \;,
\end{equation}
such that $|q|<k_F$. While $\phi_{q,\omega_m}^{\ast} =
\phi_{-q,-\omega_m}$ directly translates into $V_{q,\omega_m}^{\ast}
= V_{-q,-\omega_m}$, a similar relation for $\Delta_{q,\omega_m}$ does
only hold if $4k_F$ is a reciprocal lattice
vector. $\Delta_{q,\omega_m}^{\ast} = \Delta_{-q,-\omega_m}$ is
therefore only true for a half-filled band for which $\pi/a = 2k_F$. We
will refer to this case as the {\em commensurate case}. The more
general case for which $k_F a/\pi$ is an other fractional number is
also called commensurate but will
not be discussed here. In the {\em incommensurate case}, for which $k_F
a/\pi$ is well separated from any simple fractional number,
all $\Delta_{q,\omega_m}^{\ast}$ and
$\Delta_{-q',-\omega_{m'}}$ are independent. We will see in this work
that commensurate and incommensurate Peierls systems can have very
different physical properties.

Defining the matrices $\mathbf{G}_0^{-1}$ and $\mathbf{V}$ via
\begin{eqnarray}
  \label{eq:matrix:G^-1}
  \left(\mathbf{G}_0^{-1}\right)_{k,k',\tilde \omega_n,\tilde
    \omega_{n'}} & \hspace{-2.3mm} \equiv & \hspace{-2.3mm}
  \delta_{k,k'}\, \delta_{\tilde\omega_n,\tilde\omega_{n'}}
  \mathbf{G}_{0}^{-1}(k,\tilde \omega_n) \;, \\
  \left(\mathbf{V}\right)_{k,k',\tilde \omega_n,\tilde
    \omega_{n'}} & \hspace{-2.3mm} \equiv & \hspace{-2.3mm}
    \mathbf{V}_{k-k',\tilde \omega_n-\tilde \omega_{n'}} \;,
\end{eqnarray}
our action 
turns into
\begin{equation}
  \label{eq:action3}
  S\{\bar \psi^{\dagger},\bar \psi;V,\Delta,\Delta^{\ast}\} = S_{\rm el-ph}
  \{\bar \psi^{\dagger},\bar \psi;V,\Delta,\Delta^{\ast}\} + 
  S_{\rm ph} \{V,\Delta,\Delta^{\ast}\} \;,
\end{equation}
where
\begin{eqnarray}
  \label{S_el_ph}
  S_{\rm el-ph} \{\bar \psi^{\dagger},\bar
  \psi;V,\Delta,\Delta^{\ast}\} & \hspace{-2.3mm} = & \hspace{-2.3mm}  
  \beta \! \! \! \! \! \! \sum_{k,k',\tilde \omega_n,\tilde
    \omega_{n'},\sigma}  \! \! \! \! \! \!
  \bar \psi_{k,\tilde \omega_n,\sigma}^{\dagger}
  \left(\mathbf{G}_0^{-1} - \mathbf{V} \right)_{k,k',\tilde
      \omega_n,\tilde \omega_{n'}}
  \bar \psi_{k',\tilde \omega_{n'},\sigma} \;, \\
  S_{\rm ph} \{V,\Delta,\Delta^{\ast}\} & \hspace{-2.3mm} = &
  \hspace{-2.3mm} \frac{1}{2}\, \beta L \sum_{q,\omega_m} \frac{1}{|g_q|^2}
  \left[\frac{\omega_m^2 + \omega_q^2}{
      \omega_q} \right] V_{q,\omega_m}^{\ast} V_{q,\omega_m} \nonumber
  \\ 
  & & \hspace{-10mm} {} +  \frac{1}{c} \, \beta L
  \sum_{q,\omega_m} \frac{1}{|g_{2k_F+q}|^2} 
  \left[\frac{\omega_m^2 + \omega_{2k_F+q}^2}{
      \omega_{2k_F+q}} \right] \Delta_{q,\omega_m}^{\ast}
  \Delta_{q,\omega_m}   
  \;,
   \label{eq:Sphc}
\end{eqnarray}
and
\begin{equation}
  \label{eq:def:c}
  c \equiv \left \{ 
    \begin{array}{cl}
      2\;, & \mbox{commensurate case (half-filled band)}\;, \\
      1\;, & \mbox{incommensurate case}\;.
    \end{array} \right.
\end{equation}
While in the incommensurate case $\pm 2k_F$ lie (up to a reciprocal
lattice vector) inside the first Brillouin zone, $\pm 2 k_F$ lie
directly on the border of the first Brillouin zone in the commensurate
case. In this case the factor of $1/2$ in the last line in Eq.\
(\ref{eq:Sphc}) avoids overcounting.

\subsection{Generalized Ginzburg-Landau functional}

Since the action $S\{\bar \psi^{\dagger},\bar \psi;V,\Delta,\Delta^{\ast}\}$ 
is only Gaussian in both the Fermion and the phonon fields, either of them 
can easily be integrated out.
To derive a time-independent (generalized) Ginzburg-Landau theory which 
will enable us 
to determine the statistics of the phonon field, we ignore quantum
fluctuations, 
i.e.\ we ignore all terms involving finite bosonic frequencies. Because 
we will be only interested in static properties of the Peierls system, this
should be a reasonable approximation for not too small temperatures. 
Integrating out the fermionic degrees of freedom, the
action $S\{\bar \psi^{\dagger},\bar \psi;V,\Delta,\Delta^{\ast}\}$ turns 
into the free energy
functional 
$\beta F \{V,\Delta,\Delta^{\ast}\}$.

For the derivation of a generalized Ginzburg-Landau functional in which 
the free energy functional is expressed in terms of gradients of the order parameter field 
\begin{equation}
 \label{eq:def:FT:Delta:x}
  \Delta (x) = \sum_q e^{iqx}
  \Delta_q \;,
\end{equation}
we refer the reader to the PhD thesis of the author \cite{Bartosch00PhD}
or to Refs.\
 \cite{Werthamer63,Werthamer69,Tewordt63,Tewordt64,Kosztin98,Kos99,Bartosch99c} where the gradient-expansion is developed
in the context of superconductivity.
Up to terms of second order in the gradient, we find
\begin{eqnarray}
  \label{eq:F:gradexp}
  F \{ \Delta,\Delta^{\ast} \} & \hspace{-2.3mm} = & \hspace{-2.3mm}
  F^{(0)}  \{ \Delta,\Delta^{\ast} \} 
  + F^{(2)}  \{ \Delta,\Delta^{\ast} \} \;, \\
  \label{eq:F:gradexp0}
  F^{(0)} \{ \Delta,\Delta^{\ast} \} & \hspace{-2.3mm} = &
  \hspace{-2.3mm}  {s \rho_0} 
  \int_0^L dx\,
  \left(
    - \frac{2 \pi}{\beta} 
    \sum_{0<{\tilde \omega}_n \lesssim
 \epsilon_0} 
    \left[
      \sqrt{{\tilde \omega}_n^2 + |\Delta|^2} - {\tilde \omega}_n \right]
    + \frac{|\Delta|^2}{2\lambda} \right)
  \; , \\
  \label{eq:F:gradexp2}
  F^{(2)} \{ \Delta,\Delta^{\ast} \} & \hspace{-2.3mm} = &
  \hspace{-2.3mm}    {s \rho_0}
  \int_0^L dx\, \frac{2 \pi}{\beta} \, \sum_{{\tilde \omega}_n > 0} 
  \left[\frac{1}{8}\, \frac{|\partial_x
      \Delta|^2}{\left({\tilde \omega}_n^2 + |\Delta|^2 
      \right)^{\frac{3}{2}}} - \frac{1}{32}\,
    \frac{\left[ \partial_x 
        |\Delta|^2 \right]^2}{\left({\tilde \omega}_n^2 + |\Delta|^2
      \right)^{\frac{5}{2}}}
  \right]
  \; .
\end{eqnarray}
For $F^{(0)} \{\Delta,\Delta^{\ast}\}$ to be finite and to avoid
logarithmic divergences, the sum in Eq.\ (\ref{eq:F:gradexp0}) needs
to be regularized by an ultraviolet cutoff $\epsilon_0$.
Although the ultraviolet cutoff $\epsilon_0$ was introduced here in
the sum over Matsubara frequencies instead of as a cutoff in the
momentum integral, expanding Eqs.\ (\ref{eq:F:gradexp0}) and
(\ref{eq:F:gradexp2}) in the regime $\beta |\Delta| \ll 1$, 
we get the usual Ginzburg-Landau functional,
\begin{eqnarray}
  \label{eq:Ginzburg-Landau2}
   F \{\Delta,\Delta^{\ast}\} =  
  \frac{s \rho_0}{2} \int_0^L dx\, \Big[\, a(T)\,
    |\Delta(x)|^2 + b(T)\, |\Delta(x)|^4 + c(T)\, |\partial_x
    \Delta(x)|^2 
   \Big] \;, 
\end{eqnarray}
where the coefficients $a(T)$, $b(T)$ and $c(T)$
are given by
\begin{eqnarray}
  \label{eq:def:a}
  a(T) & \hspace{-2.3mm} = & \hspace{-2.3mm} \ln\frac{T}{T_{c}^{\rm
      MF}}\;,\quad  T_{c}^{\rm MF} = 1.134\, \epsilon_0
  \exp\left(-1/\lambda\right)\;, \\
  b(T) & \hspace{-2.3mm} = & \hspace{-2.3mm} \left(\frac{1}{4\pi 
      T}\right)^2 7 \zeta(3) \;, \\
  c(T) & \hspace{-2.3mm} = & \hspace{-2.3mm} \left(\frac{v_F}{4\pi 
      T}\right)^2 7 \zeta(3) \;.
\end{eqnarray}
Here, $\zeta(3)\approx 1.2$ is the Riemann zeta-function of $3$ and
\begin{equation}
  \lambda \equiv \frac{c\,s}{2}\, \frac{\rho_0 |g_{2k_F}|^2}{\omega_{2k_F}}
\end{equation}
is the dimensionless coupling constant. $T_{c}^{\rm MF}$ is the critical 
mean-field temperature for the Peierls distortion.
If we include spin, $s=2$, otherwise we have $s=1$.

\subsection{Breakdown of the mean-field picture}

The experimentally observed Peierls transition, including the static 
lattice distortion, 
the charge-density wave and the occurrence of a single-particle
gap can qualitatively 
already be understood in a mean-field picture. However, 
due to reduced dimensionality, fluctuations of the order parameter field
are very important and dramatically change this scenario.
The Mermin-Wagner theorem states
that these fluctuations lead to the absence of long-range order, even
at very low temperatures \cite{Mermin66}. This 
precludes a spontaneously
broken continuous symmetry.
But how can one explain the experimentally
observed charge-density wave which breaks a continuous translational
symmetry in strongly anisotropic 
materials like blue bronze \cite{Gruener94}? The answer is simply this:
These materials are quasi one-dimensional, but {\em not} strictly
one-dimensional. As we will see below, in a strictly one-dimensional
material, the correlation length $\xi(T)$ increases with decreasing
temperature, but for any finite temperature cannot approach
infinity. At very low temperatures, however, even a very weak
interchain-coupling can lead to the onset of three-dimensional order
such that the system can undergo a Peierls transition. Of course, the
transition temperature is not the mean-field
transition temperature $T_c^{\rm MF}$. Lee, Rice and Anderson
\cite{Lee73} pointed out that one should expect $T_c^{\rm 3D} \approx
\frac{1}{4} T_c^{\rm MF}$. For a derivation of an adequate
three-dimensional microscopic theory see McKenzie \cite{McKenzie95a}
and references given therein. Here, we will especially be interested 
in the temperature regime above the Peierls transition so that it 
suffices to consider only the strictly
one-dimensional case. 

\subsubsection{Correlation functions of the order parameter field}

We will now consider the fluctuations of the order parameter field
$\Delta(x)$ and calculate the correlation functions of $\Delta(x)$
which describe the phonon statistics.

\subsubsection*{Gaussian approximation}

At temperatures far above the mean-field critical temperature
$T_c^{\rm MF}$, the coefficient $a(T)$ becomes large enough such that anharmonic
corrections to the free energy functional may be
neglected. Truncating the free energy functional
(\ref{eq:Ginzburg-Landau2}) at the second order, we find
\newline
\parbox{0mm}
{\begin{eqnarray*}
\\
\end{eqnarray*}}
\parbox{7cm}
{\begin{eqnarray*}
  {\setlength{\fboxsep}{2mm} \fbox {$ \ \displaystyle  
  \begin{array}{rcl}  
    \langle \Delta(x) \rangle 
  \hspace{-2.3mm} & =  \hspace{-2.3mm} & 
 0 \;, \\
  \langle \Delta(x) \Delta^{\ast} (x') \rangle
  \hspace{-2.3mm} & = \hspace{-2.3mm} & 
  \Delta_s^2 (T)\, e^{-|x-x'|/\xi(T)} 
  \end{array} \;, $}}
\end{eqnarray*}}
\parbox{59mm}
{\begin{eqnarray}
    \label{eq:correlationDelta1} \\
   \rule[-3mm]{0mm}{5mm}
   \label{eq:correlationDelta2}
\end{eqnarray}}
\newline
where $\Delta_s^2 (T) = \frac{ T}{s \rho_0 \sqrt{a(T)\, c(T)}}$ and 
$\xi^{-1}(T) = \left( \frac{a(T)}{c(T)}\right)^{1/2}$.
While $\langle \Delta(x) \Delta (x') \rangle$ vanishes for complex $\Delta$,
it is equal to $\langle \Delta(x) \Delta^{\ast} (x') \rangle$ for real 
$\Delta$.
In the considered Gaussian approximation, higher 
correlation functions are simply
given by Wick's theorem.
The Gaussian approximation is good at sufficiently high temperatures
and is the usual approximation made when higher correlation 
functions are too complicated or not known.

\subsubsection*{Anharmonic corrections and the case of only phase 
fluctuations}

\begin{figure}[tb]
\begin{minipage}[b]{0.6\linewidth}
\begin{center}
\psfrag{n}{\small $\hspace{-3.8mm} \rho_s(T)/\rho_0$}
\psfrag{T}{\small $\hspace{-6.5mm}T/T_c^{\rm MF}$}
\psfrag{0.0}{\hspace{0.0mm}\small $0.0$}
\psfrag{0.5}{\hspace{0.0mm}\small $0.5$}
\psfrag{1.0}{\hspace{0.0mm}\small $1.0$}
%
%
\epsfig{file=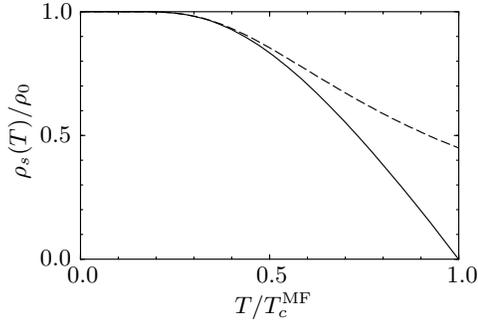,width=6.2cm}
\end{center}
\end{minipage}
\hfill
\begin{minipage}[b]{0.35\linewidth}
\caption{Plot of the superfluid density as a function of temperature
  for $\Delta_s = 1.76  T_c^{\rm MF}$ (dashed line) and $\Delta_s(T)$
  given by the BCS gap equation (\ref{eq:BCSgap}) (solid line).}
\vspace{8mm}
\label{fig:rhoofT}
\end{minipage}
\end{figure}

As the temperature is lowered and approaches the mean-field critical
temperature $T_c^{\rm MF}$, fluctuation effects become important and
the mean-field picture breaks down. However, as shown by Scalapino,
Sears and Ferrel \cite{Scalapino72} using the transfer matrix
technique, the first two moments of a one-dimensional Ginzburg-Landau
theory are still approximately given by Eqs.\
(\ref{eq:correlationDelta1}) and (\ref{eq:correlationDelta2}). For
temperatures well below $T_c^{\rm MF}$, one finds for real $\Delta(x)$
an exponential increase of the correlation length with decreasing
temperature, while for complex $\Delta(x)$ the correlation length
increases as the inverse temperature. This last result can be
understood as follows:
For small temperatures $T \ll T_c^{\rm MF}$, the generalized
Ginzburg-Landau functional has the shape of a ``Mexican hat'' and 
is dominated by its minima. Amplitude fluctuations get 
gradually frozen out and, for complex $\Delta(x)$, 
the phase of the order parameter
$\Delta(x) \approx \Delta_s e^{i\vartheta(x)}$ fluctuates at the 
bottom of the ``Mexican hat''.
Ignoring, as before, quartic terms in the gradient expansion of
the free energy, the free energy is given up to an irrelevant constant
by
\begin{equation}
  \label{eq:freeenergyphasefluctuations}
  F^{(\rm phase)} \{ V \} = F^{(\rm phase)} \{ \partial_x \vartheta/2 \} 
  =
  \frac{1}{2} \, s
  \rho_s(T) \int_0^L dx\, V^2(x) \;.
\end{equation}
In analogy to the theory of superconductivity,
\begin{equation}
  \label{eq:superfluidvelocity}
  V(x) = \partial_x \vartheta(x)/2
\end{equation}
can be interpreted (up to a constant $1/m^{\ast}$) as the superfluid
velocity and 
\begin{equation}
  \label{eq:nT}
  \rho_s (T) = \rho_0\, \frac{2 \pi}{\beta} \, \sum_{{\tilde \omega}_n > 0} 
  \frac{\Delta_s^2}{\left({\tilde \omega}_n^2 + |\Delta_s|^2 
    \right)^{\frac{3}{2}}}
\end{equation}
can be interpreted as the superfluid density. Formally, 
the free energy is identical to the kinetic energy of a superflow. A
two-dimensional analogue of Eq.\
(\ref{eq:freeenergyphasefluctuations}) has been used by Emery and
Kivelson \cite{Emery95a,Emery95b} in their theory describing
superconductors with a small phase-stiffness.
At $T=0$, the sum in Eq.\ (\ref{eq:nT}) turns into an integral
which can be done analytically and gives $\rho_s (0) = \rho_0$,
i.e.\ at $T=0$, the superfluid density is equal to the density of
states. Plots of $\rho_s(T)$ for $\Delta_s (T)$ given by 
the BCS gap equation \cite{Abrikosov63,Schrieffer64,Tinkham96,Bartosch00PhD} 
\begin{equation}
  \label{eq:BCSgap}
  \frac{1}{\lambda} = \frac{2\pi}{\beta} \sum_{0<\tilde \omega_n \lesssim
    \epsilon_0} \frac{1}{\sqrt{\tilde \omega_n^2 + \Delta_0^2(T)}} \;,
\end{equation}
and for $\Delta_s = 1.76
 T_c^{\rm MF}$ 
are shown in Fig.\ \ref{fig:rhoofT}.

Since $F^{(\rm phase)} \{ V \} $ is only quadratic in $V(x)$, correlation
functions of $\Delta(x)$ can easily be calculated. The first two 
moments are given by Eqs.\ (\ref{eq:correlationDelta1}) and 
(\ref{eq:correlationDelta2}), where
\begin{equation}
\label{eq:xiofT:phase}
  {\setlength{\fboxsep}{2mm} \fbox {$ \ \displaystyle 
  \xi(T) = \frac{s \rho_s(T)}{2 T} \;.$}}
\end{equation}
For 
$T \lesssim T_c^{\rm MF}/4$ 
we have $\rho_s(T) \approx \rho_s(0) = 1/\pi$, such that in this strictly 
one-dimensional theory we find $\xi (T) = s/2\pi  T \propto 1/T$
which for fermions with spin $1/2$ agrees with Gr\"uner's
\cite{Gruener94} result $\xi (T) = 1/ \pi  T$.

\subsection{The Hamiltonian of the fluctuating gap model}

The fluctuating gap model (FGM) describes electrons subject to a static 
disorder potential which can be seen as an approximation of a 
phonon field.
For a particular realization of the disorder, the electronic
part of the action given in Eq.\ (\ref{S_el_ph}) corresponds to the 
Hamiltonian
\begin{equation}
  \label{eq:HamiltonianFGM2nd}
  \mathcal{H}_{\rm } =
  \sum_{k,k'} \left(c_{+,k}^{\dagger} \, , \,  c_{-,k}^{\dagger}
  \right) H_{k,k'} 
  { \, c_{+,k'} \,  \choose \, c_{-,k'}\, } \;,
\end{equation}
where
\begin{equation}
 H_{k,k'} =  \left(
    \begin{array}{cc}
      v_F k \, \delta_{k,k'} + V_{k-k'} & \Delta_{k-k'} \\
      \Delta_{-(k-k')}^{\ast} & - v_F k \, \delta_{k,k'} + V_{k-k'}
    \end{array}
  \right) \;.
\end{equation}
A Fourier transformation leads to
\begin{equation}
  \label{eq:HamiltonianFGM2nd:FT}
  \mathcal{H}_{\rm } =
  \int_0^L dx\, \left(\psi_{+}^{\dagger}(x) \, , \, \psi_{-}^{\dagger} (x)
  \right) \hat H (x,-i\partial_x) 
  { \, \psi_{+} (x) \,  \choose \, \psi_{-} (x) \, } \;,
\end{equation}
with
\begin{equation}
 \hat{H} (x,-i\partial_x) =   
 -  i v_F  \partial_x \sigma_3 + V ( x ) \sigma_0
 +
 \Delta ( x ) \sigma_{+} + \Delta^{\ast} ( x ) \sigma_{-} 
 \label{eq:Hamiltonian}
 \; .
\end{equation}
This is the Hamiltonian of the FGM. 
$\sigma_{i}$ are the usual Pauli matrices, $\sigma_0$ is the
$2 \times 2$ unit matrix, and
$\sigma_{\pm} = \frac{1}{2}(\sigma_{1} \pm i \sigma_{2})$. 
Recall
that we have linearized the energy dispersion such that the FGM can
only describe the low-energy physics of Peierls chains in the
weak-coupling regime. As a further approximation, we have considered
the phonon field to be static. It will now be our aim to calculate
disorder-averaged quantities for the model described by this
Hamiltonian. As we will 
discuss in Section \ref{section:finite_xi}, instead of averaging over
the disorder, it is also possible to consider a typical realization of
the disorder potential.

\subsubsection{The fluctuating gap model in other physical contexts} 

In this section, the fluctuating gap model (FGM) emerged as an effective
low-energy model to describe quasi one-dimensional materials which
undergo a Peierls transition. Our strictly one-dimensional theory
applies to temperatures above the Peierls transition before
three-dimensional fluctuations become important and eventually lead to
a phase transition. Formally, the Hamiltonian of the FGM is of the 
Dirac type and describes electrons in a disordered potential. In the
language of relativistic quantum field theory, the backscattering
potential can be interpreted as a random (and complex) mass.
The FGM has also applications in other fields of physics.
As shown in Refs.\ \cite{McKenzie96a,Bunder99}, the 
Hamiltonian of disordered spin Peierls systems
\cite{Gruener94,Gruener94b,Fabrizio97,Steiner98,Steiner99} can be mapped
by a Jordan-Wigner transformation onto the Hamiltonian of the FGM. 
In a semiclassical approximation of superconductivity, it is also
possible to replace the original three-dimensional problem by a
directional average over effectively one-dimensional problems
\cite{Waxman93} which in the weak coupling limit are described by the
FGM. This method has been used in Refs.\
\cite{Bartosch99c,Kos99,Kosztin98} to derive the gradient expansion of
a clean superconductor.
A generalization of the FGM towards higher dimensions to describe the
phase above the phase-transition in underdoped high-$T_c$
superconductors by anti-ferromagnetic short-range order fluctuations was 
considered in Refs.\ \cite{Schmalian98,Schmalian99,Sadovskii01}.

 \section{The Green function of the fluctuating gap model 
and related quantities} 
\label{section:formalism}

{\em In this section, we will introduce different concepts to calculate
  the Green function and related quantities of the fluctuating gap
  model. The density of states and the
  localization length will be of special interest. In particular, we
  will develop a non-perturbative method which allows to 
  calculate these quantities simultaneously for arbitrary given
  disorder potentials.}

\subsection{The retarded Green function}

In the following, we are going to consider the retarded Green function
$\mathcal{G}^R (x,x^{\prime};\omega)$ of the
fluctuating gap model. This retarded Green function is of special 
interest because
it can be related to several quantities which are in principle
experimentally accessible. The trace of the imaginary part of the Green
function at coinciding space points determines the {\em local} density
of states (DOS),
\begin{equation}
  \label{def:localDOS}
   \rho ( x , \omega ) = - {\pi}^{-1}
   {\rm Im} {\rm Tr} [ {\cal{G}}^{R} ( x , x ; \omega ) ] \; .
\end{equation}
Averaging $\rho ( x , \omega )$ over all space points
gives the DOS $\rho(\omega)$, which is the fundamental
quantity that determines the whole thermodynamics of the FGM. It will
turn out that the trace of the  
energy-integrated space averaged Green function at coinciding space
points $\Gamma(\omega)$ will be easier to calculate than its
non-integrated form. While its imaginary part 
is proportional to the integrated DOS $\mathcal{N}(\omega)$, the
Thouless formula states that $\textrm{Re}\, \Gamma(\omega)$ 
is equal to the inverse localization length $\ell^{-1}(\omega)$.
As in the following sections, we will usually only consider the DOS
and the inverse localization length for positive frequencies
$\omega$. 
Due to particle-hole
symmetry, the DOS and the localization length are symmetric with
respect to the Fermi energy so that 
after setting the Fermi energy equal to zero we have $\rho(\omega) =
\rho(-\omega)$ and $\ell^{-1}(\omega)=\ell^{-1}(-\omega)$. It
therefore suffices to 
consider the case $\omega > 0$.

The retarded $2 \times 2$ matrix Green function 
${\cal{G}}^{R} ( x , x' ; \omega)$ to the Schr\"odinger operator
$\omega - \hat H$
satisfies the differential equation
\begin{equation}
  \label{Greenfunction}
  [\omega + i0^{+} - \hat H(x,-i\partial_x)]\ \mathcal{G}^R
  (x,x^{\prime};\omega) = \delta ( x -
  x^{\prime}) \sigma_0 \; .
\end{equation}
The positive but infinitesimal imaginary part added to the frequency
$\omega$ indicates that we have to impose the correct boundary conditions applying to
a retarded Green function.

\subsubsection{Free fermions}

It is easy to calculate the Green function for free fermions. 
In this case, $V(x) = \Delta(x) = 0$, such that the system is
translational invariant, and Eq.\ (\ref{Greenfunction}) simplifies 
to\footnote{Besides $\hbar$ and $k_B$, we also set the Fermi velcity
$v_F$ equal to one.}
\begin{equation}
  [\omega +i0^{+} + i\sigma_3 \partial_x 
  ] \ \mathcal{G}_{0}^{R} (x-x';\omega) =
  \sigma_0 \delta (x-x') \; .
\end{equation}
Taking the Fourier transform of this equation from real space to
momentum space gives
\begin{equation}
[\omega +i0^{+} - k \sigma_3 ]
\ \mathcal{G}_{0}^{R} (k;\omega) =
\sigma_0 \; .
\end{equation}
$\mathcal{G}_{0}^{R} (k;\omega) \equiv \int dx\ e^{-ikx}
\mathcal{G}_{0}^{R} (x;\omega)$ can now be found by a simple
matrix inversion. If $\alpha = 1$ accounts for right- and $\alpha =
-1$ for left-moving fermions, the matrix elements of
$\mathcal{G}_{0}^{R} (k;\omega)$ are given by 
\begin{equation}
  \left(\mathcal{G}_{0}^{R}\right)_{\alpha \alpha'} (k;\omega) =
  \frac{\delta_{\alpha,\alpha'}}{\omega - \alpha k +i0^{+}} \; . 
\end{equation}
A simple Fourier transformation back to real space now gives the
retarded propagator of free 
fermions in real space,
\begin{equation}
  i \label{G_0,realspace}
  \left(\mathcal{G}_{0}^{R}\right)_{\alpha \alpha'} (x;\omega) =
  \delta_{\alpha,\alpha'} \theta(\alpha x) e^{i\alpha \omega x} \; .
\end{equation}
Here, $\theta(x)$ is the Heaviside step function
\begin{equation}
  \label{def:Heaviside}
  \theta(x) = \left\{ \begin{array}{ll} 0\, & , \quad x<0 \\ 1\,  & , \quad 
      x>0 \end{array} 
    \right. \; .
\end{equation}
For concreteness, let us also define $\theta(0) = \lim_{x \to 0}
[\theta(x) + \theta(-x)]/2 = 1/2$.
While the matrix elements at $x=0$ are sensitive to the definition
$\theta(0) = 1/2$ which amounts to defining 
$
  \mathcal{G}^{R} (x=0;\omega) \equiv \frac{1}{2} \lim_{x \to 0^{+}}
  \left[\mathcal{G}^{R} (+x;\omega) + \mathcal{G}^{R} (-x;\omega)
  \right]$, 
the local DOS
$\rho(x,\omega) = - {\pi}^{-1} 
{\rm Im} {\rm Tr} [ {\cal{G}}_0^{R} (0 ; \omega ) ]$ does not depend on
this definition because it only involves
the harmless quantity 
$\theta(x)+\theta(-x) \equiv 1$. Due
to translational symmetry, the total DOS is equal to the space-independent
local DOS, 
\begin{equation}
  \label{eq:freeDOS}
  \rho_0 (\omega ) = {\pi}^{-1} \; .
\end{equation}
Note that the DOS of free fermions is independent of the frequency
because we have linearized the energy dispersion.

A further Fourier transformation of Eq. (\ref{G_0,realspace}) from
frequency to time gives 
\begin{equation}
  \label{G_0,spacetime}
  i\left(\mathcal{G}_{0}^{R}\right)_{\alpha \alpha'} (x;t) =
  \delta_{\alpha,\alpha'} \theta(t) \delta (\alpha x - t) \; .
\end{equation}
This free retarded Green function in space and time allows for a
simple interpretation: A fermion put into the system at $t'=0$ as a
right- or left-mover will at time $t>0$ have traveled a distance
$|x|=t=v_F t$ in the positive or negative direction, respectively. The
fermion can not be observed in the system at times $t<0$.

\subsection{Dyson equation and perturbation theory}

\begin{figure}[tb]
\begin{center}
\epsfig{file=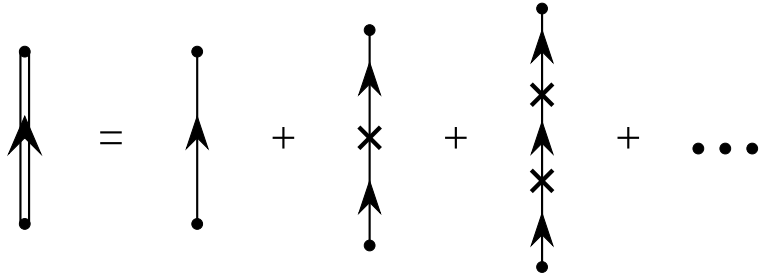,width=10cm}
\end{center}
\caption{Diagrammatic expansion of the matrix Green function. While the
  single line represents the Green function of free fermions, the
  double line is a graphical representation of the full Green function
  $\mathcal{G}^R (x,x';\omega)$. The crosses denote the disorder
  potential ${\bf V}(x)$.} 
\label{fig:Feynman}
\end{figure}

One way to handle the disorder is to consider the disorder potential
as a perturbation and expand the Green function in powers of this
potential. Defining ${\bf V}(x) \equiv V(x) \sigma_0 + \Delta(x) \sigma_{+}
+ \Delta^{\ast} (x) \sigma_{-}$, Eq.\ (\ref{Greenfunction}) may be
written as
\begin{equation}
  [i\sigma_3 \partial_x + \omega + i0^{+} ]\ \mathcal{G}^R
  (x,x^{\prime};\omega) = \delta ( x -
  x^{\prime}) \ \sigma_0 + {\bf V}(x) \ \mathcal{G}^R
  (x,x^{\prime};\omega) \; .
\end{equation}
Substituting $x$ by $x_1$, multiplying the resulting equation from the
left with the free Green function $\mathcal{G}_{0}^{R} (x-x_1;\omega)$ and
then integrating over $x_1$ gives the {\em Dyson equation}
\begin{equation}
  \label{Eq:Dyson}
  \mathcal{G}^R (x,x^{\prime};\omega)  = \mathcal{G}_0^R
  (x-x';\omega)  + \int dx \ \mathcal{G}_0^R (x - x_1;\omega) \ {\bf V}(x_1)
  \ \mathcal{G}^R (x_1,x^{\prime};\omega) \; .
\end{equation}
Iterating this Dyson equation, the exact Green function can be
expressed in terms of the free Green function and the disorder
potential:
\begin{equation}
  \label{Greenfunctionexpansion}
  \mathcal{G}^R (x,x^{\prime};\omega)  = \sum_{n=0}^{\infty}
  \mathcal{G}_{n}^R (x,x^{\prime};\omega) \; ,
\end{equation}
where $\mathcal{G}_{0}^R (x,x^{\prime};\omega) = \mathcal{G}_{0}^R
(x-x^{\prime};\omega)$ is the free Green 
function calculated above, and for $n \ge 1$ the functions
$\mathcal{G}_{n}^R (x,x^{\prime};\omega)$ are given by 
\begin{eqnarray}
 \mathcal{G}_{n}^R (x,x^{\prime};\omega) & \hspace{-2.3mm} = & \hspace{-2.3mm} 
 \int dx_1 \dots \int dx_n  \nonumber \\ 
 & & \hspace{-2.2cm} \label{expansion_nth-order}
  \quad \mathcal{G}_0^R (x - x_n;\omega) \ {\bf V}(x_n)
 \  \mathcal{G}_0^R (x_n - x_{n-1};\omega) \ \dots \ {\bf V}(x_1) \  
 \mathcal{G}_0^R (x_1 - x';\omega) \; . 
 \label{eq:Gn}
\end{eqnarray}
Recall that the right-hand side of this equation involves the product
of $2 \times 2$-matrices.
The perturbative expansion of the full Green function can be
visualized by using Feynman diagrams (see Fig. \ref{fig:Feynman}).

The physical interpretation of the perturbation expansion is simple:
The perturbative expansion takes into account all possibilities of a
particle moving through the sample getting scattered
at the various static impurities. While $\Delta(x)$ changes the
direction in which the particle travels and therefore can be
interpreted as a backscattering potential, $V(x)$ does not change the
direction of the particle, so that it only leads to forward
scattering.

\subsubsection{Boundary conditions of the retarded Green function}

Below, we will consider a non-perturbative approach to
calculate the Green function of the FGM. The above
perturbative expansion can be used to obtain the 
correct boundary conditions of the full retarded Green function: Let
us consider $\left(\mathcal{G}_n^R\right)_{\alpha \alpha'}
(x,x';\omega)$. According to Eq.\ (\ref{eq:Gn}), its expansion in a
product of free Green functions and the static impurities starts with
$\left(\mathcal{G}_0^R\right)_{\alpha \alpha} (x - x_n;\omega)$ and
ends with $\left(\mathcal{G}_0^R\right)_{\alpha' \alpha'} (x_1 -
x';\omega)$. These terms are proportional to $\theta(\alpha 
(x-x_n))$ and $\theta(\alpha' (x_1 -x'))$, respectively, so that 
$\left(\mathcal{G}_n^R\right)_{\alpha \alpha'} (x,x';\omega)$ has to
vanish as $\alpha x \to -\infty$ or $\alpha' x' \to \infty$. Since
this reasoning applies to all orders in perturbation theory, it also
applies to the full Green function.
If we demand the potentials to vanish outside the interval
$[-\Lambda,L+\Lambda]$, the boundary condition can also be written as  
\begin{equation}
  \label{BCfiniteLambda}
  \left(\mathcal{G}^R\right)_{\alpha \alpha'} (-\alpha
  \Lambda,x';\omega) = 0 \;, \quad 
  \left(\mathcal{G}^R\right)_{\alpha \alpha'} (x,\alpha'
  (L+\Lambda);\omega)  = 0 \; . 
\end{equation}

\subsubsection{Second order Born approximation}

\begin{figure}[t]
\begin{center}
\epsfig{file=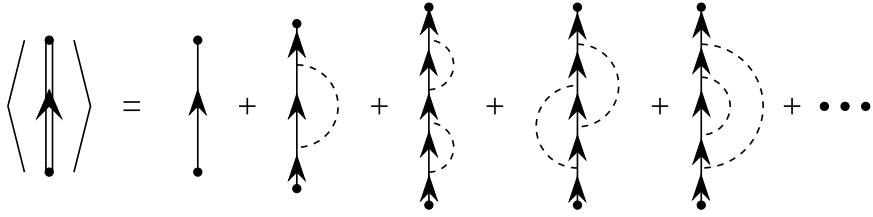,width=11.50cm}
\end{center}
\caption{Diagrammatic representation of the averaged matrix Green
  function. The 
  single line represents the Green function of free fermions and in
  this case the (averaged) 
  double line is a graphical representation of the full (averaged) Green
  function 
  $\langle \mathcal{G}^R (x,x';\omega) \rangle$. The dashed line
  denotes the disorder average $\langle {\bf V}(x) {\bf V}(x') \rangle$.} 
\label{fig:averagedG}
\end{figure}

\begin{figure}[b]
\begin{center}
\psfrag{S}{\Huge $\hspace{-1mm}\vspace{0mm} \Sigma$}
\epsfig{file=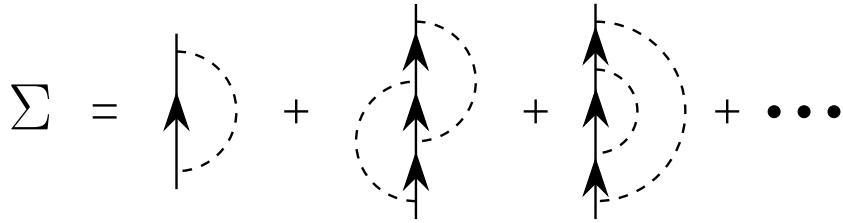,width=11cm}
\end{center}
\caption{Diagrammatic representation of the (irreducible)
  self-energy. As in the above Figure \ref{fig:averagedG},
  the single line represents the Green
  function of free fermions and the dashed line
  denotes the disorder average $\langle {\bf V}(x) {\bf V}(x') \rangle$.} 
\label{fig:selfenergy}
\end{figure}

Let us now consider the disorder-averaged Green function. As discussed in
Section \ref{chap:FGM}, above the Peierls transition, the first two
moments of the order parameter field $\Delta(x)$ are
given by $\langle \Delta(x) \rangle = 0$ and
$\langle \Delta(x) \Delta^{\ast} (x')\rangle = \Delta_s^2
  e^{-|x-x'|/\xi}$.
Following Lee, Rice and Anderson \cite{Lee73}, we ignore the forward
scattering disorder, i.e.\ set $V(x) \equiv 0$. For 
a perturbative approach we assume Gaussian statistics for the higher
moments of the order parameter field such that these moments can be
separated according to Wick's 
theorem. A diagrammatic representation of the averaged Green function
is shown in Fig.\ \ref{fig:averagedG}.
An infinite number of diagrams can be summed up by introducing
{\em irreducible diagrams} which by definition cannot be separated into two
disconnected diagrams by cutting a single propagator. The
corresponding {\em amputated diagram} is obtained by eliminating all
outer propagators. The sum of all amputated irreducible diagrams is
known as the {\em self-energy} and is diagrammatically presented in
Fig.\ \ref{fig:selfenergy}.
In terms of the self-energy, the averaged Green function reads in
momentum space
\begin{equation}
  \label{eq:GreenfunctionSelfenergy}
  \left\langle\mathcal{G}^{R}(k;\omega) \right\rangle =
  \left[\left(\mathcal{G}_0^{R}(k;\omega)\right)^{-1} - 
    \Sigma(k;\omega) \right]^{-1} \;. 
\end{equation}
The simplest approximation to take into account fluctuation effects of
the order parameter is to consider only the first diagram in
Fig.\ \ref{fig:selfenergy}. This approximation is known as the second
order Born approximation and is essentially the approximation made by
Lee, Rice and Anderson in their seminal paper \cite{Lee73} in which
fluctuations of the order parameter of the FGM were taken into account
for the first time. A special non-Gaussian probability
distribution of $\Delta(x)$ involving only phase fluctuations for
which the second order Born 
approximation turns out to be exact is presented in Ref.\ \cite{Bartosch00b}.

\subsubsection{The self-energy}

Since $\langle {\bf V}(x) \rangle = 0$, the self-energy in the second order
Born approximation is given by
\begin{equation}
  \label{eq:selfenergyBorn}
  \Sigma_B (x-x';\omega) = \langle {\bf V}(x) \mathcal{G}_0^{R} (x-x';\omega) {\bf V}(x')
  \rangle  \;.
\end{equation}
Placing Eq.\ (\ref{G_0,realspace}) into this equation, we get
\begin{eqnarray}
  \label{eq:selfenergyBornII}
  \left(\Sigma_B \right)_{\alpha \alpha'} (x-x';\omega) & \hspace{-2.3mm} = &
  \hspace{-2.3mm}  
  \delta_{\alpha,\alpha'} \, \Delta_s^2 \, e^{-|x-x'|/\xi}
  \left(\mathcal{G}_0^{R} \right)_{\bar \alpha, \bar \alpha} (x-x';\omega)
  \;. \nonumber \\
  & \hspace{-2.3mm} = & \hspace{-2.3mm}  
  - i \delta_{\alpha,\alpha'} \, \Delta_s^2 \, \theta(-\alpha(x-x'))
  e^{-i\alpha [\omega+i/\xi](x-x')} \;.
\end{eqnarray}
As one should expect, the process of averaging restored translational
invariance. Taking the Fourier transform of Eq.\
(\ref{eq:selfenergyBornII}), we arrive at 
\begin{eqnarray}
  \label{eq:selfenergyBornIII}
  \left(\Sigma_B \right)_{\alpha \alpha'} (k;\omega) =
  \int dx \,
  e^{-ikx} \left(\Sigma_B \right)_{\alpha \alpha'} (x;\omega) =
  \delta_{\alpha,\alpha'}\, \frac{\Delta_s^2}{\omega + \alpha k +
    i/\xi} \;. 
\end{eqnarray}
Within the second order Born approximation, we therefore find for the
one-particle Green 
function
\begin{equation}
  \label{eq:G2ndBorn}
  \left(\mathcal{G}_B^{R}\right)_{\alpha,\alpha'} (k;\omega) =
  \frac{\delta_{\alpha,\alpha'}}{
    \omega-\alpha k -
    \frac{\Delta_s^2}{\omega +\alpha k +i/\xi}} \;.
\end{equation}
This result was first obtained by Lee, Rice and Anderson \cite{Lee73}.

\subsubsection{The density of states and the inverse localization length}

\begin{figure}[tb]
\begin{center}
\psfrag{omega}{\small \hspace{-0.4mm}$\omega / \Delta_s$}
\psfrag{dos}{\hspace{-1.0mm}\small $\rho(\omega)/\rho_{0}$}
\psfrag{ll}{\small \hspace{-5.5mm}$\ell^{-1}(\omega)/\Delta_s$}
\psfrag{0.2}{\tiny $0.2$}
\psfrag{0.50}{\tiny $0.5$}
\psfrag{1.0}{\tiny $1.0$}
\psfrag{2.0}{\tiny $2.0$}
\psfrag{10.0}{\tiny $10$}
\psfrag{inf}{\tiny $\infty$}
\psfrag{0}{\small $0$}
\psfrag{0.5}{\small $0.5$}
\psfrag{1}{\small $1$}
\psfrag{1.5}{\small $1.5$}
\psfrag{2}{\small $2$}
\psfrag{3}{\small $3$}
\epsfig{file=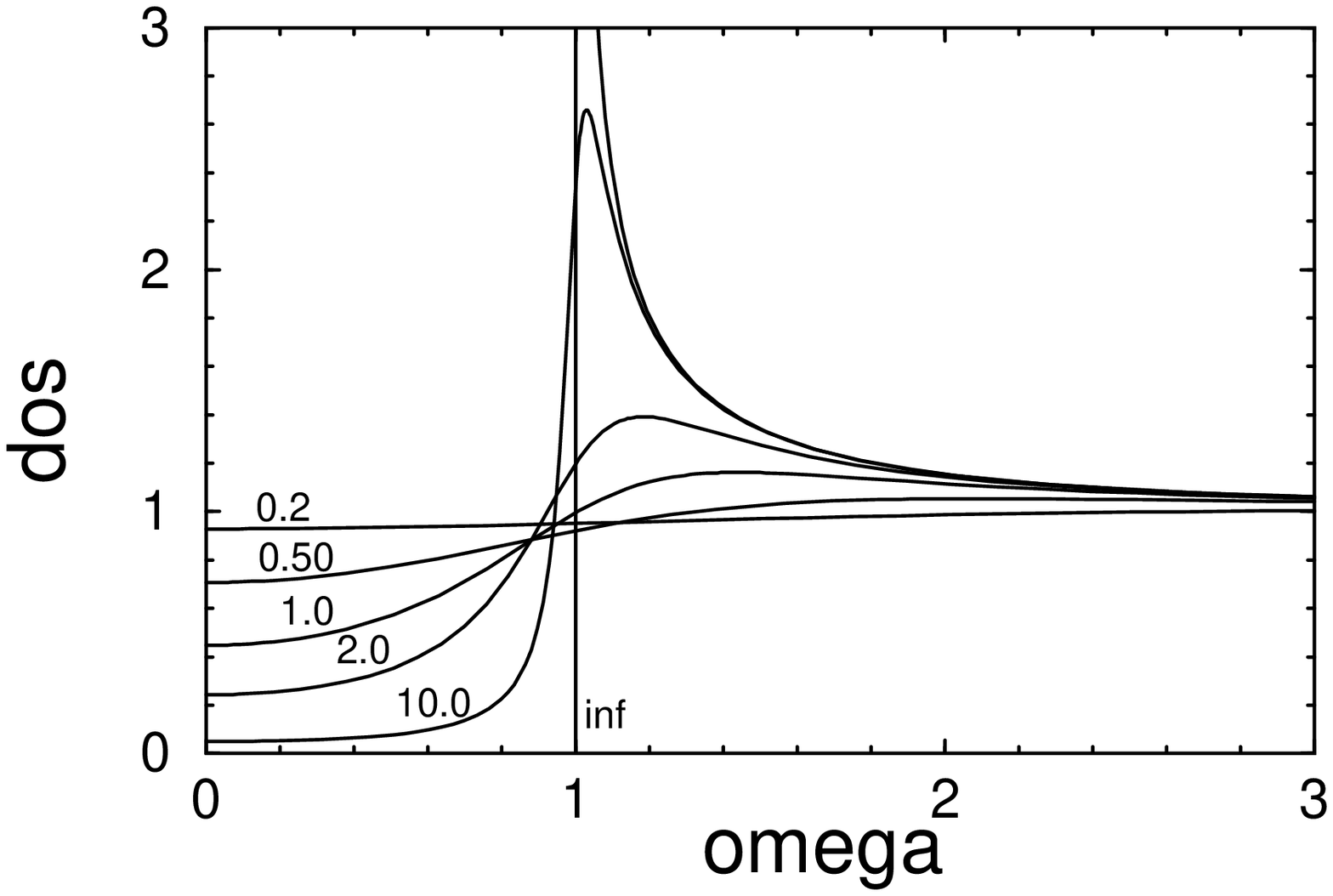,width=6.20cm}
\hfill
\epsfig{file=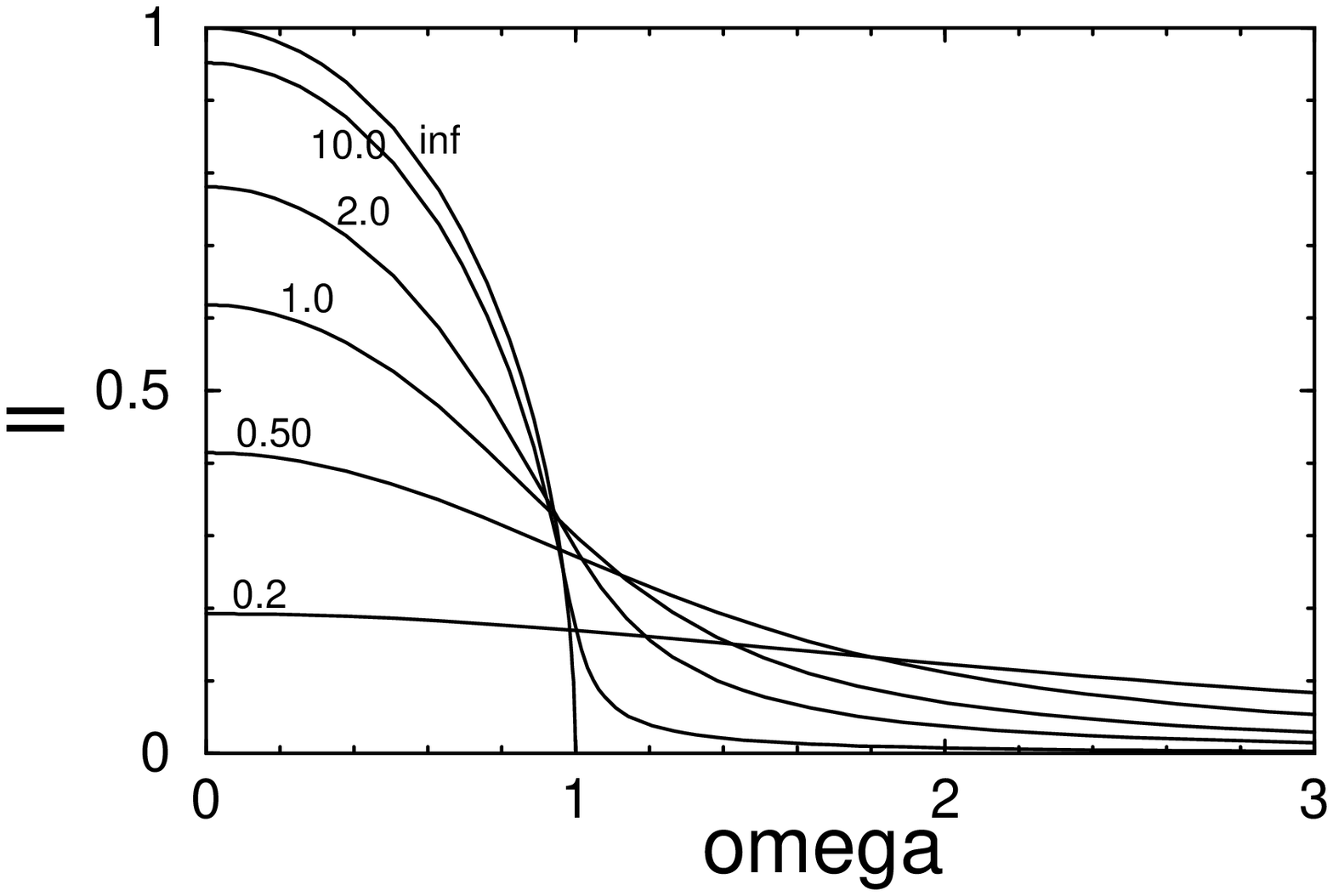,width=6.20cm}
\end{center}
\caption{Plot of the DOS $\rho_B(\omega)$ and the inverse localization length
  $l_B^{-1}(\omega)$ calculated for 
   $\Delta_s \xi=0.2,0.5,1.0,2.0,10.0$, and $\infty$ (mean-field
   result) in the 
   second order Born approximation.}
\label{fig:DOS_Born}
\end{figure}

Integrating Eq.\ (\ref{eq:G2ndBorn}) over $k$ and taking the trace, we
obtain 
\begin{equation}
  \label{eq:TrG_B}
  \textrm{Tr}\, \mathcal{G}_B^{R}(x,x;\omega) = -i
  \frac{\omega+i/2\xi}{\sqrt{(\omega+i/2\xi)^2 -\Delta_s^2}} \;,
\end{equation}
where $\sqrt{z}$ is defined as the principal part of the square root
with the cut chosen along the negative real axis. 
The imaginary part of Eq.\ (\ref{eq:TrG_B}) gives the (averaged) DOS,
\begin{equation}
  \label{eq:averagedDOS_B}
  \rho_B (\omega) = \rho_0\, \textrm{Re}\, \frac{\omega +
    i/2\xi}{\sqrt{(\omega+i/2\xi)^2 -\Delta_s^2}} \;.
\end{equation}
As we will show in Subsection \ref{sec:Thouless}, the real part of the Green
function is equal to the derivative of the inverse localization
length $\ell^{-1}(\omega)$.
Integrating this equation with respect to $\omega$ and setting the
integration constant at infinity equal to zero, we obtain
\begin{equation}
  \label{ReG_BII}
  \ell_B^{-1} (\omega) = \textrm{Im}\, \sqrt{(\omega+i/2\xi)^2
    -\Delta_s^2} -1/2\xi \;.
\end{equation}
A plot of both $\rho_B(\omega)$ and $\ell_B^{-1} (\omega)$ is shown for
different values of the dimensionless parameter 
$\Delta_s \xi$ in Fig.\ \ref{fig:DOS_Born}. 
With increasing correlation length $\xi$, the DOS gets more and more
suppressed for $\omega \lesssim \Delta_s$. However, instead of a real
gap the 
fluctuations can only create a pseudogap. At $\omega=0$, the DOS is
given by 
\begin{equation}
  \label{eq:DOS_B:0}
  \rho_B(0) = \rho_0 \frac{1}{\sqrt{1+(2\Delta_s \xi)^2}}\;,
\end{equation}
such that for $\Delta_s \xi \gg 1$, the DOS vanishes as
\begin{equation}
  \label{eq:DOS_B:0:asymptotics}
  \rho_B(0) \sim \frac{\rho_0}{2\Delta_s \xi} \propto \frac{1}{\Delta_s
    \xi} \;.
\end{equation}
As the
correlation length approaches infinity, the DOS assumes the mean-field result
\begin{equation}
    \rho^{\rm MF} (\omega) = 
  \rho_0 \frac{|\omega|}{\sqrt{\omega^2 -\Delta_0^2}} \,
  \theta \left(\omega^2 -\Delta_0^2 \right) \;,
  \label{eq:DOSmeanfield}
\end{equation}
At zero frequency, the inverse localization length
is given by 
\begin{equation}
  \label{eq:LL0_B}
  \ell_B^{-1}(0) = \Delta_s \left[\sqrt{1+\left({1}/{2\Delta_s
          \xi}\right)^2} - {1}/{2\Delta_s
          \xi} \right] \;.
\end{equation}

\subsubsection{The spectral function}

\begin{figure}[tb]
\begin{center} 
\psfrag{omega}{\hspace{-0.2mm}\small$\omega / \Delta_s$}
\psfrag{spectralf}{\small\hspace{-3.5mm}$\rho_B(k_F;\omega)/\rho_0
  \Delta_s^{-1}$} 
\psfrag{spectralfII}{\small\hspace{-5.5mm}$\rho_B(k_F+k;\omega)/\rho_0
  \Delta_s^{-1}$} 
\psfrag{0.2}{\tiny $0.2$}
\psfrag{0.50}{\tiny $0.5$}
\psfrag{1.0}{\tiny $1.0$}
\psfrag{2.0}{\tiny $2.0$}
\psfrag{10.0}{\tiny $10.0$}
\psfrag{0}{\small $0$}
\psfrag{1}{\small $1$}
\psfrag{2}{\small $2$}
\psfrag{3}{\small $3$}
\psfrag{4}{\small $4$}
\psfrag{5}{\small $5$}
\psfrag{6}{\small $6$}
\psfrag{-1}{\hspace{-2mm}\small $-1$}
\psfrag{-2}{\hspace{-2mm}\small $-2$}
\psfrag{-3}{\hspace{-2mm}\small $-3$}
\epsfig{file=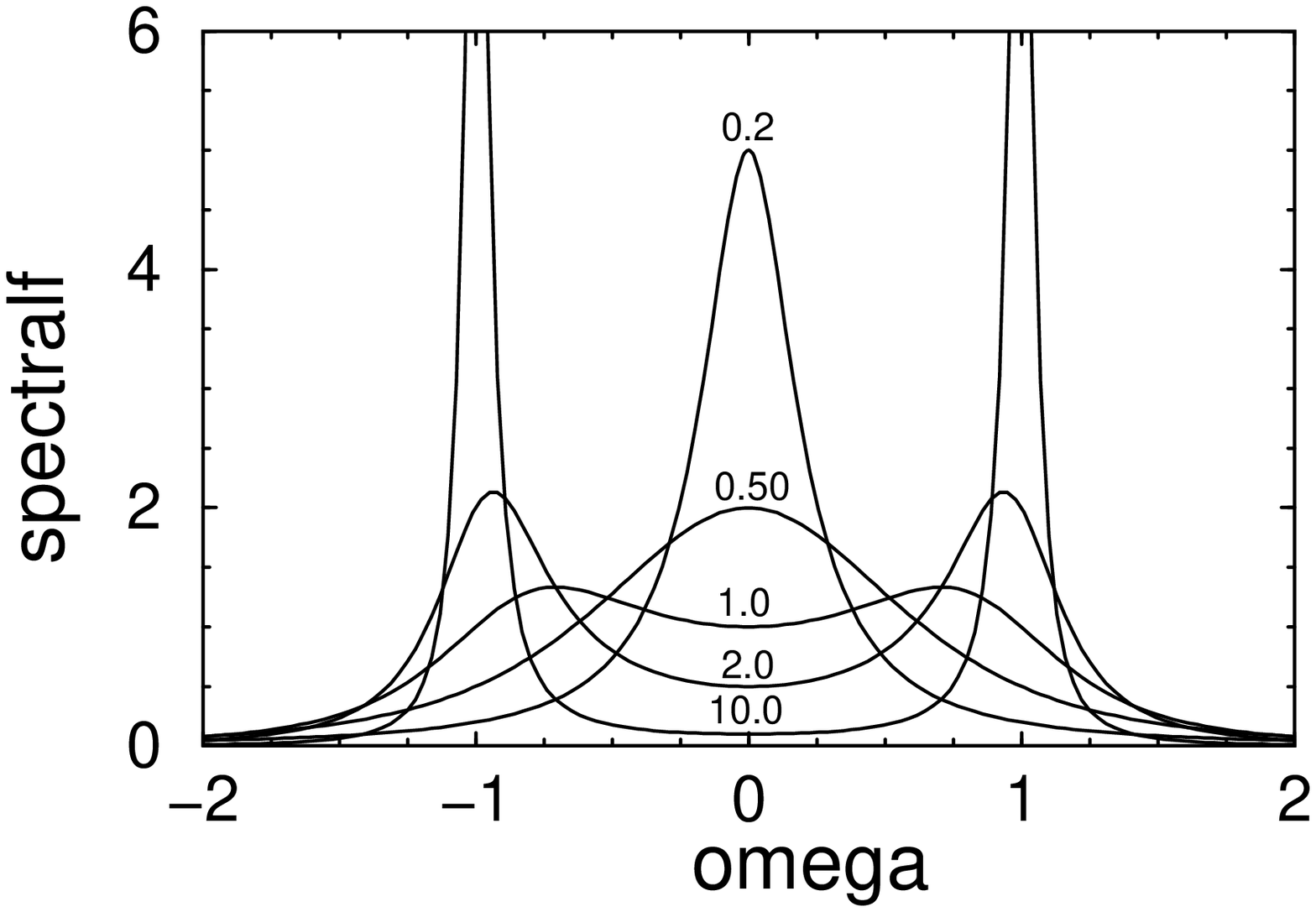,width=6.2cm}
\hfill
\epsfig{file=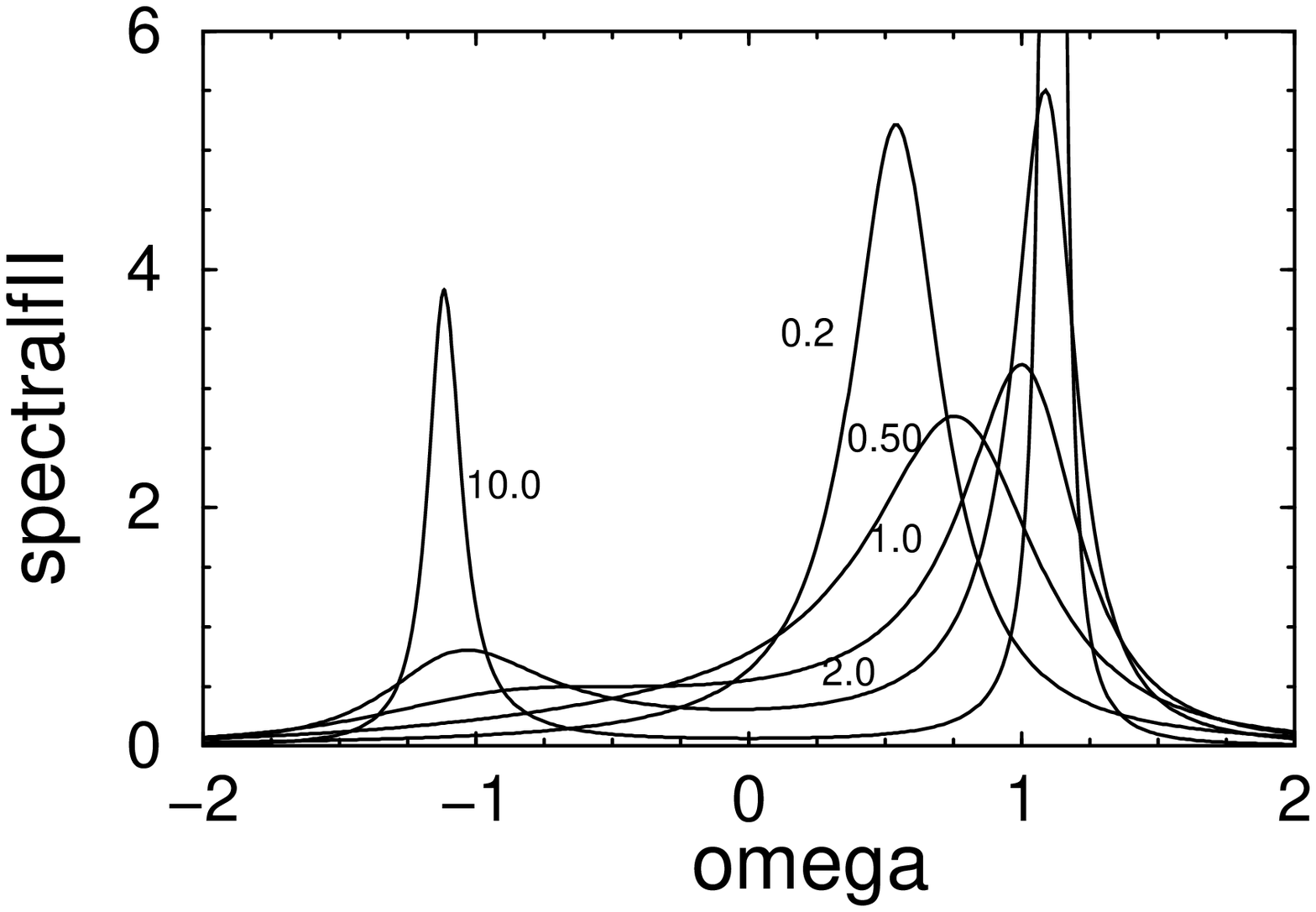,width=6.2cm}
\end{center}
\vspace{-5mm}
\caption{Plot of the spectral function $\rho_B(k_F + k;\omega)$ 
  with $\Delta k = 0.0$ (left-hand side) and $0.5$ (right-hand side)
  calculated for $\Delta_s \xi=0.2,0.5,1.0,2.0,10.0$ in the 
  second order Born approximation.}
\label{fig:spectralfunction_Born}
\end{figure}

Another interesting quantity related to the single-particle Green
function is the spectral function
\begin{equation}
  \label{eq:def:spectralfunction}
  \rho(\alpha k_F +k ;\omega) = -\frac{1}{\pi} \, \textrm{Im} \left(
    \mathcal{G}^{R} (k;\omega) \right)_{\alpha,\alpha} \;.
\end{equation}
Experimentally, the spectral function can be measured by angular
resolved photoemission spectroscopy (ARPES).
It follows from Eq.\ (\ref{eq:G2ndBorn}) that in the Born
approximation, the spectral function is given by
\begin{equation}
  \label{eq:spectralfunction_Born}
  \rho_B (\alpha k_F +k ;\omega) = \rho_0 \, 
  \frac{\Delta_s^2 \xi}{\left(\Delta_s^2 -(\omega^2-k^2) \right)^2
    \xi^2 + (\omega - \alpha k)^2} \;.
\end{equation}
Plots of the spectral function $\rho_B (\alpha k_F ;\omega)$ and
$\rho_B (\alpha k_F + k ;\omega)$ with $k = 0.5
\Delta_s$ as functions of $\omega$ are shown for different values of
$\Delta_s \xi$ in Figs\ \ref{fig:spectralfunction_Born}.
While for small correlation lengths, i.e.\ $\Delta_s \xi \ll 1$ the
spectral function exhibits a maximum near $\omega = \alpha k$, for
large correlation lengths we find two maxima near $\omega = \pm
\sqrt{\Delta_s^2 + k^2}$, the closest one to $\alpha k$ having the
larger weight.

\subsubsection*{Corrections
  to the second order Born approximation and Sadovskii's solution} 

An attempt to sum up all diagrams in the perturbative expansion of the
averaged Green function was made by Sadovskii in the late seventies
\cite{Sadovskii79}. 
The first diagram in Fig.\ \ref{fig:selfenergy} which represents the 
contribution given by Eq.\ (\ref{eq:selfenergyBornIII}) can be calculated
by attributing a factor of $\Delta_s^2$ to the phonon line and a
factor $\left(\mathcal{G}_{0}^{R}\right)_{\bar\alpha \bar\alpha} 
(k;\omega+i/\xi)$ to the fermion line.
Sadovskii conjectured that all other non-vanishing diagrams could
be calculated using a similar recipe. (In the general case, the
imaginary part added to $\omega$ has to be multiplied by the number of
phonon lines above a given fermion line.)
This recipe enabled Sadovskii to give a
continued fraction representation of the self-energy (or Green function).
Sadovskii's solution was known as the only available exact
solution of the pseudo-gap state (see Ref.\ \cite{Tchernyshyov99}) and was
therefore also used by other authors
\cite{McKenzie96b,Schmalian98,Schmalian99}. However, only recently,
Tchernyshyov \cite{Tchernyshyov99} discovered an unfortunate error in
Sadovskii's solution which also turned the work based on it into
question. 
As pointed out by Tchernyshyov, Sadovskii's conjecture already breaks
down for terms contributing to the next to leading order in the 
self-energy. While
Sadovskii neglects the second term in Fig.\ \ref{fig:selfenergy}, 
which is correct for 
the incommensurate case, the last term in this figure already gives a different
contribution than conjectured by Sadovskii.
Instead of trying to correctly sum up all diagrams in the
perturbative expansion of the Green function we will now develop a  method
which will allow for a non-perturbative calculation of the Green
function.

\subsection{Non-Abelian Schwinger-ansatz}

We base our non-perturbative approach to calculate the Green
function of the FGM on a matrix generalization of the 
Schwinger-ansatz \cite{Schwinger62}. To make the differential operator
$-i\partial_x$ 
proportional to 
the unit 
matrix, we first factor out a Pauli matrix $\sigma_3$, so that the
retarded Green function 
\begin{equation}
\label{eq:Gtilde}
\tilde \mathcal{G}^R (x,x^{\prime};\omega)=\sigma_3
\mathcal{G}^R (x,x^{\prime};\omega)
\end{equation} 
satisfies
\begin{equation}
 \label{GreenfunctionII}
 [ i \partial_x - M(x,\omega + i0^{+}) ]
 \tilde \mathcal{G}^R ( x , x^{\prime} ; \omega)
 = \delta ( x - x^{\prime}) \sigma_0
 \; ,
\end{equation}
where
\begin{equation}
  \label{M}
  M (x,\omega) = [V(x) -\omega + \Delta ( x ) \sigma_{+} + \Delta^{\ast}
  (x) \sigma_{-} ] \sigma_{3}
  \;
\end{equation}
is a traceless matrix.
We now try to solve Eq.\ (\ref{GreenfunctionII}) by making the ansatz
 \begin{equation}
 \tilde \mathcal{G}^{R} (x,x',\omega  ) = U (x,\omega) \,
 \tilde \mathcal{G}_{0}^{R} ( x - x'  ) \, U^{-1} (x',\omega)
 \; ,
 \label{eq:ansatz}
 \end{equation}
where $U (x,\omega)$ is an invertible $2 \times 2 $ matrix and
$\tilde \mathcal{G}_{0}^{R} (x)$ is the Green function to
the operator $i\partial_{x} +i0^{+}\sigma_{3}$, i.e.
\begin{equation}
  \label{G_0^R}
  \tilde \mathcal{G}^{R}_{0} (x) = -i 
  \left( \begin{array}{cc}
   \theta(x) & 0 \\ 
   0 & -\theta(-x) 
 \end{array} \right) \; . 
\end{equation}
The ansatz (\ref{eq:ansatz}) resembles the transformation law for the
comparator in 
non-Abelian Gauge theory (see, for example, the book on quantum field
theory by Peskin and Schr\"oder \cite{Peskin95}), and since it is also
similar to the scalar Schwinger-ansatz \cite{Schwinger62} which is
sometimes used in 
functional bosonization of interacting fermions
\cite{Kopietz97book},
we will refer to it as the non-Abelian Schwinger-ansatz
\cite{Bartosch99a,Bartosch99c,Bartosch99d}. But note that in contrast
to earlier formulations of the non-Abelian Schwinger-ansatz
\cite{Bartosch99a,Bartosch99c}, we use the zero-frequency free
retarded Green function, such that the whole $\omega$ dependence is
included in $U (x,\omega)$. 

In the following we are going to suppress the parameter $\omega$.
The ansatz (\ref{eq:ansatz}) indeed solves Eq.\ (\ref{GreenfunctionII})
if $U(x)$ satisfies
 \begin{equation}
 \left[ i \partial_x - M(x) \right] U(x) = 0 \; .
 \label{eq:Udif}
 \end{equation}
To establish our formalism, let us first restrict the disorder
potentials to the interval $(-\Lambda,L+\Lambda)$. While the potentials
are assumed to be constant in the intervals $(-\Lambda,0)$ and
$(L,L+\Lambda)$, they are allowed to fluctuate in between.
Later we can let $\Lambda \to \infty$ (or we will set $\Lambda =
0$ and let $L \to \infty$). The boundary conditions for the retarded
Green function given in 
Eq.\ (\ref{BCfiniteLambda}) now renders into\footnote{In this work we
  will identify matrix elements as $U_{ij}$ 
with $U_{\alpha \alpha'}$, where $i,j=1,2$ corresponds to
$\alpha,\alpha'=+,-$, e.g. $U_{12} \equiv U_{+-}$.\label{fn:alpha_ij}}
\begin{equation}
  \label{BC}
  U_{12}(-\Lambda) = U_{21}(L+\Lambda) = 0 \; .
\end{equation}
Two different solutions of Eq.\ (\ref{eq:Udif}) are given by
\begin{eqnarray}
  \label{U+}     
  U_{+}(x) \hspace{-2.3mm} & = & \hspace{-2.3mm} 
 T\textrm{exp}\left[-i\int_{-\Lambda}^{x} M(y) dy\right] 
 \; ,
  \label{U-} \\
  U_{-}(x) \hspace{-2.3mm} & = & \hspace{-2.3mm}
  T^{-1}\textrm{exp}
  \left[i\int_{x}^{L+\Lambda} M(y) dy\right]  \; ,
\end{eqnarray}
where $T\textrm{exp}$ is the path-ordered and $T^{-1}\textrm{exp}$ is the
anti-path-ordered exponential function. Both, 
$U_{+}(x)$ and $U_{-}(x)$ can be expressed in terms of the $S$-matrix,
\begin{equation}
  \label{eq:S-matrix}
  S(x,x') = 
 T\textrm{exp}\left[-i\int_{x'}^{x} M(y) dy\right] 
 \; .
\end{equation}
By definition, $U_{+}(x) = S(x,-\Lambda)$ and $U_{-}(x) = S^{-1}
(L+\Lambda,x)$. 
Because 
$M^{\dagger} = \sigma_3 M \sigma_3$ and
${\rm Tr}\, M = 0$,
the $S$-matrices satisfy
$S^{\dagger}  = \sigma_3 S^{-1} \sigma_3$ and
${\rm det}\, S =1$, which means
that they belong to the
non-compact group $SU(1,1)$. It follows that the elements of $S$
satisfy 
$S_{22}=S_{11}^{\ast}$,
$S_{21}=S_{12}^{\ast}$, and
$| S_{11}|^2 - 
| S_{12}|^2  =1 $.
While each 
$U_{\alpha}(x)$ only obeys one of the two conditions (\ref{BC}), 
the combination
\begin{equation}
  \label{eq:U}
  U(x) \equiv \frac{1}{\sqrt{u}}\left( \begin{array}{cc}
   U_{-11}(x) & U_{+12}(x) \\ 
   U_{-21}(x) & U_{+22}(x)
 \end{array} \right)
\end{equation}
satisfies both boundary conditions. 
Here, $u = S_{22} (L+\Lambda,-\Lambda) = U_{-11}(-\Lambda) =
U_{+22}(L+\Lambda)$, so that 
${\rm det}\, U(x) = 1$. 
Defining\footnote{Note, that this definition deviates from
  the definition used in Ref.\ \cite{Bartosch99d}. Here, the vectors
$\mathbf{u}_{\alpha}$ and $\mathbf{v}_{\alpha}$ are given by the first
and second column of the matrix $U_{\alpha}$ {\em multiplied from the
  left by $\pm \sigma_3$.}}
\begin{eqnarray}
{\bf u}_{\alpha} \equiv (U_{\alpha 11},-U_{\alpha 21})^{T} \;, \quad
{\bf v}_{\alpha} \equiv (-U_{\alpha 12},U_{\alpha 22})^{T} \;, 
\end{eqnarray}
(such that ${\bf v}_{\alpha} = \sigma_{1} {\bf
  u}_{\alpha}^{\ast}$),
we obtain from Eqs. (\ref{eq:Gtilde}), (\ref{eq:ansatz}) and (\ref{eq:U})
\begin{equation}
  \label{GreenfunctionIII}
  {\setlength{\fboxsep}{2mm} \fbox {$ \ \displaystyle 
  i {\cal{G}}^{R} ( x , x^{\prime} ; \omega) = 
  \theta(x-x')\, \frac{{\bf{u}}_{-}(x) {\bf{u}}_{+}^{\dagger}(x')}{u}
  + 
     \theta(x'-x)\,
     \frac{{\bf{v}}_{+}(x) {\bf{v}}_{-}^{\dagger}(x')}{u} 
       \; .\ $}}
\end{equation}
${\bf{u}}_{+}^{\dagger}$ and ${\bf{v}}_{-}^{\dagger}$ are the
adjungated row vectors to the column vectors ${\bf{u}}_{+}$ and
${\bf{v}}_{-}$, so that ${\bf{u}_{-}}{\bf{u}}_{+}^{\dagger}$ and
${\bf{v}_{+}}{\bf{v}}_{-}^{\dagger}$ are $2 \times 2$-matrices. 
Note that Eq.\ (\ref{GreenfunctionIII}) involves only $U_{+12}$,
$U_{+22}$, $U_{-12}^{\ast}$ and $U_{-22}^{\ast}$, but not its complex
conjugates. In principle,
Eq.\ (\ref{GreenfunctionIII}) allows to determine the full
Green function of the FGM by evaluating time-ordered exponential
functions. 
Equivalent but more complicated forms of this equation 
were first
derived by Abrikosov and Ryzhkin \cite{Abrikosov76}. 
Of special interest will be the trace of the Green function at coinciding
space points,
\begin{equation}
  \label{eq:TraceG}
  \textrm{Tr}[{i\cal{G}}^{R} (x,x;\omega)] = \frac{U_{+22}(x)
  U_{-22}^{\ast}(x) + {U_{+12}}(x) U_{-12}^{\ast}(x)}{U_{+22}(x)
  U_{-22}^{\ast}(x) - {U_{+12}}(x) U_{-12}^{\ast}(x)} \;.
\end{equation}
It immediately follows from Eq.\ (\ref{def:localDOS}) that the local
DOS is given by
\begin{equation}
  \label{eq:localDOS}
  \rho(x,\omega) = \frac{1}{\pi} \textrm{Re} \left( \frac{U_{+22}(x)
  U_{-22}^{\ast}(x) + {U_{+12}}(x) U_{-12}^{\ast}(x)}{U_{+22}(x)
  U_{-22}^{\ast}(x) - {U_{+12}}(x) U_{-12}^{\ast}(x)} \right) \;.
\end{equation}

\subsubsection{Riccati equation}
\label{sec:Riccati}

Since Eq.\ (\ref{eq:localDOS}) only depends on the
ratios\footnote{Note also that the definition of
  $\Phi_{\alpha}(x)$ does not involve the 
  extra factor $\pm i$ used in Ref.\ \cite{Bartosch99d}.}
\begin{equation}
\Phi_{\alpha}(x)
\equiv U_{\alpha12}(x)/U_{\alpha22}(x) \;,
\end{equation}
we may also write the DOS as
\begin{equation}
  \label{localDOSRic}
  \rho(x,\omega) = \frac{1}{\pi} \textrm{Re} \left(
  \frac{1+\Phi_{+}(x)\Phi_{-}^{\ast}(x)}{1-\Phi_{+}(x)\Phi_{-}^{\ast}(x)} 
  \right) \; .
\end{equation}
More generally, it follows from Eq.\ (\ref{GreenfunctionIII}) that
the whole matrix Green function at coinciding space
points can be written in terms of $\Phi_{+}(x)$ and $\Phi_{-}^{\ast}(x)$:
\begin{eqnarray}
  i \mathcal{G}^R (x,x;\omega) 
  \hspace{-2.3mm}& \equiv &\hspace{-2.3mm} \frac{i}{2}
  \left[\mathcal{G}^R (x+0^+,x;\omega) + \mathcal{G}^R
    (x,x+0^+;\omega) \right] \nonumber \\
   & & \hspace{-23mm} = \frac{1}{1-\Phi_{+}(x) \Phi_{-}^{\ast}(x)} \left( 
    \begin{array}{cc}
      \frac{1}{2} \left(1+\Phi_{+}(x) \Phi_{-}^{\ast}(x) \right) & -
      \Phi_{+}(x)  \\ 
      - \Phi_{-}^{\ast}(x) & \frac{1}{2} \left(1+\Phi_{+}(x)
        \Phi_{-}^{\ast}(x) \right) 
    \end{array}
\right) \; . \qquad \quad
\end{eqnarray}
Using  Eq.\ (\ref{eq:Udif}),  we find that
the $\Phi_{\alpha}(x)$ are both solutions of the same Riccati equation,
\begin{equation}
  \label{Riccati}
  - i \partial_{x} \Phi_{\alpha}(x) =
   2 \tilde \omega(x) \Phi_{\alpha}(x) + \Delta(x) + \Delta^{\ast}(x)
  \Phi_{\alpha}^{2}(x)  \;. 
\end{equation}
Here, we have introduced $\tilde \omega(x) = \omega - V(x)$.
Similar Riccati equations have
recently been obtained by Schopohl \cite{Schopohl98} from the
Eilenberger equations of superconductivity.
To specify the initial conditions, let us assume that
outside the interval $[0, L]$  the potentials
$V(x)$ and $\Delta(x)$ are real constants, 
$V_{\textrm{\scriptsize BC}}$ and
$\Delta_{\textrm{\scriptsize BC}} \ge 0$. This amounts to taking the limit $\Lambda \to
\infty$, but keeping $L$ constant. The initial values
$\Phi_{+} (0 ) = \lim_{\Lambda \to \infty}
S_{12}(0,-\Lambda)/S_{22}(0,-\Lambda)$ and $\Phi_{-} 
( L ) = \lim_{\Lambda \to \infty}
(S^{-1})_{12}(L+\Lambda,L)/(S^{-1})_{22}(L+\Lambda,L)$ can be obtained
by evaluating the $S$-matrix for constant potentials.

\subsubsection{The $S$-matrix for constant potentials}

For constant potentials $V_n$ and $\Delta_n$, the time-ordering
operator $T$ may be omitted in Eq.\ (\ref{eq:S-matrix}), and the $S$-matrix
is given by 
\begin{eqnarray}
  \label{eq:S-matrix_const}
  S_n (x-x') \hspace{-2.3mm}&=&\hspace{-2.3mm} 
  \cosh[\sqrt{|\Delta_{n}|^2-\tilde \omega_{n}^2}\,(x-x')]
  \sigma_{0} \nonumber \\
  \hspace{-2.3mm}&\ &\hspace{-2.3mm} \quad  {}+i    
  \sinh[\sqrt{|\Delta_{n}|^2-\tilde \omega_{n}^2}\,(x-x')]
  \; \frac{\tilde \omega_{n} \sigma_{3} + \Delta_{n} \sigma_{+}
    -\Delta_{n}^{\ast} \sigma_{-} }{\sqrt{|\Delta_{n}|^2-\tilde
      \omega_{n}^2}}  \; . \qquad
\end{eqnarray}
For $|\Delta_{n}|^2 < \tilde \omega_{n}^2$, the argument of the square
root is negative, so that in this case we write the $S$-matrix
as
\begin{eqnarray}
  \label{eq:S-matrix_constII}
  S_n (x-x') 
  \hspace{-2.3mm}&=&\hspace{-2.3mm} \cos[\sqrt{\tilde \omega_{n}^2 - |\Delta_{n}|^2}\,(x-x')]
  \sigma_{0} \nonumber \\
  \hspace{-2.3mm}&\ &\hspace{-2.3mm} \quad  {}+i   
  \sin[\sqrt{\tilde \omega_{n}^2 - |\Delta_{n}|^2}\,(x-x')]
  \; \frac{\tilde \omega_{n} \sigma_{3} + \Delta_{n} \sigma_{+}
    -\Delta_{n}^{\ast} \sigma_{-} }{\sqrt{\tilde \omega_{n}^2 -
      |\Delta_{n}|^2}}  \; . \qquad 
\end{eqnarray}
For notational simplicity, let us introduce
$\Delta_n^{\textrm{\scriptsize red}} \equiv
\sqrt{|\Delta_{n}|^2 - \tilde \omega_{n}^2}$. To calculate $\Phi_{+}
(0)$ and $\Phi_{-} (L)$, we take the limit $x \to \pm \infty$ of the ratio
$S_{n12}(x)/S_{n22}(x)$ and obtain
\begin{equation}
  \label{eq:limitratio}
  \lim_{x \to \pm \infty} \frac{S_{n12}(x)}{S_{n22}(x)} = 
  \frac{\pm i \Delta_n^{\textrm{\scriptsize red}} - \tilde
    \omega_n}{\Delta_{n}^{\ast}} \; . 
\end{equation}
If we keep in mind that the
frequency $\omega$ involves a small imaginary part, we see that this
result is not restricted to the 
case $|\Delta_{n}|^2 > \tilde\omega_n^2$: For $|\Delta_n|^2 \le \tilde
\omega_n^2$ the square root has to be taken such that the right-hand
side of Eq. (\ref{eq:limitratio}) vanishes as $\Delta_n \to 0$. This
follows directly from the definition of the $S$-matrix.

\subsubsection{Initial conditions}

It follows from Eq.\ (\ref{eq:limitratio}) that the Riccati equation
(\ref{Riccati}) should be integrated with the initial conditions 
\begin{eqnarray}
  \label{BCPhi}
  \Phi_{+} (0) \hspace{-2.3mm} & = & \hspace{-2.3mm} \Phi_{-}^{\ast}
  (L) \nonumber \\
  \hspace{-2.3mm} & = & \hspace{-2.3mm} \left\{\begin{array}{ll} 
      \rule[-2mm]{0mm}{5mm} 
      \frac{i \sqrt{\Delta_{\textrm{\tiny BC}}^{2} -(\omega-V_{\textrm{\tiny BC}})^{2}} -
        (\omega-V_{\textrm{\tiny BC}})}{\Delta_{\textrm{\tiny BC}}} & ,\  \Delta_{\textrm{\scriptsize BC}}^{2} >
      (\omega-V_{\textrm{\scriptsize BC}})^{2} \; , \\
      \rule[0mm]{0mm}{8mm}
      \frac{\textrm{sgn} (\omega
        -V_{\textrm{\tiny BC}})\sqrt{(\omega-V_{\textrm{\tiny BC}})^{2} - \Delta_{\textrm{\tiny BC}}^{2}}
        -(\omega-V_{\textrm{\tiny BC}})}{\Delta_{\textrm{\tiny BC}}} & ,\  \Delta_{\textrm{\scriptsize BC}}^{2} \le
      (\omega-V_{\textrm{\scriptsize BC}})^{2} \; , \end{array} \right.
\end{eqnarray}
where $\textrm{sgn} (x)$ is equal to $1$ for positive $x$ and equal
to $-1$ for negative $x$.
While for $\Delta_{\textrm{\scriptsize BC}} = 0$ the initial conditions are given
by $\Phi_{+}(0) = \Phi_{-}^{\ast}(L)=0$, for $\Delta_{\textrm{\scriptsize BC}} \to \infty$ one
gets $\Phi_{+} (0) =\Phi_{-}^{\ast} (L) = i$.
Note that for arbitrary potentials $V_{\textrm{\scriptsize BC}}$ and
$\Delta_{\textrm{\scriptsize BC}}$, 
the initial values are  
simply given by the stable stationary solution of the Riccati equation
(\ref{Riccati}) with $V(x) = V_{\textrm{\scriptsize BC}}$ and $\Delta(x) =
\Delta_{\textrm{\scriptsize BC}}$.

\subsubsection{The case of a discrete spectrum}
Defining the (complex) phase $\varphi_{\alpha} (x)$ via 
\begin{equation}
\Phi_{\alpha}(x) = e^{i\varphi_{\alpha}(x)} \;,
\end{equation}
and decomposing $\Delta(x)$ into its amplitude $|\Delta(x)|$ and
its phase $\vartheta(x)$,
\begin{equation}
  \label{eq:decomposition:Delta}
  \Delta ( x ) = | \Delta (x ) | e^{ i \vartheta ( x ) } \;,
\end{equation}
the Riccati equation\ (\ref{Riccati}) turns into  
\begin{equation}
  \label{eq:phi}
  {\setlength{\fboxsep}{2mm} \fbox {$ \ \displaystyle 
  \partial_{x} \varphi_{\alpha}(x) = 2\tilde \omega(x) +
  2|\Delta(x)|\cos \left[\varphi_{\alpha}(x)-\vartheta(x) \right] \;.$}}
\end{equation}
Note that this equation of motion is of the Langevin type
\cite{Gardiner83,vanKampen81} 
\begin{equation}
  \label{eq:gen_Langevin}
  \partial_x v(x) = - a + b_1 V(x) + b_2^{\ast}(v) \Delta(x) + b_2(v)
  \Delta^{\ast}(x) \;.
\end{equation}
Let us consider the case $( \omega -  V_{\textrm{\scriptsize BC}})^2 <
\Delta_{\textrm{\scriptsize BC}}^2$: It directly follows from Eq.\ (\ref{BCPhi}) that
$|\Phi_{+} (0)| = |\Phi_{-} ( L )| =1$, such that
the initial values $\varphi_{+} ( 0 )$ and $\varphi_{-} ( L )$ are
real. Hence,
the solutions of Eq.\ (\ref{eq:phi}) remain real, which implies
$ | \Phi_{\alpha} ( x ) | = 1$ for all $x$. As can be seen from 
Eq.\ (\ref{BCPhi}), $\varphi_{+}(0)$ and $\varphi_{-}(L)$ can be
chosen to fulfill 
$\varphi_{+}(0) = -\varphi_{-}(L) \in [0,\pi]$,
so that the initial values $\varphi_{+}(0)$ and $\varphi_{-}(L)$ are
uniquely determined by
\begin{equation}
 \label{phiBC}
 \cot \varphi_{+} (0) = -\cot \varphi_{-} (L) =
 -\frac{\omega-V_{\textrm{\scriptsize BC}}}{\sqrt{\Delta_{\textrm{\scriptsize BC}}^{2}-
 (\omega-V_{\textrm{\scriptsize BC}})^{2}}}   \;.
 \end{equation}
For the phases $\varphi_{\alpha}(x)$ to be continuous, they have to be
unreduced phases which are not limited to take values between $0$
and $2\pi$. 
In terms of the $\varphi_{\alpha}(x)$, the local DOS can
be written as
\begin{eqnarray}
  \label{localDOS}
  \rho(x,\omega) \hspace{-2.3mm} & = & \hspace{-2.3mm}
  -\frac{1}{\pi} \textrm{Im} \cot 
  \left[\frac{\varphi_{+}(x)-\varphi_{-}(x)}{2}+i0\right] \nonumber \\
  \hspace{-2.3mm} & = & \hspace{-2.3mm} 2 \sum_{m = - \infty}^{\infty} \delta
  \left(\varphi_{+}(x)-\varphi_{-}(x) - 2 \pi m \right) \; .
\end{eqnarray}
It is easy to show that for $x \ge 0$ the phase $\varphi_{+}(x,\omega)$
is a monotonic increasing function of $\omega$.
For the initial value at $x=0$,
this follows from Eq.\ (\ref{phiBC}):
\begin{equation}
  \label{eq:phiderivative0}
  \partial_{\omega} \varphi_{+} (0) = -\partial_{\omega} \varphi_{-} (L) =
 \frac{1}{\sqrt{\Delta_{\textrm{\scriptsize BC}}^{2}-(\omega-V_{\textrm{\scriptsize BC}})^{2}}} > 0 \;.
\end{equation}
It can be seen directly from Eq.\ (\ref{eq:phi}) that
$\partial_{\omega} \varphi_{+} (x) > 0$ is true for every $x \ge
0$. More formally, differentiating the equation of motion
(\ref{eq:phi}) with respect to $\omega$ gives
\begin{equation}
  \label{eq:phi:differential}
    \partial_{x} 
    \partial_{\omega} \varphi_{\alpha}(x,\omega) 
    = 2 - 2|\Delta(x)|\sin
    \left[\varphi_{\alpha}(x)-\vartheta(x) \right]\,
    \partial_{\omega} \varphi_{\alpha}(x,\omega) 
    \;. 
\end{equation}
Solving this first-order differential equation for $\partial_{\omega}
\varphi_{\alpha}(x,\omega)$, we obtain
\begin{eqnarray}
   & &  \partial_{\omega} \varphi_{\alpha}(x,\omega) 
    = \partial_{\omega} \varphi_{\alpha}(x_0,\omega) \nonumber \\
   & & \quad \qquad {}+ 2 \int_{x_0}^{x}
    \exp \left[-2 \int_{x'}^{x} |\Delta(x'')|\sin
    \left[\varphi_{\alpha}(x'')-\vartheta(x'') \right]\,
    dx''\right] \, dx' \;. \qquad \qquad
  \label{eq:phi:integrated}
\end{eqnarray}
Since both terms on the right-hand side of Eq.\
(\ref{eq:phi:integrated}) are positive for $\alpha = +$ and $x_0 = 0$,
we have $\partial_{\omega} \varphi_{+}(x,\omega) > 0$ for $x \ge 0$.
Analogously we find $\partial_{\omega} \varphi_{-}(x,\omega) < 0$ for
$x \le L$.
For arbitrary $\omega$, $\varphi_{+}(x,\omega)$ integrated with the
initial condition $\varphi_{+}(0,\omega)$ given in Eq.\ (\ref{phiBC})
will usually not be equal to $\varphi_{-}(L,\omega)$ up to a multiple
of $2\pi$ at $x=L$. For certain discrete frequencies $\omega_m$,
however, this is the case. Since $\varphi_{+}(x,\omega)$ is a
monotonic function of $\omega$, we can uniquely define $\omega_m$ by
the condition 
\begin{equation}
  \label{eq:def:omega_m}
  \varphi_{+}(L,\omega_m) = \varphi_{-}(L,\omega_m) + 2\pi m \;.
\end{equation}
Note that $\omega_m$ is only well-defined if it turns out that
$(\omega_m-V_{\textrm{\scriptsize BC}})^2 < \Delta_{\textrm{\scriptsize BC}}^2$. 
Since the right-hand side of Eq.\ (\ref{eq:phi}) is a $2 \pi$-periodic
function of $\varphi_{\alpha}(x)$, we see that
$\varphi_{+}(x,\omega_m)$ and $\varphi_{-}(x,\omega_m)$ are equal
up to the constant $2\pi m$ for every $x \in [0,L]$,
\begin{equation}
  \label{eq::omega_m}
  \varphi_{+}(x,\omega_m) - \varphi_{-}(x,\omega_m) = 2\pi m \;.
\end{equation}
Of course, the $\omega_m$ are the discrete eigenvalues of the
system. This follows immediately from the fact that the local DOS
$\rho(x,\omega)$ is equal to zero if not $\omega = \omega_m$ for
one $m$. Therefore, the total DOS is given by
\begin{equation}
  \label{eq:totalDOS}
  \rho({\omega}) = \frac{1}{L} \sum_m \delta(\omega -\omega_m) \;.
\end{equation}
Using the well-known formula $\delta\left(f(\omega)\right) 
= \sum_m \frac{1}{|f'(\omega_m)|} \delta(\omega-\omega_m)$,
where $\omega_m$ are the zeros of $f(\omega)$, we can write Eq.\
(\ref{localDOS}) as
\begin{equation}
  \label{eq:localDOSII}
  \rho(x,\omega) = 
  \sum_m \frac{2
    \delta(\omega-\omega_m)}{|\partial_{\omega} \varphi_{+}(x,\omega)
    - \partial_{\omega} \varphi_{-}(x,\omega)|} \;.
\end{equation}
Expressing $\partial_{\omega} \varphi_{\alpha}(x,\omega)$ by the right-hand side of 
Eq.\ (\ref{eq:phi:integrated}) with $\alpha = +$ and $x_0 = 0$ or
$\alpha = -$ and $x_0 = L$, respectively, we arrive for 
$\Delta_{\textrm{\scriptsize BC}} = \infty$ at
\begin{equation}
  \label{eq:localDOSIII}
  \rho(x,\omega) = 
  \sum_m \frac{
    \delta(\omega-\omega_m)}{\int_0^L  dx'\, \exp \left(-2
      \int_{x'}^{x} dx'' \,
      |\Delta(x'')|\sin 
    \left[\varphi_{\alpha}(x'',\omega_m)-\vartheta(x'') \right] \right)} \;.
\end{equation}
Once the eigenvalues $\omega_m$ have been determined, this equation in
principle allows to calculate the local DOS for arbitrary potentials
$V(x)$ and $\Delta(x)$.

\subsubsection{Integrated averaged Green function $\Gamma(\omega)$}

In the last subsection, 
we have seen that the integrated DOS
can be obtained by solving a simple initial value problem for the
phase $\varphi(x)$, which is a functional of the disorder.
To implement the correct boundary conditions for a system of length
$L$, we have first assumed that outside the interval $(0,L)$ the
potentials are constant over a range $\Lambda$, but then
drop to zero. Finally, we have let $\Lambda$ go to infinity. However,
if we are only interested in the limit $L \to \infty$, i.e. the
bulk properties, we can set $\Lambda = 0$ at the beginning of
our calculations. The physical meaning of this is that bulk properties
should be independent of the boundary conditions. 

By setting the potential equal to zero outside the interval $(0,L)$,
we will not only be able to recover the equation of
motion (\ref{eq:phi}) satisfied by the phase $\varphi(x)$ which
determines the 
integrated DOS, we will also be able to derive an additional equation
which allows to calculate the inverse localization length.

It follows from Eq.\ (\ref{eq:TraceG}) that the trace of the
space-averaged diagonal element of the retarded Green function
is given by
\begin{eqnarray}
  \left< \textrm{Tr}[{\cal{G}}^{R} (x,x;\omega)] \right>_{x}
  \hspace{-2.3mm} & \equiv & \hspace{-2.3mm}
  \frac{1}{L} \int_0^L dx\ \textrm{Tr}[{\cal{G}}^{R} (x,x;\omega)]
  \nonumber \\
  & & \hspace{-28mm} =  \frac{1}{L S_{22}(L,0)} \left[ -i \int_0^L dx\ \left( S_{22}(L,x) S_{22}(x,0) -
  S_{21}(L,x) S_{12}(x,0) \right) \right] \;. \qquad \quad
  \label{eq:spaceaveragedG}
\end{eqnarray}
The term in angular brackets can easily be identified to be equal to
$\partial_{\omega} S_{22}(L,0)$.
We can therefore rewrite Eq.\ (\ref{eq:spaceaveragedG}) as
\begin{equation}
  \label{eq:spaceaveragedGII}
\left< \textrm{Tr}[{\cal{G}}^{R} (x,x;\omega)]
    \right>_{x} = 
  \frac{\partial_{\omega} S_{22}(L,0)}{L S_{22}(L,0)} = 
    \frac{1}{L}
  \partial_{\omega} \ln \left[S_{22}(L,0) \right] \; .
\end{equation}
To describe the bulk properties, one should now take the limit $L \to
\infty$. If we introduce $\Gamma(\omega)$ as the trace of the 
energy-integrated space-averaged Green function at coinciding space
points,
\begin{equation}
  \label{eq:Gamma}
  {\setlength{\fboxsep}{2mm} \fbox {$ \ \displaystyle \Gamma(\omega)
      \equiv \lim_{L \to \infty} 
    \frac{1}{L} 
    \ln \left[ S_{22}(L,0;\omega) \right] \; ,\ $}}
\end{equation}
we have
\begin{equation}
  \left< \textrm{Tr}[{\cal{G}}^{R} (x,x;\omega)] \right> =
  \partial_{\omega} \Gamma(\omega) \;.
\end{equation}
While the integrated DOS is given by $\mathcal{N} (\omega) = -\pi^{-1}
\textrm{Im}\, \Gamma(\omega)$, we will see in the next subsection that
$\textrm{Re}\, \Gamma(\omega)$ is equal to the inverse localization
length $\ell^{-1}(\omega)$. The decomposition of $\Gamma(\omega)$ into its
real and imaginary part can
therefore be written as
\begin{equation}
  \label{eq:Gammadecomposition}
  \Gamma (\omega) = \ell^{-1} (\omega) -i\pi\mathcal{N}(\omega) \;.
\end{equation}
Both $\mathcal{N}(\omega)$ and $\ell^{-1} (\omega)$ can simultaneously be
calculated 
by determining the logarithm of the $S$-matrix element $S_{22}$. 
It is
convenient to express the $S$-matrix elements in terms of their phases.
Let us define $\varphi_{\alpha \alpha'} (x)$ via
\begin{equation}
  \label{eq:phaseSmatrix}
  S_{\alpha \alpha'} (x,0)\equiv e^{-i \varphi_{\alpha \alpha'} (x)} \;.
\end{equation}
$\Gamma (\omega)$ is then given by 
\begin{equation}
  \label{eq:Gammaphase}
  i \Gamma (\omega) = \lim_{L \to \infty} \frac{\varphi_{22}(L)}{L} \;.
\end{equation}
Introducing $\bar \alpha \equiv - \alpha$, the properties of the
$S$-matrix $S_{\alpha \alpha'} = S_{\bar \alpha \bar \alpha'}^{\ast}$
render into $\varphi_{\alpha \alpha'} = -\varphi_{\bar \alpha
  \bar \alpha'}^{\ast}$. The
$S$-matrix can
be expressed in terms of $\varphi_{12}$, $\varphi_{22}$ and its
complex conjugates.\footnote{Recall that instead of $\alpha,\alpha' = +,-$
we also use $i,j = 1,2$ (see
footnote \ref{fn:alpha_ij} on page \pageref{fn:alpha_ij})} 
It follows from $i\partial_x S(x,0) = M(x) S(x,0)$ that the
$\varphi_{\alpha \alpha'}$ satisfy 
\begin{equation}
  \label{eq:phi:motion}
  \partial_x \varphi_{\alpha \alpha'} (x)= M_{\alpha \alpha}(x) + M_{\alpha
    \bar \alpha}(x) \exp\left[{i(\varphi_{\alpha \alpha'}(x) -
    \varphi_{\bar \alpha \alpha'} (x))}\right] 
  \;.
\end{equation}
Recalling that $M(x) = -\tilde\omega(x) \sigma_3 - \Delta(x)
\sigma_{+} + \Delta^{\ast}(x) \sigma_{-}$, the two equations for
$\varphi_{12}$ and $\varphi_{22}$ read
\begin{eqnarray}
  \label{eq:phi22}
  \partial_x \varphi_{22}(x)\hspace{-2.3mm} & = & \hspace{-2.3mm} \tilde \omega (x) +
  \Delta^{\ast}(x) 
  \exp \left[{i(\varphi_{22}(x) - \varphi_{12}(x))} \right]\;,\\
  \label{eq:phi12}
  \partial_x \varphi_{12}(x) \hspace{-2.3mm} & = & \hspace{-2.3mm} - \tilde \omega (x) - \Delta(x)
  \exp \left[{-i(\varphi_{22}(x) - \varphi_{12}(x))} \right] \;.
\end{eqnarray}
If we now introduce 
\begin{eqnarray}
  \label{def:phi}
  \varphi (x) \hspace{-2.3mm} & \equiv & \hspace{-2.3mm} \varphi_{22} (x) -\varphi_{12} (x)
  \;,\\
  \label{def:theta}
  \zeta (x) \hspace{-2.3mm} & \equiv & \hspace{-2.3mm} -i \left(\varphi_{22} (x) +
    \varphi_{12} (x) \right) \;,
\end{eqnarray}
we arrive at the following system of equations of motion:
\newline
\hfill
\parbox{0mm}
{\begin{eqnarray*}
\\
\end{eqnarray*}}
\parbox{11cm}
{\begin{eqnarray*}
  {\setlength{\fboxsep}{2mm} \fbox {$ \ \displaystyle  
  \begin{array}{rcl}  \partial_{x} \varphi (x) \hspace{-2.3mm} & = & \hspace{-2.3mm} 2
    \tilde \omega(x) + 2 
  |\Delta(x)|\cos \left[\varphi (x)-\vartheta(x) \right] \;, \\
   \rule[0mm]{0mm}{6mm}
 \partial_{x} \zeta (x) \hspace{-2.3mm} & = & \hspace{-2.3mm} 2 |\Delta(x)|\sin \left[\varphi
    (x)-\vartheta(x) \right] \;. \end{array} $}}
\end{eqnarray*}}
\parbox{19mm}
{\begin{eqnarray}
    \label{eq:phiII} \\
   \rule[-3mm]{0mm}{8mm}
    \label{eq:zeta}
\end{eqnarray}}
\newline
Note that Eq.\ (\ref{eq:phiII}), which determines $\varphi(x)$, is
exactly the same Langevin equation that we 
derived before from the Riccati equation [see Eq.\ (\ref{eq:phi})] and
is independent from Eq.\ 
(\ref{eq:zeta}). After having found a solution to Eq.\
(\ref{eq:phiII}), $\zeta(x)$ can in principle be obtained by 
integrating Eq.\ (\ref{eq:zeta}). In terms of $\varphi(x)$ and
$\zeta(x)$, $\Gamma(\omega)$ is now given by
\begin{equation}
  \label{eq:GammmaII}
  i \Gamma (\omega) = \lim_{L \to \infty} \left[\varphi(L) +i \zeta(L)
  \right]/2L \;.
\end{equation}
The initial condition $S(0,0) = \sigma_0$ could be mapped on the
initial conditions for $\varphi(0)$ and $\zeta(0)$ which are,
strictly speaking, singular but integrable.
In the limit $L \to \infty$, the initial conditions finally drop out, but 
to describe bulk properties for finite $L$, it is better to choose
$\varphi(0) = \zeta(0) = 0$,
such that $\varphi(x)$ and $\zeta(x)$ are real for all $x$ and there
is no contribution to $\Gamma(\omega)$
which as $L$ becomes large only dies out as
$1/L$. The integrated
DOS and the inverse localization length 
can now be expressed as
\newline
\hfill
\parbox{0mm}
{\begin{eqnarray*}
\\
\end{eqnarray*}}
\parbox{11cm}
{\begin{eqnarray*}
  {\setlength{\fboxsep}{2mm} \fbox {$ \ 
  \begin{array}{rcl}  
  \displaystyle  \mathcal{N} (\omega) \hspace{-2.3mm} & \displaystyle  = \hspace{-2.3mm} &
  \displaystyle  \rho_0 \lim_{L \to \infty} 
  \varphi(L)/2L \;, \\ 
  \rule[0mm]{0mm}{6mm}
  \displaystyle
  \ell^{-1} (\omega) \hspace{-2.3mm} & \displaystyle
  = \hspace{-2.3mm} & 
  \displaystyle  \lim_{L \to \infty} \zeta(L)/2L 
  \;. \end{array} $}}
\end{eqnarray*}}
\parbox{19mm}
{\begin{eqnarray}
    \label{eq:IDOS-phi} \\
   \rule[-4.0mm]{0mm}{10.5mm}
    \label{eq:l-phi}
\end{eqnarray}}
\newline
These two equations in combination with the equations of motion
(\ref{eq:phiII}) and (\ref{eq:zeta}) allow for simultaneous exact
numerical computations of the (integrated) DOS and the inverse
localization length for arbitrary given disorder potentials. Since the
(integrated) DOS and the inverse localization length are
self-averaging quantities \cite{Lifshits88}, it is sufficient to
consider just one typical realization of the disorder potential.
We will do this for various interesting cases in Section
\ref{section:finite_xi}.

In analytical calculations one does not usually work with a certain
realization of the disorder. Instead, one tries to calculate {\em
  averaged} quantities by using the given statistical properties of
the disorder potentials. Taking the average of Eq.\
(\ref{eq:GammmaII}) with respect to the 
distribution of the random potentials $V(x)$ and $\Delta(x)$, we obtain
\begin{equation}
  \label{eq:Gamma:average}
  i \langle \Gamma (\omega) \rangle = \lim_{L \to \infty} \left[
    \langle \varphi(L)\rangle +i \langle \zeta(L) \rangle \right]/2L
  \;. 
\end{equation}
Integrating the equations of motion (\ref{eq:phiII}) and
(\ref{eq:zeta}) with respect to $x$ from $0$ to 
$L$, $\langle \Gamma (\omega)\rangle$ can be rewritten as
\begin{equation}
  \label{eq:Gamma:averageII}
  i \langle \Gamma (\omega) \rangle = \lim_{L \to \infty} \frac{1}{L}
  \int_{0}^{L} dx\ \langle \tilde
  \omega(x) + \Delta^{\ast}(x) \exp\left[i\varphi(x) \right] \rangle \;. 
\end{equation}
The average in Eq.\ (\ref{eq:Gamma:averageII}) has to be taken with respect
to the probability distribution involving the
disorder at $x$ and via $\varphi(x)$ also at all space points between
$0$ and $x$, which can also be considered as an average with respect to 
the joint probability distribution of the random potentials at $x$ and 
the {\em unreduced} phase $\varphi(x)$. However, since
$\exp\left[i\varphi(x)\right]$ is a $2\pi$-periodic function in
$\varphi(x)$, it is also sufficient to use 
the joint probability distribution of
the random potentials and the {\em reduced} phase $\varphi(x) \in [\,
0,2\pi)$. 
For large $x$, the joint probability distribution becomes stationary,
and since
$\langle V(x) \rangle=0$, Eq.\ (\ref{eq:Gamma:averageII}) reduces to
\begin{equation}
  \label{eq:<Gamma>}
  {\setlength{\fboxsep}{2mm} \fbox {$ \ \displaystyle   i \left<
        \Gamma (\omega) \right> = \omega + \left< 
    \Delta^{\ast}(x) \exp\left[i \varphi(x) \right] \right> \;. $}}
\end{equation}
 We will use Eq.\ (\ref{eq:<Gamma>}) in Section 
\ref{chap:whitenoise} to calculate both $\mathcal{N}(\omega)$
and $\ell^{-1}(\omega)$ for the FGM in the white noise limit.

\subsection{Gauge invariance}

It turns out that fluctuations of the forward scattering disorder have
similar effects on the DOS and localization length as have phase
fluctuations of the gap parameter. To illuminate the hidden symmetry,
let us again consider the equation satisfied by the retarded Green
function $\mathcal{G}^R(x,x';\omega)$,
\begin{equation}
  \left( 
 \begin{array}{cc}
 \omega - V ( x ) 
 + i  \partial_x & -\Delta ( x ) \\
 -\Delta^{\ast} ( x ) & \omega - V ( x ) 
 - i  \partial_x 
 \end{array}
 \right ) \mathcal{G}^{R} ( x , x^{\prime} ; \omega ) 
 = \delta ( x - x^{\prime} ) \sigma_0
 \label{eq:satisfied_by_G}
 \; . \qquad
\end{equation}
A crucial point of this equation is that its form is left invariant under
the gauge transformation \cite{Brazovskii76}
\begin{eqnarray}
  {\cal{G}}^{R} ( x , x^{\prime} ; \omega )  & \hspace{-2.3mm} \to &
  \hspace{-2.3mm} 
  \exp\left[{- ({i}/{2}) \chi ( x ) \sigma_3 }\right]
  {\cal{G}}^{R} ( x , x^{\prime} ; \omega )  
  \exp\left[{ ({i}/{2}) \chi ( x^{\prime} ) \sigma_3 }\right]
  \label{eq:Gbardef}
  \; , \\
  V(x) & \hspace{-2.3mm} \to & \hspace{-2.3mm} V(x) + {\textstyle
    \frac{1}{2}} \partial_x \chi (x) \;, \\
  \Delta(x) & \hspace{-2.3mm} \to & \hspace{-2.3mm} \Delta(x)
  \exp[-i\chi(x)] \;. 
\end{eqnarray}
Here, $\chi(x)$ is a local phase rotation which is allowed to vary
arbitrarily from point to point.
Since the trace of a product of matrices is invariant under cyclic
permutations of the matrices, $\textrm{Tr}\, {\cal{G}}^{R}
(x,x;\omega)$ and therefore the local DOS and the inverse
localization length
are also invariants under the above gauge 
transformation.\footnote{Strictly speaking, it follows only that
  $\partial_{\omega} \ell^{-1}(\omega)$ is left invariant. The integration
  constants at $\omega = \infty$ are, however, the same, so that
  $\ell^{-1}(\omega)$ is invariant under the gauge transformation.}

Instead of directly looking at the trace of the Green function at
coinciding space points, we can also consider the equations of motion
(\ref{eq:phiII}) and 
(\ref{eq:zeta}). Their form is gauge invariant under the combined
transformation
\begin{eqnarray}
  \varphi(x) & \hspace{-2.3mm} \to & \hspace{-2.3mm} \varphi(x) - \chi(x)
  \; , \\
  \zeta(x) & \hspace{-2.3mm} \to & \hspace{-2.3mm} \zeta(x) 
  \; , \\
  V(x) & \hspace{-2.3mm} \to & \hspace{-2.3mm} V(x) + {\textstyle
    \frac{1}{2}} \partial_x \chi (x) \;, \\
  \vartheta (x) & \hspace{-2.3mm} \to & \hspace{-2.3mm} \vartheta (x) -
  \chi(x) \;. 
\end{eqnarray}
It follows from Eqs.\ (\ref{eq:IDOS-phi}) and (\ref{eq:l-phi}) that
both the integrated DOS $\mathcal{N}(\omega)$ and
the inverse localization length $\ell^{-1} (\omega)$ are invariant under
the considered gauge transformations. Only in the unusual case of a
finite limit $\lim_{L\to \infty} \chi(L)/L$ we get an irrelevant shift in
the additive constant of $\mathcal{N}(\omega)$. 
The choices for which either $V(x)$ or
$\vartheta(x)$ vanishes are especially convenient.
\begin{enumerate}
\item
  {\em Effectively vanishing phase fluctuations:}
  If $\vartheta(x)$ is
  differentiable, we can define
  $\chi(x) \equiv \vartheta(x)$,
  such that
  there are no phase fluctuations of $\Delta(x)$ left and $\Delta(x)$ can
  be taken to be real and positiv. We will use the resulting transformations
  $V(x) \to V(x) + \partial_x \vartheta (x)/2$ and
  $\Delta(x) \to |\Delta(x)|$ at the
  end of Section \ref{section:finite_xi} to find an exact solution for
  the FGM involving only phase fluctuations.
\item 
  {\em Effectively vanishing forward scattering potential:}
  Choosing the phase $\chi ( x )$ such that 
  $\partial_x \chi (x)/2 = - V ( x )$,
  the forward scattering potential $V(x)$ can be eliminated by
  renormalizing the phase fluctuations $\vartheta(x)$ of the backscattering
  potential $\Delta(x)$.
\end{enumerate}

\subsection{Lyapunov exponent and localization length}
\label{sec:Thouless}

{\em In this subsection, we will explicitly show that the Thouless
formula holds for the FGM and} that $\textrm{Re} \, \Gamma (\omega)$ {\em can
indeed be identified with the inverse localization length
$\ell^{-1}(\omega)$.}
\newline
\newline
One of the striking properties of disordered systems
treated in the independent electron approximation is the fact that
the disorder can lead to a macroscopic large number of localized
states. These states are eigenfunction of the Schr\"odinger equation
falling off exponentially with distance from one point in space which is
characteristic for the particular solution.
In one dimension, an arbitrary weak disorder suffices to localize all
eigenstates (excluding perhaps states at isolated energy
values) \cite{Mott61}. As a 
consequence, the diffusion coefficient and the dc 
conductivity vanish. Therefore, there is no metal-insulator transition
in one dimension.

\subsubsection{Thouless formula}

Since the energy dispersion of the FGM is linear, the Schr\"odinger
equation of the FGM, $\hat H \psi(x) = \omega \psi (x)$,
is a linear first 
order differential equation. Fixing the 
two-component wave function $\psi (x) \equiv (\psi_1 (x),\psi_2
(x))^T$ at one space point $x_0$ therefore constitutes the wave
function at all space points $x$. As we will see explicitly
below, for large 
distances $|x-x_0|$ the envelope of the wave function will grow
exponentially with probability one, i.e. $||\psi (x)|| \sim
||\psi_0||\exp\left(+\gamma   |x-x_0| \right)$, where $||\psi (x)||^2 \equiv
|\psi_1 (x)|^2 + |\psi_2 (x)|^2$. 
Of course, $\psi(x)$ cannot be an eigenfunction of the Hamiltonian
$\hat H$ satisfying the right boundary conditions.
The proportionality
factor $\gamma$ in the 
exponential function is called the Lyapunov exponent. 
The (mean) localization length is usually defined as the inverse
Lyapunov exponent \cite{Lifshits88},
\begin{equation}
  \label{def:localizationlength}
  \ell^{-1}(\omega) \equiv \gamma (\omega) \equiv \lim_{L \to \infty}
  \frac{1}{L} \ln \left(\frac{||\psi(L)||}{||\psi_0||} \right) \;.
\end{equation}
Note that since we take the limit $L \to \infty$, the initial value
$||\psi_0||$ drops out.

Let us now rewrite the Schr\"odinger equation such that the
differential operator $\partial_x$ is proportional to the 
unit matrix. As in the non-Abelian Schwinger-ansatz we factor out a
$\sigma_3$-matrix so that the wave function $\tilde \psi(x) \equiv
\sigma_3 \psi(x)$ satisfies
\begin{equation}
  \label{eq:SchroedingerM}
  \left[i\partial_x - M(x)\right] \tilde\psi(x) = 0 \;.
\end{equation}
Here, $M(x)$ is the matrix defined in Eq.\ (\ref{M}). The solution to
the Schr\"odinger equation (\ref{eq:SchroedingerM}) is therefore given by
\begin{equation}
  \label{eq:psi_x}
  \tilde \psi (x) = S(x,x_0) \tilde \psi_0 \;.
\end{equation}
It follows with $||\tilde\psi(x)|| = ||\psi(x)||$ that the
Lyapunov exponent can be expressed in terms of the $S$-matrix,
\begin{equation}
  \label{eq:LyapunovS}
  \gamma (\omega) = \lim_{L \to \infty}
  \frac{1}{L} \ln \left|\left|S(L,0)\, \tilde \psi_0 \right|\right| \;.
\end{equation}
Due to $|S_{11}|^2 = |S_{12}|^2 + 1$, $S_{22} = S_{11}^{*}$ and
$S_{21} = S_{12}^{*}$, all $S$-matrix elements grow equally fast and we
have\footnote{It should be noted that the above reasoning is not true in the 
unprobable case where $\tilde\psi_0$ is very close to the eigenvector
to the smallest eigenvalue
of $S$. However, since the definition of the Lyapunov exponent involves
the limit $L \to \infty$, this only happens with zero probability.}
\begin{equation}
  \label{eq:LyapunovThouless}
  \gamma
       (\omega) = \lim_{L \to \infty} 
  \frac{1}{L} \ln \left|S_{22}(L,0)\right| \;.
\end{equation}
Comparing this equation with Eq.\ (\ref{eq:Gamma}), we see that the
inverse localization length
$\ell^{-1}(\omega) \equiv \gamma(\omega)$ is in fact equal to
$\textrm{Re}\, \Gamma(\omega)$. The equation
\begin{equation}
  \label{eq:Thouless}
  {\setlength{\fboxsep}{2mm} \fbox {$ \ \displaystyle     
      \ell^{-1}(\omega) = \textrm{Re}\, \Gamma(\omega)  $}} 
\end{equation}
is known as the {\em Thouless formula} \cite{Thouless72} and was
first shown to be valid for the FGM by Hayn and John \cite{Hayn87} in
a different and more complicated way. But note that these authors have 
only derived
Eq.\ (\ref{eq:Thouless}) up to an integration constant which they had to
determine by different means.

\subsubsection{Localization length at $\omega=0$ for real $\Delta(x)$}

Although we are going to postpone detailed calculations of the DOS and
the localization length to the next sections, let us now consider the
localization length at frequency $\omega=0$ for a real disorder
potential $\Delta(x)$ and $V(x)=0$. In this case, $M(x)= -i\sigma_2
\Delta(x)$, so that the $S$-matrix may be expressed without the
path-ordering operator as
\begin{eqnarray}
  S(x,x') = \cosh
  \left[\int_{x'}^{x}\Delta(y)\ dy \right] \sigma_0 - \sinh
  \left[\int_{x'}^{x}\Delta(y)\ dy \right] \sigma_2 \;. \qquad
  \label{eq:S-matrix-0omega}
\end{eqnarray}
It follows from Eq.\ (\ref{eq:LyapunovThouless}) that the
inverse localization length at $\omega = 0$ is given by 
\begin{eqnarray}
  \label{eq:localizationlengthforomega-0} 
  \ell^{-1} (0)
  = \lim_{L \to \infty} \frac{1}{L} \ln
  \left(\cosh \left[\int_{0}^{L}\Delta(x)\ dx \right] \right)
  =
  \left| \langle \Delta (x) \rangle_x \right| = \left|
    \Delta_{\textrm{\scriptsize av}} \right| \;, \quad
\end{eqnarray}
where we have used the fact that the average
 $\langle \Delta (x) \rangle_x \equiv \lim_{L \to \infty} \frac{1}{L} 
\int_{0}^{L}\Delta(x)\ dx$ is equal to the 
expectation value
$\Delta_{\textrm{\scriptsize av}}$.
The inverse localization length at frequency $\omega=0$ is equal to
the absolute value of the expectation value of the backscattering
potential $\langle \Delta 
(x) \rangle$ and does not depend on the random fluctuations around
this average value. Note that this result is valid for arbitrary
higher correlation functions of 
$\Delta(x)$. In the case $\langle \Delta (x) \rangle = 0$,
the localization length diverges for $\omega=0$ which clearly
distinguishes the point $\omega=0$. The $\omega = 0$ eigenstate is
delocalized!

 \section{Exact results}
{\em In this section, we review exact results for the density of states
  (DOS) and the inverse localization length
  of the fluctuating gap model in the white noise limit and the 
  limit of infinite correlation lengths $\xi$.}
\label{chap:whitenoise}

\subsection{The white noise limit}
For small correlation lengths $\xi$, the disorder of the fluctuating gap
model (FGM) may be approximated by Gaussian white noise. This $\xi
\to 0$ limit is of special interest because in this case the disorder
at different space points is uncorrelated which basically admits for an
exact analytic solution of the model. Various methods may now be
applied to find analytic results for the density of states
(DOS). Ovchinnikov and Erikhman 
\cite{Ovchinnikov77} were the first to solve the commensurate case for
which the random backscattering potential $\Delta(x)$ is real. They showed that
in the symmetric phase for which $\langle \Delta(x) \rangle = 0$,
the DOS has a Dyson singularity previously
found by Dyson \cite{Dyson53}. In the case of
$\langle \Delta(x) \rangle \neq 0$ which models a phase below a phase
transition, the DOS either exhibits a
singularity or a pseudogap near the 
Fermi energy depending on the ratio of the disorder and the static
gap. Using the technique of 
$S$-matrix summation, Golub and 
Chumakov \cite{Golub79} confirmed the results by Ovchinnikov and
Erikhman and were also able to solve the incommensurate
case. In the incommensurate case which is described by a complex
backscattering 
potential, there is no singularity and the disorder can only
lead to a filling up of the pseudogap. The incommensurate case was
also considered by Abrikosov and Dorotheyev \cite{Abrikosov82}.

In recent years, the method of
supersymmetry developed by Efetov \cite{Efetov97} has been
established as a powerful tool to describe disordered systems in the
white noise limit. 
First, Hayn and John \cite{Hayn87} re-derived the Ovchinnikov
and Erikhman result for the DOS and were also able to give an
analytic expression for the localization length.
Later, Hayn and Fischbeck \cite{Fischbeck90,Hayn89} used
the method of supersymmetry to generalize the known results for the
integrated DOS to the case of three 
independent disorder parameters which describe forward, backward, and
umklapp scattering.
These solutions include both the commensurate and the incommensurate
case as special cases. 
Finally, both the integrated DOS and the localization length were
calculated by Hayn and Mertsching \cite{Hayn96} in the most general
case with a complex static gap parameter and three disorder
parameters.

In this section, we use the formalism developed in the previous
section and follow the ideas of Lifshits, Gredeskul and
Pastur \cite{Lifshits88} to show that the probability density
for the distribution of the reduced phase $\varphi$ satisfies a
continuity equation. The stationary probability flux of the continuity
equation turns out to be equal to the integrated DOS. We then derive an
equation closely related to a 
Fokker-Planck equation which allows to rederive the integrated DOS
for the most general case exactly. Before considering the most general
case we discuss the commensurate case, i.e. the Ovchinnikov and
Erikhman limit which we can solve in 
analogy to Halperin's calculation of the integrated DOS of a particle
with an effective mass in a white noise disorder potential
\cite{Halperin65}. The treatment of the general case is similar
but more awkward than the Ovchinnikov and Erikhman limit
because instead of a linear differential equation of second order one
has to face a linear differential equation of fourth order. The
equations to determine the integrated DOS and the localization length
are, however, the same as 
those derived by Hayn and Mertsching using the method of supersymmetry
\cite{Hayn96} so that we recover their general results which also 
include the incommensurate case which we will discuss afterwards.

\subsubsection{Equality of the integrated density of states and the
  stationary probability flux}

As shown in the previous section, the averaged integrated DOS can be
written in the thermodynamic limit as [see Eq.\ (\ref{eq:<Gamma>})]
\begin{equation}
  \label{eq:NF}
    \mathcal{N}(\omega) = \langle F_{\omega} (V,\Delta,\varphi)
    \rangle /2\pi \;. 
\end{equation}
$F_{\omega} (V,\Delta,\varphi)$ is linear
in the disorder $V$, $\textrm{Re}\,\Delta$, $\textrm{Im}\,\Delta$ and
given by
\begin{equation}
  \label{eq:F}
  F_{\omega} (V,\Delta,\varphi) = 
  2 (\omega - V) + 
  2\, \textrm{Re}\, \Delta \cos 
  \varphi + 2\, \textrm{Im}\, \Delta \sin 
  \varphi \; .
\end{equation}
The unreduced phase $\varphi(x,\omega)$ satisfies the 
equation of motion (\ref{eq:phiII}) which can also be written as
\begin{equation}
  \label{eq:phiprime}
  \partial_{x} \varphi(x,\omega) = F_{\omega}
  (V(x),\Delta(x),\varphi(x,\omega)) \; . 
\end{equation}

\subsubsection*{Continuity equation}

The space-dependent probability distribution can be defined as
\begin{equation}
P_{\omega}(x,\varphi) = \langle \delta_{2\pi} (\varphi -
\varphi(x,\omega)) \rangle \;,
\end{equation}
where
$\delta_{2\pi}(x) \equiv \sum_{m=-\infty}^{\infty} \delta(x-2\pi m)$
is the $2\pi$-periodic delta function.
A continuity equation may be derived by partially differentiating
$P_{\omega}(x,\varphi)$ with respect to $x$ which 
results in $\partial_x P_{\omega}(x,\varphi) = - \partial_{\varphi} \langle
  \delta_{2\pi} (\varphi - \varphi(x,\omega)) \partial_x \varphi(x,\omega)
  \rangle $.
Replacing $\partial_{x} \varphi(x,\omega)$ by the right-hand side of
Eq.\,(\ref{eq:phiprime}), the {\em continuity equation} reads
\begin{equation}
  \label{eq:continuity}
  {\setlength{\fboxsep}{2mm} \fbox {$ \ \displaystyle 
  \partial_x
  P_{\omega}(x,\varphi) + \partial_{\varphi} 
  J_{\omega}(x,\varphi) = 0 \; ,$}}
\end{equation}
where the {\em probability flux} $J_{\omega}(x,\varphi)$ is given by
\begin{equation}
  \label{eq:J}
  J_{\omega}(x,\varphi) = \langle
  \delta_{2\pi} (\varphi - \varphi(x,\omega)) F_{\omega}
  (V(x),\Delta(x),\varphi(x,\omega)) 
  \rangle \; .
\end{equation}
Letting $x$ go to infinity, the probability distribution and the
probability flux become stationary, i.e. independent of $x$. Due to the
continuity equation (\ref{eq:continuity}), $J_{\omega}$ becomes also
independent of $\varphi$. Integrating the stationary form of
Eq.\,(\ref{eq:J}) with respect to 
$\varphi$ from $0$ to $2\pi$ therefore leads to $2\pi J_{\omega} =
\langle F_{\omega} (V,\Delta,\varphi) \rangle$, so that together with
Eq.\,(\ref{eq:NF}) we find the remarkable
relationship \cite{Lifshits88}
\begin{equation}
  \label{eq:NJ}
  {\setlength{\fboxsep}{2mm} \fbox {$ \ \displaystyle 
  \mathcal{N}(\omega) = J_{\omega} \; . $}}
\end{equation}
The integrated DOS, i.e. the number of states in the energy interval
$(\, 0,\omega]$ per unit length, is equal to the stationary probability
flux. Note that this result is valid for any
disorder potential and is not restricted to the white noise limit
which we will consider in the following.

\subsubsection{White noise and the Fokker-Planck equation}

While for arbitrary finite correlation lengths $\xi$ and 
Gaussian statistics, it does not seem
to be possible to find an exact analytic expression for the
(integrated) DOS, the white noise limit $\xi \to 0$, $V_{\sigma}^2 \xi
\to D_V$, $(\textrm{Re}\,\Delta_{\sigma})^2 \xi \to D_R$, and
$(\textrm{Im} \, \Delta_{\sigma})^2 \xi \to D_I$ admits for an exact
solution. This is due to the fact that in this case the disorder at
different space points is uncorrelated.
In the white noise limit, the disorder is characterized by the
correlation functions
\begin{eqnarray}
  \label{disorder:V}
  \langle V(x) \rangle = 0 \;,\hspace{-2.3mm} &  & \hspace{-2.3mm} 
  \quad  \langle V(x)
  V(x')\rangle = 2 D_V \, \delta (x-x') \; ,\\
  \label{disorder:Re}
  \langle \, \textrm{Re} \, \Delta(x) \rangle =  \textrm{Re} \, \Delta_0 
\;,\hspace{-2.3mm}  &  & \hspace{-2.3mm} \quad
  \langle \, \textrm{Re}\,  \tilde\Delta(x) \, \textrm{Re}\,  \tilde\Delta(x') \rangle
  = 2 D_R \, \delta (x-x') \; ,\\ 
  \langle \, \textrm{Im} \, \Delta(x) \rangle =  \textrm{Im} \, \Delta_0 
 \;,\hspace{-2.3mm}  &  & \hspace{-2.3mm} \quad
  \langle \, \textrm{Im} \, \tilde\Delta(x) \, \textrm{Im} \, \tilde\Delta(x') \rangle
  = 2 D_I \, \delta (x-x')\; ,
\end{eqnarray}
where $\tilde\Delta(x) \equiv \Delta(x) -\Delta_{0}$. It is no loss of
generality to assume that the first moment of the forward scattering
potential $V(x)$ vanishes because a finite value would only lead to a
renormalization of $\omega$.
Since the probability distribution of the disorder is assumed to be
Gaussian, 
higher correlation functions are 
simply given by Wick's theorem.

To cast the continuity equation into a Fokker-Planck equation, we make
use of the Gaussian nature of the disorder, so that for a functional
$f\{V(y)\}$ of the disorder $V(y)$ we have
$\langle V(x) f\{V(y)\} \rangle = \int dx' \ \langle V(x) V(x') \rangle
  \  \left\langle \frac{\delta f \{ V(y)\} }{\delta V(x')} \right\rangle$,
where in the last term we take the functional derivative of
$f\{V(y)\}$ with respect to $V(x')$. Using
Eq.\,(\ref{disorder:V}), this simplifies to
\begin{equation}
  \label{eq:GaussianaverageII}
  \langle V(x) f\{V(y)\} \rangle = 2 D_V \left\langle \frac{\delta f \{
    V(y)\} }{\delta V(x)} \right\rangle \; . 
\end{equation}
If we want to apply this
relation to Eq.\,(\ref{eq:J}), $f\{V(y)\}$ 
has to be replaced by the $2\pi$-periodic delta-function
$\delta_{2\pi} (\varphi - \varphi(x,\omega))$ whose phase
$\varphi(x,\omega)$ is a functional of the disorder involving the
disorder at all space points $y \le x$. 
Using the chain rule for the functional derivative, we get 
\begin{equation}
\langle
\delta_{2\pi} (\varphi(x,\omega) - \varphi)\, V(x) \rangle = - 2 D_V
\partial_{\varphi} \left\langle
\delta_{2\pi} (\varphi(x,\omega) - \varphi)\, \frac{\delta
  \varphi(x,\omega)}{\delta V(x)} \right\rangle \;.
\end{equation}
Writing $\varphi(x,\omega)$ as $\varphi(x,\omega) - \varphi(0,\omega) =
\int_0^x dx' \ F_{\omega} (V(x'),\Delta(x'),\varphi(x',\omega))$,
we find $
\frac{\delta \varphi(x,\omega)}{\delta V(x')} = -2
\theta(x-x')$.
Since in our case $x=x'$, $\theta(0)$ needs to be defined carefully
(see also Itzykson and Drouffe \cite{Itzykson89}).
Recalling that we have introduced the delta function
$\delta(x)$ as the limit $\xi \to 0$ of a symmetric function of $x$, we
see that to maintain this symmetry, we have to define $\theta(0)= 1/2$.
It therefore follows with $P_{\omega}(x,\varphi) = \langle
\delta_{2\pi} (\varphi - \varphi(x,\omega)) \rangle$ that
\begin{equation}
  \langle
\delta_{2\pi} (\varphi - \varphi(x,\omega))\, V(x) \rangle =  2 D_V
\partial_{\varphi} P_{\omega}(x,\varphi) \; .
\end{equation}
Similarly, we can show that 
\begin{eqnarray}
\langle \delta_{2\pi} (\varphi - \varphi(x,\omega))\,
\textrm{Re} \,\tilde\Delta(x) \rangle & \hspace{-2.3mm} = \hspace{-2.3mm} & - 
2 D_R \partial_{\varphi} ( \cos 
\varphi \, P_{\omega}(x,\varphi)) \; , \\
\langle \delta_{2\pi} (\varphi - \varphi(x,\omega))
\textrm{Im} \, \tilde\Delta(x) \rangle & \hspace{-2.3mm} = \hspace{-2.3mm} & -2 D_I
\partial_{\varphi} ( 
\sin 
\varphi \, P_{\omega}(x,\varphi)) \; . 
\end{eqnarray}
Making use of these relations, the probability flux
$J_{\omega}(x,\varphi)$ is given by
\begin{eqnarray}
 \lefteqn{
 J_{\omega}(x,\varphi) = 2[\omega + 
 \textrm{Re} \, \Delta_{0} \cos 
 \varphi
  + \textrm{Im} \,  \Delta_{0} \sin 
  \varphi] P_{\omega}(x,\varphi) }
  \nonumber \\
   & \qquad & {} - 4 \left[ D_V \partial_{\varphi} P_{\omega}(x,\varphi) 
    + 
     D_R \cos 
     \varphi \, \partial_{\varphi} \left(\cos 
       \varphi \,
      P_{\omega}(x,\varphi) \right) \right. \nonumber \\
  & \qquad & \left. \hspace{0.0mm} {} +
    D_I \sin 
    \varphi \, \partial_{\varphi} \left(\sin 
      \varphi \,
      P_{\omega}(x,\varphi) \right) \right] \; . 
\end{eqnarray}
The probability flux can also be written as
\begin{equation}
  \label{eq:JII}
  J_{\omega}(x,\varphi) = A_{\omega}(\varphi)\, P_{\omega}(x,\varphi) -
  \frac{1}{2} \partial_{\varphi} 
  \left[B(\varphi) \, P_{\omega}(x,\varphi) \right] \; ,
\end{equation}
where
\begin{eqnarray}
  \label{eq:A}
  A_{\omega}(\varphi) &\hspace{-2.3mm} = &\hspace{-2.3mm} 2\left[\omega + 
    \textrm{Re} \, \Delta_{0} \cos 
    \varphi
  + \textrm{Im} \,  \Delta_{0} \sin 
  \varphi\right] - 4(D_R - D_I) \cos \varphi \,
  \sin \varphi \; , \qquad \\
  B(\varphi) &\hspace{-2.3mm} = &\hspace{-2.3mm} 8 \left[ D_V + D_R \cos^2 
    \varphi + D_I \sin^2 
    \varphi
  \right] \; .
\end{eqnarray}
Note that the forward scattering disorder $D_V$ only leads to a
renormalization of $D_R$ and 
$D_I$. We therefore define $\tilde D_R \equiv D_R + D_V$ and
$\tilde D_I \equiv D_I + D_V$.

Eq.\,(\ref{eq:JII}) together with the continuity equation explicitly
shows that the probability distribution satisfies the following
one-dimensional {\em Fokker-Planck equation} \cite{Gardiner83,vanKampen81}:
\begin{equation}
  \label{eq:fokkerplanck}
  {\setlength{\fboxsep}{2mm} \fbox {$ \ \displaystyle 
  \partial_x P_{\omega}(x,\varphi) = - \partial_{\varphi}
  \left[A_{\omega}(\varphi) P_{\omega}(x,\varphi) \right] + \frac{1}{2}
  \partial_{\varphi}^{\, 2}
  \left[B(\varphi) P_{\omega}(x,\varphi) \right] \; . $}}
\end{equation}
To find an analytic expression for the integrated DOS, we use the fact
that, as shown above, 
$\mathcal{N}(\omega)$ is equal to the stationary probability flux.
A good starting point to calculate the integrated DOS is therefore the
stationary form of Eq.\,(\ref{eq:JII}), which is nothing but the 
integrated stationary Fokker-Planck equation with the constant of
integration being the stationary probability flux that is equal to
the integrated DOS 
\begin{equation}
  \label{eq:N}
  \mathcal{N} (\omega) = A_{\omega}(\varphi)\, P_{\omega}(\varphi) -
  \frac{1}{2} \partial_{\varphi} 
  \left[B(\varphi)\, P_{\omega}(\varphi) \right] \; .
\end{equation}
In principle, one could first find a solution for $\left[B(\varphi) \,
  P_{\omega}(\varphi) \right]$ to this first-order differential 
equation subject to the boundary condition $[B(2\pi) \,
P_{\omega}(2\pi)] = [B(0) \, P_{\omega}(0)]$ with $\mathcal{N} (\omega)$
as a parameter and could then use 
the normalization condition 
$\int_{0}^{2\pi} d\varphi \, P_{\omega}(\varphi) = 1$ to determine
$\mathcal{N} (\omega)$. While this procedure works quite well in the
incommensurate case with $D_R = D_I$ and after a variable transformation
also in the
commensurate case (without forward scattering) for which $D_I = D_V =
0$, the 
most general case considered here seems to defy such a treatment.
Here, we therefore present an alternative method which allows to
recover all known results in the white noise limit.

\subsubsection{The density of states in the white noise limit}

To find the (integrated) DOS for arbitrary parameters $D_R, D_I, D_V$
and complex $\Delta_{0}$, we first use the variable transformation
\begin{equation}
  \label{eq:variabletransformation}
  z = \tan \left(\frac{\varphi}{2} - \frac{\pi}{4} \right) \; . 
\end{equation}
The trigonometric functions $\sin \varphi$ and $\cos \varphi$ can now
be expressed in terms of $z$ as $\sin \varphi = \frac{1-z^2}{1+z^2}$ 
, and
$\cos \varphi = -\frac{2z}{1+z^2}$.
Since
$\frac{d \, z}{d \, \varphi} = 
\frac{1}{2} (1+z^2)$, we find $\partial_{\varphi} = \frac{1}{2} (1+z^2)
\partial_{z}$, and the probability distribution $P(\varphi)$ has to be
replaced by $\frac{1}{2} (1+z^2) P(z)$.
Using these relations, Eq.\,(\ref{eq:N}) turns into
\begin{eqnarray}
  \label{eq:NII}
  \mathcal{N} (\omega)  =  \left[(\omega + \textrm{Im} \, \Delta_{0}) -
  2(\textrm{Re} \,\Delta_{0} - (2\tilde D_R - \tilde D_I)) z 
  + (\omega -
  \textrm{Im} \, \Delta_{0}) z^2 \right. \nonumber \\ 
  \left. {} + 2 \tilde D_I z^3 \right] P(z) 
  - \tilde D_I \partial_{z} \left( \left[1+2 \tilde D_I^{-1} (2 \tilde
      D_R-\tilde D_I) z^2 + z^4 
    \right] P(z)
  \right) \; .
\end{eqnarray}
Taking the Fourier transform of this equation leads to
\begin{eqnarray}
  \label{eq:NFT}
  \lefteqn{ 2 \pi \mathcal{N} (\omega) \delta(k) =
    (\omega + \textrm{Im} \, \Delta_{0}) \tilde P(k) -2 i (\textrm{Re} \,\Delta_{0}
    - (2\tilde D_R -\tilde D_I)) \tilde P'(k) } \nonumber \\  
    & \qquad \quad & {} - (\omega -
    \textrm{Im} \, \Delta_{0}) 
    \tilde P''(k) -2i \tilde D_I \tilde P'''(k) \nonumber \\  
  & \qquad \quad & {} - i \tilde D_I k \left[\tilde P(k)
    - 2 \tilde D_I^{-1} (2 \tilde D_R-\tilde D_I) \tilde P''(k) +
    \tilde P''''(k) 
    \right] \; , \qquad
\end{eqnarray}
where 
\begin{equation} 
\tilde P(k) = \int_{-\infty}^{\infty} e^{-ikz} P(z) \, dz
\end{equation}
 is the Fourier transform of
$P(z)$ which is also known as the {\em characteristic
  function} \cite{Feller71}. Normalization of the probability
distribution $P(z)$ implies $\tilde P(k=0) = 1$.
Assuming higher derivatives of
$P(z)$ to be integrable, the other boundary con\-di\-tions are given by
demanding that $\tilde P(k) \to 0$ sufficiently rapidly
as $|k| \to \infty$.

\subsubsection{The commensurate case without forward scattering}

\begin{figure}[thb]
\begin{center}
\psfrag{omega}{\hspace{-0.1mm}\large$\omega / \Delta_0$}
\psfrag{dos}{\hspace{-0.5mm}\large$\rho(\omega)/\rho_{0}$}
\psfrag{ll}{\large\hspace{-4.6mm}$\ell^{-1}(\omega)/D$}
\psfrag{omega2}{\hspace{1.5mm}$\omega / D$}
\psfrag{dos2}{\hspace{1.0mm}$\rho(\omega)/\rho_{0}$}
\psfrag{ll2}{\hspace{-2.6mm}$\ell^{-1}(\omega)/D$}
\psfrag{top}{}
\psfrag{0i}{\hspace{0.5mm}\Large $0$}
\psfrag{1i}{\hspace{0.5mm}\Large $1$}
\psfrag{2i}{\hspace{0.5mm}\Large $2$}
\psfrag{3i}{\hspace{0.5mm}\Large $3$}
\psfrag{4i}{\hspace{0.5mm}\Large $4$}
\psfrag{6i}{\hspace{0.5mm}\Large $6$}
\psfrag{0}{$0$}
\psfrag{0.5}{$0.5$}
\psfrag{1}{$1$}
\psfrag{2}{$2$}
\psfrag{3}{$3$}
\psfrag{0.1}{\small $0.1$}
\psfrag{0.3}{\small $0.3$}
\psfrag{0.50}{\small $0.5$}
\psfrag{1.0}{\small $1.0$}
\psfrag{1.50}{\small $1.5$}
\psfrag{2.0}{\small $2.0$}
\psfrag{2.5}{\small $2.5$}
\psfrag{3.0}{\small $3.0$}
\psfrag{10.0}{\small $10.0$}
\psfrag{30.0}{\small $30.0$}
\psfrag{100.0}{\small $100.0$}
\psfrag{inf}{$\infty$}
\epsfig{file=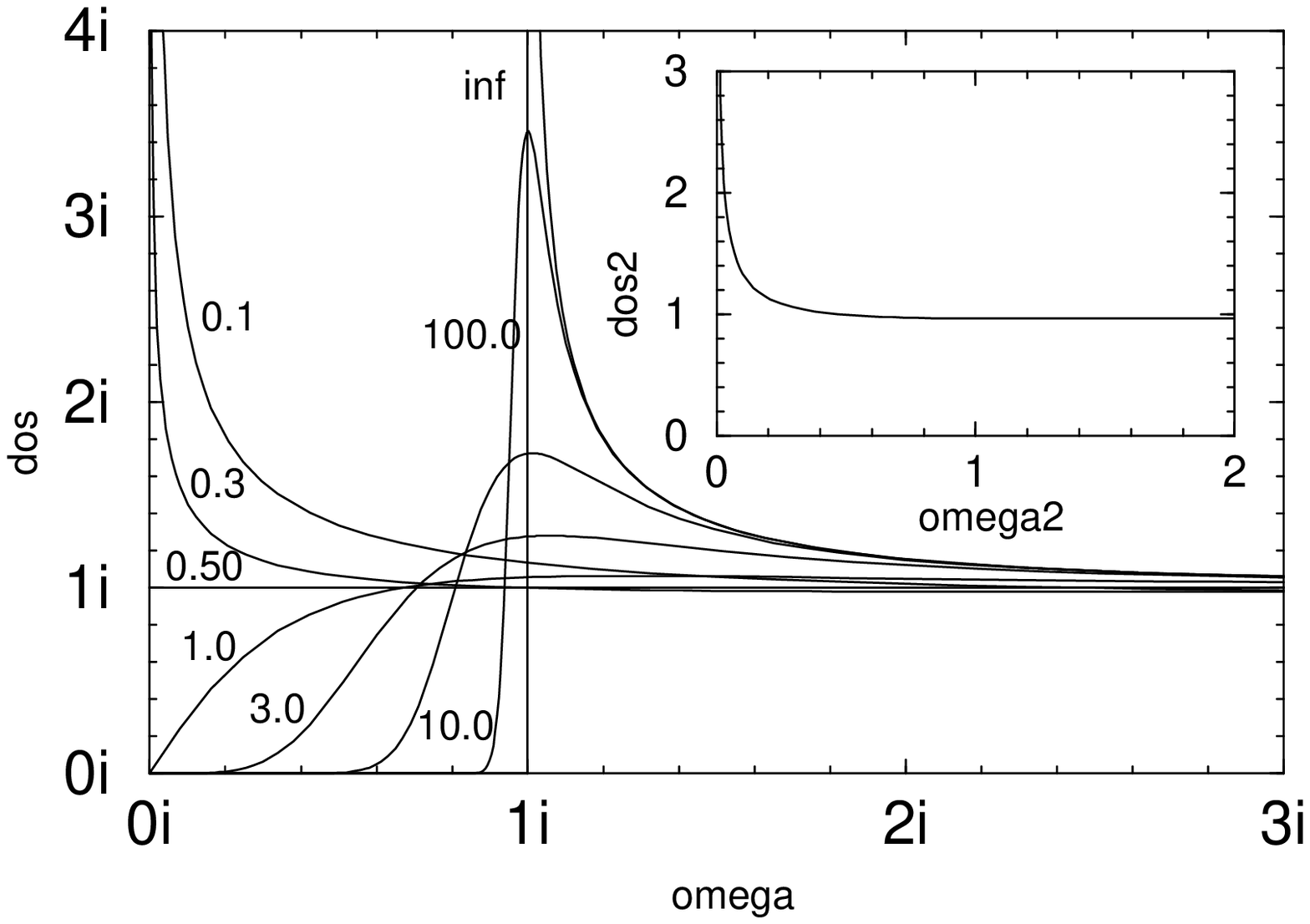,width=11.50cm}
\epsfig{file=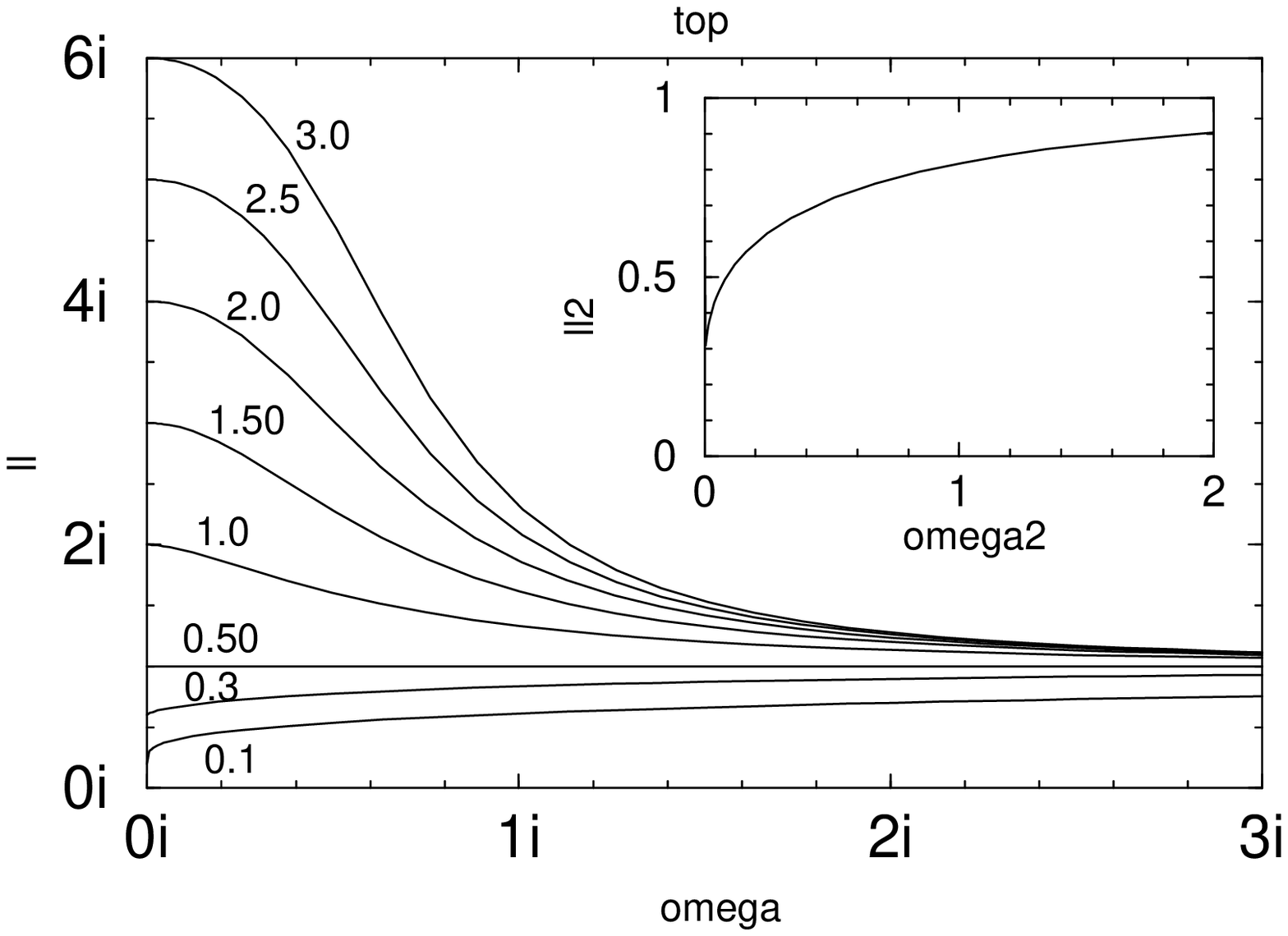,width=11.50cm}
\end{center}
\caption{The DOS $\rho(\omega)$ and the inverse localization length
  $\ell^{-1}(\omega)$ in the white
  noise limit for the commensurate case plotted versus $\omega /
  \Delta_0$ for $\nu=\Delta_0/2D = 
  0.1,0.3,0.5,1.0,3.0,10.0,100.0,\infty$ or $\nu=\Delta_0/2D = 
  0.1,0.3,0.5,1.0,1.5,2.0,2.5,3.0$, respectively. The inserts show the 
respecting graphs for the case $\Delta_0 =0$, i.e.\ $\nu =0$.}
\label{fig:wn-commensurate}
\end{figure}

Before we proceed with the most general case, let us first consider
the Ovchinnikov and Erikhman limit, i.e. the commensurate case without 
forward scattering. In this case, $\Delta_{0}$ is real and may be
assumed to be positive, $D_I = D_V =
0$ and we set $D \equiv D_R$. Eq.\ (\ref{eq:NFT}) reduces to a second order
differential equation 
with the boundary conditions $\tilde P(0) = 1$ and $\tilde P(k) \to
0$ as $|k| \to \infty$:
\begin{equation}
    \label{eq:NFT2ndorder}
  2 \pi \mathcal{N} (\omega) \delta(k) = \omega \tilde P(k) +4 i D
  \left(1-\frac{\Delta_{0}}{2D}\right) 
  \tilde P'(k) + 4 i D \left(k+\frac{i\omega}{4D}\right) \tilde
  P''(k)\; .
\end{equation}
Integrating this equation from $-\epsilon$ to $+\epsilon$ with
$\epsilon \to 0^{+}$ leads to 
\begin{equation}
2\pi \mathcal{N}(\omega) =
-\omega\left[\tilde P'(0^{+}) - \tilde P'(0^{-})\right] \;.
\end{equation}
Since $P(z)$
is real, $\tilde P(-k) = \tilde P^{\ast} (k)$, and $\tilde P'(-k) =
-\tilde P^{\prime\ast} (k)$, which implies 
\begin{equation}
\mathcal{N}(\omega) = -
\frac{\omega}{\pi} \, \textrm{Re} \, \tilde P'(0^{+}) \;.
\end{equation}
$\tilde P'(0^{+})$
can be determined by first finding a solution to the differential
equation (\ref{eq:NFT2ndorder}) for $k>0$ which vanishes as $k$ approaches
infinity and then normalizing this solution such that $\tilde P(0)
=1$. Because Eq. (\ref{eq:NFT2ndorder}) is homogeneous for $k>0$,
it therefore follows that if $g(k)$ is any solution to the
homogeneous differential equation which obeys $g(k) \to 0$ as $k \to
\infty$, then $\mathcal{N}(\omega) = -\frac{\omega}{\pi} \, \textrm{Re} \,
\left[g'(0)/g(0)\right]$. 
We introduce $y(t) \equiv g(k)$ with $t=-i(k+i\omega/4D)$. 
The integrated DOS is now given by 
\begin{equation}
  \label{IDOS2ndorder}
  \mathcal{N}(\omega) = \frac{\omega}{\pi} \, \textrm{Im} 
  \left(\frac{y'\left(\frac{\omega}{4D}\right)}{y\left(\frac{\omega}{4D}\right)}\right) \; ,           
\end{equation}
where $y(t)$ has to satisfy the differential equation
\begin{equation}
  \label{eq:almostBessel}
  t \, y''(t) + \left(1-\frac{\Delta_0}{2D}\right) y'(t)
  + \frac{\omega}{4D} y(t) = 0 \;,
\end{equation}
with the only restriction that $y(t)$ should approach zero as $t$ goes
to $i \infty$.
The general solution of this differential equation can be expressed in
terms of 
a linear combination of Bessel functions of the first and second kind,
$J_{\nu}(x)$ and $N_{\nu}(x)$ (Abramowitz and Segun (A\&S)
\cite{Abramowitz65}). The only solution which 
satisfies the boundary condition involves the Hankel function of the first
kind, $H_{\nu}^{(1)}(x) = J_{\nu}(x)+iN_{\nu}(x)$.
Introducing 
\begin{equation}
\nu =\frac{\Delta_0}{2D}\;,
\end{equation}
we find $y(t)= \left(\nu/2\right)^t
H_{\nu}^{(1)} \left(\left(\omega t/D\right)^{1/2}\right)$. Because the
prefactor is real and we only need the imaginary part of the quotient
$y'(\omega/4D) / y(\omega/4D)$, Eq.\ (\ref{IDOS2ndorder}) turns into
\begin{equation}
   \label{IDOS2ndorderII}
  \mathcal{N}(\omega) = \frac{\omega}{\pi} \, \textrm{Im}
  \left(\frac{\left[H_{\nu}^{(1)} \left(\frac{\omega}{2D}\right)
      \right]'}{H_{\nu}^{(1)} \left(\frac{\omega}{2D}\right)} \right)
  \;.
\end{equation}
Note that this equation is valid for arbitrary $\omega$ and is even
analytic in the upper half plane. In the following discussion of the
Ovchinnikov and Erikhman limit, we will again restrict ourselves to
$\omega > 0$. It follows
\begin{equation}
  \label{IDOS2ndorderIII}
    \mathcal{N}(\omega) = \frac{\omega}{\pi} \,
    \frac{J_{\nu} \left(\frac{\omega}{2D}\right) N_{\nu}^{\prime}
      \left(\frac{\omega}{2D}\right)- 
      J_{\nu}^{\prime}  \left(\frac{\omega}{2D}\right) N_{\nu}
      \left(\frac{\omega}{2D}\right)}{J_{\nu}^2
      \left(\frac{\omega}{2D}\right) + N_{\nu}^2 
      \left(\frac{\omega}{2D}\right)}   \;. 
\end{equation}
The numerator can be simplified by using the Wronski relation [A\&S,
Eq.\ (9.1.16)] 
$
J_{\nu}
\left(x\right) N_{\nu}^{\prime} \left(x\right)- J_{\nu}^{\prime}
\left(x\right) N_{\nu} \left(x\right) = \frac{2}{\pi x}$,
so that as 
our final expression for the integrated DOS we are left with
\begin{equation}
  \label{IDOSwhitenoisecommensurate}
  {\setlength{\fboxsep}{2mm} \fbox {$ \ \displaystyle 
  \mathcal{N}(\omega) = \frac{4 
    D}{\pi^2 \left[J_{\nu}^2 
      \left(\frac{\omega}{2D}\right) + N_{\nu}^2
      \left(\frac{\omega}{2D}\right)\right]} \; . $}}
\end{equation}
This result was first obtained by Ovchinnikov and
Erikhman \cite{Ovchinnikov77} in a more complicated manner. 
Differentiating Eq.\ (\ref{IDOSwhitenoisecommensurate}) with respect
to $\omega$, we find
\begin{equation}
  \label{DOSwhitenoisecommensurate}
  \rho (\omega) = - \frac{4 \left[ J_{\nu}
      \left(\frac{\omega}{2D}\right) J_{\nu}'
      \left(\frac{\omega}{2D}\right) + N_{\nu} 
      \left(\frac{\omega}{2D}\right) N_{\nu}' 
      \left(\frac{\omega}{2D}\right) \right]}{\pi^2 \left[J_{\nu}^2
      \left(\frac{\omega}{2D}\right) + N_{\nu}^2 
      \left(\frac{\omega}{2D}\right)\right]^{2}} \; ,
\end{equation}
where the derivatives of Bessel functions could in principle also be 
written in terms of Bessel functions.
For $\Delta_0 = 0$ which implies $\nu = 0$, $D$ is the only
characteristic energy scale, 
and it is useful to measure all energies in terms of
$D$. 
The insert at the top of Fig.\ \ref{fig:wn-commensurate} shows the DOS
$\rho(\omega)$ 
plotted versus
$\omega/D$. The DOS clearly exhibits a singularity near $\omega = 0$ whose
asymptotic behavior can be found by noting that $J_{0}(0)$ is finite
while $N_{0}(x)$ diverges logarithmically for small $x$. It 
follows\footnote{Note that the argument of the logarithm in the
  asymptotic form given by Ovchinnikov and Erikhman deviates by a
  factor $1/2$ from our result. Nevertheless both expressions lead to
  the same asymptotic 
  behavior. To take into account next to leading terms one has to use
  $N_0(x) = 2/\pi \ln(a x) + \mathcal{O}(x)$, where $a=e^{\gamma}/2$
  which follows from A\&S, Eq.\ (9.1.89). Here,
  $\gamma \approx 0.5772$ is Euler's 
  constant which leads to $a \approx 0.8905$. This value is closer to
  our choice $a=1$ than to Ovchinnikov's and Erikhman's choice $a = 1/2$.}
 with $N_{0}(x)
\sim (2/\pi) \, \ln x$ [A\&S, Eq.\ 9.1.8]
\begin{equation}
    \label{eq:NDysonsingularity}
  \mathcal{N}(\omega) \sim \frac{
    D}{\ln^2
    (\omega / 2D)} \;, 
\end{equation}
such that the asymptotics of the DOS for $\Delta_{0} = 0$ 
is given by
\begin{equation}
    \label{eq:Dysonsingularity}
  \rho(\omega) \sim - \frac{1}{(\omega/2D) \ln^3(\omega / 2D)} \;.
\end{equation}
Recall that to generalize this result towards arbitrary frequencies
$\omega$,  we have to replace $\omega$ by
$|\omega|$. 
The singularity described by Eq.\,(\ref{eq:Dysonsingularity}) is called a
{\em Dyson singularity} and was first found by Dyson in a different model
\cite{Dyson53} involving also off-diagonal disorder. Outside this
singularity, the DOS is almost equal to the DOS
of the disorder-free model, taking its minima
$\rho(\omega^{\ast}) = 0.9636\, \rho_0$ at $\omega^{\ast} = \pm 1.2514\, D$.

If $\Delta_0 \neq 0$, $\Delta_0$ is another characteristic energy
scale. $\nu \equiv \Delta_0 / 2D$ then basically gives the ratio of the
two relevant energy scales $\Delta_0$ and $D$. 
The DOS plotted against $\omega/\Delta_0$ for different
values of the parameter $\nu$ is shown at the top of Fig.\
\ref{fig:wn-commensurate}. 
While the singularity only survives
for $\nu < 1/2$, the DOS is constant for $\nu = 1/2$ and for $\nu > 1/2$
the effects of the constant gap $\Delta_0$ dominate 
those due to the disorder and a pseudogap emerges.

The algebraic dependence of the
DOS at small $\omega$ on the parameter $\nu$ can be found by using
again an asymptotic expansion of the Bessel functions. If $\nu > 0$ is
fixed and $x \to 0$, the Bessel function $J_{\nu}(x)$ is finite and
$N_{\nu}(x) 
\sim -(1/\pi) \Gamma (\nu) (x/2)^{-\nu}$ [A\&S, Eq.\ (9.1.9)]. It
follows 
\begin{equation}
\label{eq:IDOS:wn:neq}
\mathcal{N} (\omega) \sim \frac{4\, D}{\Gamma^2(\nu)}\,
\left(\frac{\omega}{4D}\right)^{2\nu} \;,
\end{equation}
so that
\begin{equation}
\label{eq:DOS:wn:neq}
  \rho(\omega) \sim \frac{2\nu}{\Gamma^2(\nu)}
  \left(\frac{\omega}{4D}\right)^{2\nu-1} \;. 
\end{equation}
This implies that for $\nu < 1/2$ the DOS in fact diverges algebraically
as $\omega$ approaches zero because in this case the exponent is
negative.  Although the algebraic divergence differs from
the divergence found in Dyson's model, we will nevertheless refer to
this singularity as a Dyson singularity.
For $\nu > 1/2$,
however, the exponent is positive and the DOS
vanishes algebraically.

For large $\nu$ the
disorder becomes irrelevant and the DOS reduces to the mean-field result
$\rho(\omega) = \rho_0 \, \theta(\omega^2 - \Delta_0^2) \omega /
(\omega^2 - \Delta_0^2)^{1/2}$.

\subsubsection*{Localization length}

Since $\Gamma(\omega) \equiv \ell^{-1}(\omega) - i\pi
\mathcal{N}(\omega)$ is an analytic function in the upper half
plane, we can also easily find an analytic expression for the
localization length $\ell^{-1}(\omega)$. It follows from Eq.\
(\ref{IDOS2ndorderII}) that up to a constant
\begin{equation}
   \label{IDOS2ndorderIV}
  {\setlength{\fboxsep}{2mm} \fbox {$ \ \displaystyle 
  \Gamma (\omega) = -\omega \, \frac{\left[H_{\nu}^{(1)}
        \left(\frac{\omega}{2D}\right) 
      \right]'}{H_{\nu}^{(1)} \left(\frac{\omega}{2D}\right)} 
  \;. $}}
\end{equation}
The inverse localization length is now given by
\begin{equation}
  \label{eq:inverselocalization-wn}
  \ell^{-1}(\omega) = \textrm{Re} \, \Gamma (\omega) = - \omega \frac{
    J_{\nu} 
    \left(\frac{\omega}{2D}\right) J_{\nu}'
    \left(\frac{\omega}{2D}\right) + N_{\nu} 
    \left(\frac{\omega}{2D}\right) N_{\nu}' 
    \left(\frac{\omega}{2D}\right) }{J_{\nu}^2
    \left(\frac{\omega}{2D}\right) + N_{\nu}^2 
    \left(\frac{\omega}{2D}\right)} \;. 
\end{equation}
Comparing the right-hand side of this equation with Eqs.\
(\ref{IDOSwhitenoisecommensurate}) and
(\ref{DOSwhitenoisecommensurate}), we find
\begin{equation}
  \label{eq:localizationlength-DOS}
  {\setlength{\fboxsep}{2mm} \fbox {$ \ \displaystyle 
  \ell^{-1}(\omega) = D\, \frac{\omega\,
    \rho(\omega)}{\mathcal{N}(\omega)} \;.$}}
\end{equation}
This equation is exact and can already be found Ref.\ \cite{Lifshits88}.

If $\Delta_{0}=0$, it follows from Eqs.\ (\ref{eq:NDysonsingularity})
and (\ref{eq:Dysonsingularity}) that 
$\ell^{-1}(\omega)$ vanishes logarithmically as $\omega$ approaches zero,
\begin{equation}
  \label{eq:l:asymptotics}
  \ell^{-1}(\omega) \sim - \frac{2D}{\ln\left({\omega}/{2D}\right)} \;.
\end{equation}
Using Eqs.\ (\ref{eq:IDOS:wn:neq}) and (\ref{eq:DOS:wn:neq}), we get
for arbitrary $\Delta_{0}$
\begin{equation}
  \label{eq:l:wn:omega0}
  \ell^{-1}(0) = \Delta_{0} \;, 
\end{equation}
which, as can be seen from Eq.\ (\ref{eq:l:asymptotics}) is also true
for $\Delta_0 = 0$. Eq.\ (\ref{eq:l:wn:omega0}) agrees with
Eq.\ (\ref{eq:localizationlengthforomega-0}), so 
that $\Gamma(\omega)$ involves no extra constant.
For large frequencies, $\rho(\omega) \to \rho_0$ and $\mathcal{N}(\omega)
\to \rho_0 \, \omega$, such that 
\begin{equation}
  \label{eq:localizationlengthlimitlargeomega}
  \ell^{-1}(\omega) \to D \;.
\end{equation}
A plot of the inverse localization length
$\ell^{-1}(\omega)$ for various 
values of $\nu = \Delta_0/2D$ is given at the bottom of Fig.\
\ref{fig:wn-commensurate}.

\subsubsection{Solving the general case with arbitrary parameters $D_R$,
  $D_I$, $D_V$ and $\Delta_0$}

While in the commensurate case we only had to solve a differential
equation of second order, for $D_I \neq 0$, Eq.\,(\ref{eq:NFT}) is a
differential equation of fourth order and more difficult to
solve. 
Without loss of generality we may assume that $\tilde D_R > \tilde
D_I$ (later we 
can also take the limit $\tilde D_R \to \tilde D_I$).
Integrating Eq.\,(\ref{eq:NFT}) from $-\epsilon$ to
$+\epsilon$ with $\epsilon \to 0^{+}$, we can proceed as before and
express the 
integrated DOS in terms of $\tilde P (k)$ and derivatives thereof
evaluated at $\pm 0^{+}$. Integrating the terms linear in $k$ by parts and
using again the fact that $\tilde P^{(n)}(-k) = (-1)^{n}
\tilde P^{(n)\ast}(k)$, we find
\begin{equation}
  \label{eq:NintermsofP}
  \mathcal{N}(\omega) = \frac{1}{\pi} \textrm{Im} \left[-i (\omega -
    \textrm{Im} \, \Delta_0) \tilde P'(0^{+}) + D_{I} \tilde P''(0^{+})
  \right] \;.
\end{equation}
So, if $y(k)$ is any solution to the homogeneous differential equation
\begin{eqnarray}
  \label{eq:NFTy}
  \lefteqn{(\omega + \textrm{Im} \, \Delta_{0}) y(k) -2 i (\textrm{Re} \,\Delta_{0}
    - (2 \tilde D_R -\tilde D_I)) y'(k) 
     - (\omega - \textrm{Im} \, \Delta_{0}) y''(k)}
     \nonumber \\ & \  & \hspace{-3mm}{} -2i \tilde D_I y'''(k)
     - i \tilde D_I k \left[y(k)
    - 2 \tilde D_I^{-1} (2 \tilde D_R-\tilde D_I) y''(k) + y''''(k) 
    \right] = 0 \; , 
\end{eqnarray}
which vanishes for $k \to \infty$ sufficiently rapidly,
then the integrated DOS is given by
\begin{equation}
  \label{eq:Nintermsofy}
  \mathcal{N}(\omega) = \frac{1}{\pi} \textrm{Im}\left[-i (\omega -
    \textrm{Im} \, \Delta_0) \frac{y'(0)}{y(0)} + D_{I} \frac{y''(0)}{y(0)}
  \right] \;.
\end{equation}
Again, since $\Gamma(\omega) \equiv \ell^{-1} (\omega) -i\pi
\mathcal{N}(\omega)$ is an analytic function in the upper half plane,
up to a constant we have
\begin{equation}
  \label{eq:Gamma-y}
  \Gamma (\omega) = i (\omega -
    \textrm{Im} \, \Delta_0) \frac{y'(0)}{y(0)} - \tilde D_{I} \frac{y''(0)}{y(0)}
  \;.
\end{equation}
Using the method of supersymmetry invented by Efetov \cite{Efetov97},
Hayn and Mertsching \cite{Hayn96} derived
the set of 
equations (\ref{eq:NFTy}) and (\ref{eq:Gamma-y}) by different
means\footnote{Hayn and Mertsching use a slightly different notation
  but apart from this and some irrelevant different signs their
  expressions are equal to ours.}. 
Applying the method of Laplace transforms they 
found an exact 
expression with the constant of integration chosen such that one
obtains the correct asymptotic behavior for large frequencies
determined in the Born approximation,
$\Gamma(\omega) = D -i\omega
= D_R+D_I -i\omega$. Here, instead of presenting a lengthy derivation,
we will only cite the exact result found in \cite{Hayn96}:
\begin{equation}
  \label{eq:Gamma-Hayn-Mertsching}
  {\setlength{\fboxsep}{2mm} \fbox {$ \ \displaystyle 
  \Gamma (\omega) = 2 D_I + 4 \tilde D_R \left[z(1-z) \frac{F'(z)}{F(z)} + z
    \delta_R - i (1-z) \epsilon \right] \;.$}}
\end{equation}
In this equation, $F(z)$ is the hypergeometric function
\begin{equation}
  \label{eq:hypergeometric}
  F(z)=F ({\textstyle \frac{1}{2}} - i \epsilon + i \delta_I -
    \delta_R , {\textstyle \frac{1}{2}} - i 
  \epsilon - i \delta_I - \delta_R , 1 - 2 i \epsilon ; z ) \;,
\end{equation}
with the parameters $\delta_R$, $\delta_I$, $\epsilon$, and $z$ (in our
notation) given by 
\begin{eqnarray}
  \label{eq:delta}
  \delta_R = \frac{\textrm{Re} \,\Delta_0}{4\left({\tilde D_R
        (\tilde D_R-\tilde D_I)}\right)^{1/2}} & , &   
  \delta_I = \frac{\textrm{Im} \, \Delta_0}{4\left({\tilde D_I
        (\tilde D_R-\tilde D_I)}\right)^{1/2}} \;, \\
  \epsilon = \frac{\omega}{4\left({\tilde D_R \tilde D_I}\right)^{1/2}} & , & 
  z = \frac{\tilde D_R - \tilde D_I}{\tilde D_R} \;.
\end{eqnarray}
Recall that we incorporated the parameter of the forward scattering
disorder, $D_V$, 
into $D_R$ and $D_I$ by defining $\tilde D_R \equiv D_R + D_V$ and
$\tilde D_I 
\equiv D_I + D_V$. Only the additive constant $2D_I$ in Eq.\
(\ref{eq:Gamma-Hayn-Mertsching}) does not get
renormalized by $D_V$. This is due to the fact that for large
frequencies $\Gamma(\omega) \sim D_R + D_I -i\omega$ is independent of $D_V$.
The imaginary part of Eq.\ (\ref{eq:Gamma-Hayn-Mertsching}) determines
the integrated DOS,
\begin{equation}
  {\setlength{\fboxsep}{2mm} \fbox {$ \ \displaystyle 
  \label{eq:IDOS-Hayn-Mertsching}
  \mathcal{N} (\omega) = \rho_{0} \left(\frac{\tilde D_R}{\tilde D_I}\right)^{1/2-2
    \delta_R} \frac{\omega}{|F|^2} \;.$}}
\end{equation}
Taking the confluent limit $D_I
\to 0$ of Eq.\ (\ref{eq:Gamma-Hayn-Mertsching}), one can recover
Eq.\ (\ref{IDOS2ndorderIV}) which describes
the commensurate case without forward scattering. Turning on the
forward scattering disorder $D_V$ in the general expression gradually
removes the singularity in the DOS for $\nu < 1/2$. Below, we are 
only going to discuss the incommensurate case. Other special cases can
be found in \cite{Fischbeck90,Hayn87,Hayn89,Hayn96,Mertsching92}.

\subsubsection{The incommensurate case}

\begin{figure}[tb]
\begin{center}
\psfrag{omega}{\hspace{-0.6mm}\small $\omega / \Delta_0$}
\psfrag{dos}{\hspace{-1.7mm}\small $\rho(\omega)/\rho_{0}$}
\psfrag{ll}{\small \hspace{-4.6mm}$\ell^{-1}(\omega)/D$}
\psfrag{top}{}
\psfrag{0}{\small $0$}
\psfrag{0.5}{\small $0.5$}
\psfrag{1}{\small $1$}
\psfrag{1.5}{\small $1.5$}
\psfrag{2}{\small $2$}
\psfrag{3}{\small $3$}
\psfrag{4}{\small $4$}
\psfrag{5}{\small $5$}
\psfrag{6}{\small $6$}
\psfrag{0.0}{\tiny $0.0$}
\psfrag{0.3}{\tiny $0.3$}
\psfrag{0.50}{\tiny $0.5$}
\psfrag{1.0}{\tiny $1.0$}
\psfrag{1.50}{\tiny $1.5$}
\psfrag{2.0}{\tiny $2.0$}
\psfrag{2.5}{\tiny $2.5$}
\psfrag{3.0}{\tiny $3.0$}
\psfrag{10.0}{\tiny $10$}
\psfrag{30.0}{\tiny $30$}
\psfrag{inf}{\tiny $\infty$}
\epsfig{file=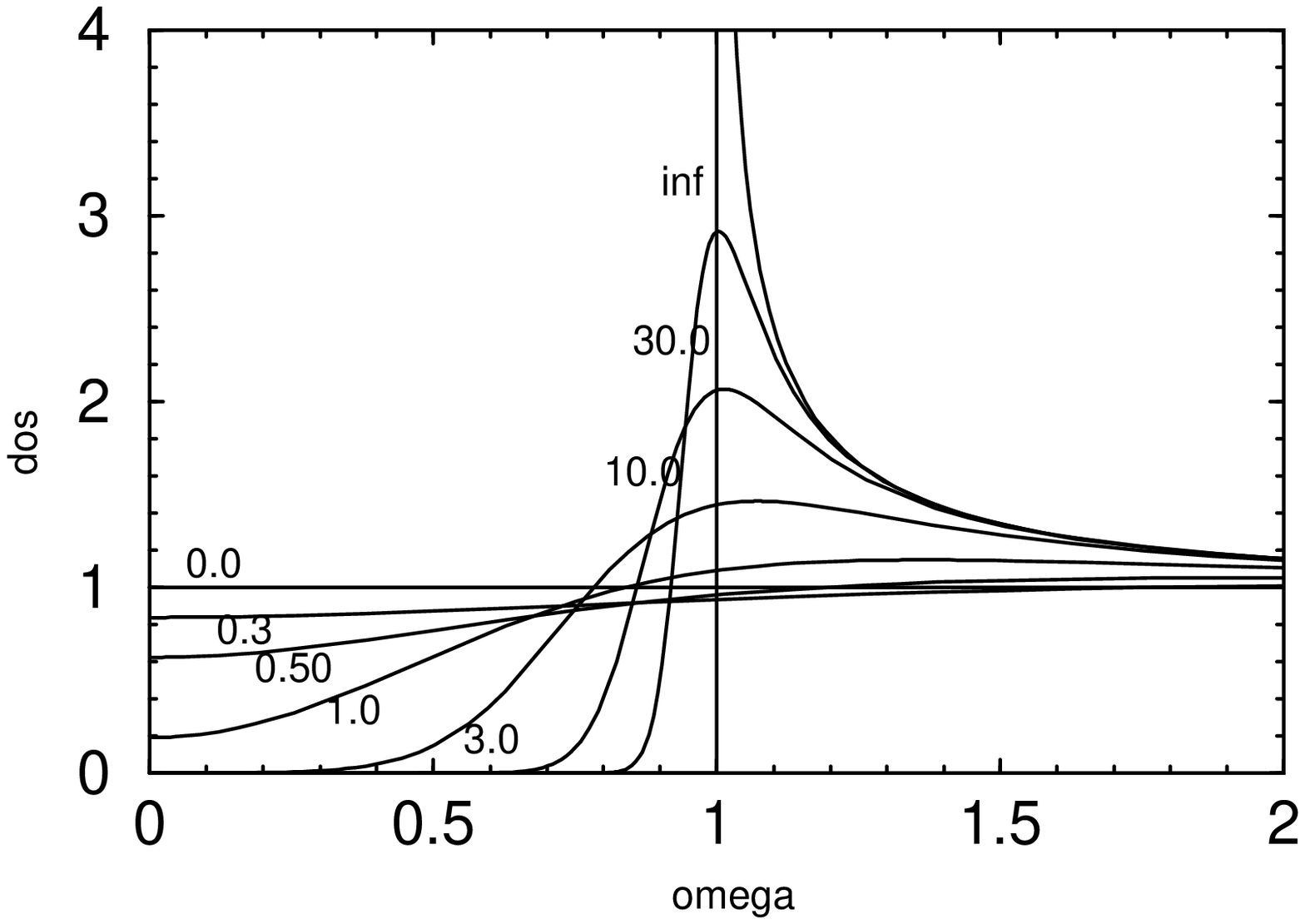,width=6.20cm}
\hfill
\epsfig{file=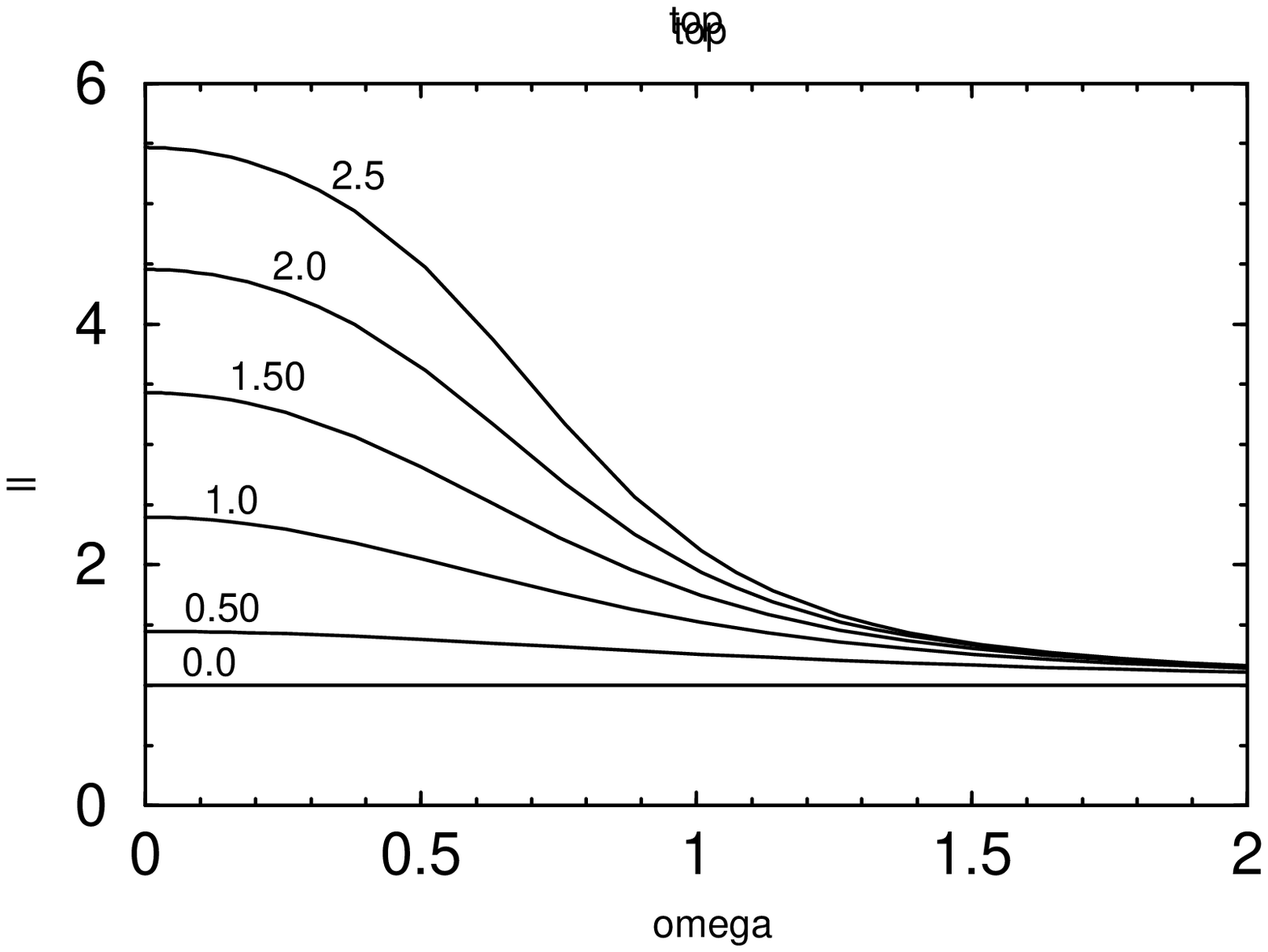,width=6.20cm}
\end{center}
\caption{The DOS $\rho(\omega)$ and the inverse localization length
  $\ell^{-1}(\omega)$ in the white
  noise limit for the incommensurate case plotted versus $\omega /
  \Delta_0$ for $\nu=\Delta_0/2D = 
  0.0,0.3,0.5,1.0,3.0,10.0,30.0,\infty$ or $\nu=\Delta_0/2D =
0.0,0.5,1.0,1.5,2.0,2.5$, respectively.
}
\label{fig:wn-incommensurate}
\end{figure}

In the incommensurate case, $D_R = D_I \equiv D/2$, and the forward
scattering potential $D_V$ only leads to a renormalization of
$D$. 
If we introduce $\tilde D \equiv D + D_V = \tilde D_R/2 +
\tilde D_I/2$ and take the confluent limit $D_I \to D_R$ in Eq.\ 
(\ref{eq:Gamma-Hayn-Mertsching}), 
we obtain
\begin{equation}
  \label{eq:Gamma:incommensurateII}
  \Gamma (\omega) = D -i\omega + \Delta_0 \, \frac{I_{1-i\omega/\tilde D}
\left(\frac{\Delta_0}{\tilde D}\right)}{I_{-i\omega/\tilde D}
\left(\frac{\Delta_0}{\tilde D}\right)} \;,
\end{equation}
which is equivalent to
\begin{equation}
  \label{eq:Gamma:incommensurateIII}
  {\setlength{\fboxsep}{2mm} \fbox {$ \ \displaystyle 
  \Gamma (\omega) = D + \Delta_0 \, \frac{I_{-i\omega/\tilde D}^{\prime}
\left(\frac{\Delta_0}{\tilde D}\right)}{I_{-i\omega/\tilde D}
\left(\frac{\Delta_0}{\tilde D}\right)} \;.$}}
\end{equation}
The imaginary part of Eq.\ (\ref{eq:Gamma:incommensurateII})
determines the integrated DOS,
\begin{equation}
  \label{eq:IDOSic}
  {\setlength{\fboxsep}{2mm} \fbox {$ \ \displaystyle 
  \mathcal{N}(\omega) = \frac{\tilde D}{\pi} \sinh\left(\frac{\pi
      \omega}{\tilde D}\right) \, \frac{\rho_0}{\left|I_{i\omega/\tilde D}
      \left(\frac{\Delta_0}{\tilde D}\right)\right|^2} \; .$}}
\end{equation}
This expression agrees with Ref.\ \cite{Golub79}.
A plot of the DOS for different values of the parameter $\nu =
\Delta_0/2D$ is shown at the left-hand side of 
Fig.\ \ref{fig:wn-incommensurate}. 
There is
no Dyson singularity, and in the absence of a static gap $\Delta_0$,
the disorder
has no effect on the DOS so that $\rho(\omega) = 
\rho_0$ for $\nu = 0$. At zero frequency, $\rho(0)$ is always finite and the
DOS vanishes with increasing $\nu$ as $\rho(0) =
\rho_0/\left[I_0(2\nu)\right]^2$. 
For a given $\Delta_0$, the disorder leads to a filling of the
gap.
As in the commensurate case, in the limit $D\to 0$, i.e.
$\nu \to \infty$, the DOS reduces to the mean-field result
$\rho(\omega) = \rho_0 \, \theta(\omega^2 - \Delta_0^2)\
|\omega|/(\omega^2 - \Delta_0^2)^{1/2}$.

Taking the real part of Eq.\ (\ref{eq:Gamma:incommensurateII}), we get
for the inverse localization length
\begin{equation}
  \label{eq:ic:wn:localizationlength}
  \ell^{-1}(\omega) = D + \Delta_0\, \frac{I_{i\omega/\tilde D}
      \left(\frac{\Delta_0}{\tilde D}\right) 
    I_{1-i\omega/\tilde D} \left(\frac{\Delta_0}{\tilde
          D}\right) + I_{-i\omega/\tilde D}
      \left(\frac{\Delta_0}{\tilde D}\right) 
    I_{1+i\omega/\tilde D} \left(\frac{\Delta_0}{\tilde
          D}\right)}
  {2 \left|I_{i\omega/\tilde D} 
      \left(\frac{\Delta_0}{\tilde D}\right)\right|^2} \;.
\end{equation}
In contrast to the commensurate case, the localization length at
$\omega = 0$ is finite for any $\Delta_0$ and given by
\begin{equation}
    \label{eq:ic:wn:localizationlengthat0}
  \ell^{-1}(0) = D + \Delta_0\, \frac{I_{1}
      \left(\frac{\Delta_0}{\tilde D}\right)}{I_{0}
      \left(\frac{\Delta_0}{\tilde D}\right)} \;.
\end{equation}
While for $\Delta_0/\tilde D \to 0$ one has $\ell^{-1}(0) \to 
D$ (and also $\ell^{-1}(\omega) \to D$ for every $\omega$), for
$\Delta_0/\tilde D \gg 1$ one finds $\ell^{-1}(0) \sim
\Delta_0 + D/2 - D_V/2$. 
For weak disorder, the disorder 
creates a few localized states with energies $|\omega| < \Delta_0$
whose inverse localization length is given by $\ell^{-1}(\omega) \sim
\sqrt{\Delta_0^2 -\omega^2}$.
A plot of the inverse localization length
for different values of the parameter $\nu$ and $D_V=0$ is given at
the right-hand side of Fig.\ 
\ref{fig:wn-incommensurate}.

\subsection{Infinite correlation lengths}

In the limit of infinite correlation lengths, $\Delta(x)$ becomes
independent of $x$ and exact results for the averaged DOS of the FGM
may be obtained. The limit of large correlation lengths $\xi$ is of
special importance for Peierls systems because the
correlation length of the order parameter diverges at the Peierls
transition. Sadovskii was the first to consider the fluctuating gap
model (FGM) with infinite correlation lengths \cite{Sadovskii74} and
calculated the one-electron Green function for the incommensurate
case by summing up all diagrams in the perturbation expansion. This
Green function leads to a DOS which exhibits a pseudogap at 
the Fermi energy. The commensurate case was later solved by
Wonneberger and Lautenschlager \cite{Wonneberger76}.

\subsubsection*{Taking the ensemble average}

The limit of infinite correlation lengths can be solved by averaging
the desired quantity calculated with a static gap $\Delta$ over an
appropriate probability distribution of $\Delta$. This amounts to
taking an ensemble average.

\subsubsection{The commensurate case}

For real $\Delta$ and Gaussian statistics we have
\begin{equation}
  \label{eq:process_of_averaging:real}
  \langle \dots \rangle = \int_{-\infty}^{\infty}
  \frac{d\Delta}{\sqrt{2\pi \Delta_s^2}} 
  e^{-\Delta^2/2\Delta_s^2} \dots \;.
\end{equation}
Calculating the DOS by averaging $\theta(\omega^2 - \Delta^2) 
\,\omega / \sqrt{\omega^2-\Delta^2}$ with respect to the above probability 
distribution, one obtains
\begin{eqnarray}
  \rho_{\infty} (\omega) =
  \rho_0 \sqrt{\frac{\pi}{2}} \, \frac{\omega}{\Delta_s} \,
  e^{-\omega^2/4\Delta_s^2} \, I_0
  \left(\frac{\omega^2}{4\Delta_s^2}\right) \;.
\end{eqnarray}
Here, $I_0(u)$ is the modified Bessel function with index $0$.

If we define the inverse
localization length $\ell_{\infty}^{-1}(\omega)$ for $\xi =
\infty$ by the Thouless formula, we have 
$\ell_{\infty}^{-1}(\omega) = \left\langle \sqrt{\Delta^2-\omega^2}\,
 \theta(\Delta^2-\omega^2) \right\rangle$, such that
\begin{eqnarray}
\ell_{\infty}^{-1}(\omega)
  \sqrt{\frac{2}{\pi}} \,\frac{\omega^2}{\Delta_s} \int_1^{\infty}
  du\, e^{-(\omega^2/2\Delta_s^2) u^2}\sqrt{u^2-1} \;.
\end{eqnarray}

\subsubsection{The incommensurate case}

For complex $\Delta(x)$ and Gaussian statistics, the process of
averaging can be written as
\begin{equation}
  \label{eq:averaging:complex}
  \langle \dots \rangle = \int \frac{d \textrm{Re} \Delta\, 
    d \textrm{Im} \Delta}{\pi \Delta_s^2} \,
  e^{-|\Delta|^2/\Delta_s^2} \dots \;.
\end{equation}
A similar calculation as above leads to
\begin{eqnarray}
  \rho_{\infty} (\omega)
  & \hspace{-2.3mm} = & \hspace{-2.3mm}
  2 \rho_0 \frac{\omega}{\Delta_s}\,e^{-(\omega^2/\Delta_s^2)}
  \,\textrm{Erfi}\, \left(\frac{\omega}{\Delta_s}\right)\;,
\\
   \ell_{\infty}^{-1} (\omega) 
  & \hspace{-2.3mm} = & \hspace{-2.3mm} 
  \Delta_s \frac{\sqrt{\pi}}{2}\, e^{-\omega^2/\Delta_s^2} \;.
  \label{eq:LLinf}
\end{eqnarray}
Here, $\textrm{Erfi}\, (u) \equiv \int_0^u e^{x^2}\,dx$
is the error function with an imaginary argument. 

Plots of $\rho_{\infty} (\omega)$ and $\ell_{\infty}^{-1} (\omega)$
can be found in the next section.
While in the
commensurate case the DOS vanishes linearly in $\omega$, it only
vanishes quadratically in the incommensurate case. This is due to the
fact that the probability distribution for complex $\Delta$ has less
weight for small $|\Delta|$ than the one for real $\Delta$.
The inverse localization length assumes for both the commensurate and the
incommensurate case a finite value at $\omega = 0$ and drops to zero
as $\omega$ increases. That $\ell_{\infty}^{-1} (0)$ is finite in
the commensurate case seems to contradict the general result 
$\ell_{\infty}^{-1} (0) = \Delta_{\rm av} = 0$ derived in Section
\ref{section:formalism}. One should keep in mind, however, that for
$\xi = \infty$ we have only defined $\partial_{\omega} 
\ell_{\infty}^{-1} (\omega)$ by
$\textrm{Re}\, \langle \mathcal{G} (x,x;\omega) \rangle$. While for
finite $\xi$ a single chain is representative for an ensemble of
chains, for $\xi = \infty$, there is no self-averaging effect. On the
other hand, it seems plausible to assume that the above results for
$\xi = \infty$ give a good approximation to the case of finite $\xi$
if $\xi$ is much larger that any microscopic length scale involved. In
particular, we have to demand $\Delta_s \xi \gg 1$ and $\omega \xi \gg
1$. The above results for the DOS and the inverse localization length
at $\omega = 0$ can therefore not be expected to hold for finite
correlation lengths. In fact, we will see in the next section that for
any finite $\xi$ we find in the commensurate case $\rho(0)=\infty$ and
$\ell^{-1}(0)=0$. For $\Delta_s \xi \gg 1$ and $\omega \xi \gg 1$,
however, we will find a remarkable agreement between the two solutions
as predicted above.

 \section{Finite correlation lengths}
\label{section:finite_xi}

{\em While in the limit of very
small and infinite correlation lengths $\xi$ of the random disorder,
the fluctuating gap model (FGM) admits for an exact analytic 
calculation of the density of states (DOS) and the inverse localization
length, in the intermediate regime of finite $\xi$ there are only
approximate solutions available. It especially turns up the question: 
``How accurate are Sadovskii's solutions \cite{Sadovskii79}, which for a
long time were thought to be exact?'' An answer to this question is of
particular interest because Sadovskii's solutions 
have 
become quite popular since the experimental discovery of a pseudogap
in the underdoped cuprates above the critical temperature $T_c$
\cite{Schmalian98,Schmalian99,Sadovskii01}. 
In this section, we will calculate the
DOS and the inverse localization length for Gaussian statistics, as
approximately done by Sadovskii with very high accuracy
numerically. Finally, we will consider the case of only phase fluctuations
for which we recently found
an exact solution  by applying a gauge transformation to the Green
function and mapping the
original problem onto a problem involving only white noise 
\cite{Bartosch00c}.}

\subsection{Singularities in the density of states}

The exact results of the FGM derived in the white noise limit in the
previous section imply under certain circumstances a Dyson singularity in
the DOS. This singularity arises only in the commensurate case
[i.e. for real $\Delta(x)$] and only if the forward scattering
potential and
$\Delta_{\textrm{\scriptsize av}} = \langle \Delta(x) \rangle$ are sufficiently
small [see Eqs.\ (\ref{eq:Dysonsingularity}) and
(\ref{eq:DOS:wn:neq})]. Since the white noise limit
describes the low-energy physics of physical systems characterized by
small correlation lengths $\xi$, this statement should also be true
for small but finite $\xi$. As far as I know, it was first shown by
myself in collaboration with Peter Kopietz that the DOS
$\rho ( \omega )$ of  the FGM
exhibits a singularity at the Fermi energy for any finite value of the 
correlation length $\xi$ if the fluctuating order
parameter field $\Delta (x)$ is real and its
average $\langle \Delta ( x ) \rangle$
is sufficiently small \cite{Bartosch99a}. To detect the singularity,
we applied the boundary condition $\Delta_{\textrm{\scriptsize BC}} =
V_{\textrm{\scriptsize BC}} = 0$, such that the complete 
spectrum turned out to be continuous [see Eq.\ (\ref{BCPhi}) and its
following remark]. In the Comment \cite{Millis00a}, Millis and Monien
showed that the local DOS $\rho(\omega=0,x)$ calculated in our Letter
\cite{Bartosch99a} is equal to the absolute square of the wave function
$\psi(x)$, which led them to claim that we have not calculated the DOS 
at all. However, as pointed out in the Reply \cite{Bartosch00a}, for the 
boundary conditions used in Ref.\ \cite{Bartosch99a}, one finds
$\rho(\omega,x)=|\psi_{\omega}|^2$, i.e.\ the local DOS is equal to the 
absolute square of the wave function. 
This clearly invalidates the argument given by Millis and Monien.
Today, I would not use the 
artificial boundary conditions which lead to a continuous spectrum 
any more: The existence of
the Dyson singularity put forward in the Letter \cite{Bartosch99a} 
can also be 
seen in the discrete case by considering the equation of motion
(\ref{eq:phiII}) for $V(x)=0$ and real $\Delta(x)$ which after the
shift $\varphi \to \varphi - \pi/2$ reads
\begin{equation}
  \label{eq:motion:phi:sin}
  \partial_{x} \varphi (x) = 2 \omega +2 \Delta(x) \sin
    \varphi (x) \;.
\end{equation}
The Dyson singularity in the DOS is due to phase resonance: If
$\omega$ is small 
(compared to $\Delta_s$, $\Delta_s^2 \xi$ and $\xi^{-1}$) but positive,
the change of $\varphi(x)$ is dominated by the fluctuating term $2
\Delta(x) \sin \varphi (x)$. Only near $\varphi(x) = n \pi$ (with $n$
an integer) we have $\partial_x \varphi (x) = 2 \omega > 0$, such that
$\varphi(x)$ can only grow on average. As $\left(\varphi(x) - n \pi
\right) \approx \omega/\Delta_s$, fluctuation effects of $\Delta(x)$
become important, driving $\varphi(x)$ from $n \pi + \omega/\Delta_s$
to $(n+1) \pi - \omega/\Delta_s$. Near $\varphi(x) = (n+1)\pi$, the
constant force $2\omega$ dominates again and the above picture repeats
itself. 

As we decrease $\omega$, the ``time'' (which corresponds to the space
coordinate $x$) to move $\varphi(x)$ from $n \pi - \omega/\Delta_s$ to
$n \pi + \omega/\Delta_s$ will not change, but fluctuations of
$\Delta(x)$ will need slightly longer to drive $\varphi(x)$ from $n
\pi + \omega/\Delta_s$ to $(n+1) \pi - \omega/\Delta_s$, implying that
$\varphi(x)$ decreases more slowly than $\omega$ as $\omega$
decreases. Now, the average DOS for frequencies between $0$ and
$\omega$ is given by
\begin{equation}
\rho(\zeta \omega) = \frac{\mathcal{N} (\omega)}{\omega} = \lim_{x \to
  \infty} \frac{\varphi_{\omega}(x)}{2\pi \omega x} \;,
\end{equation}
where $\zeta$ is a number between $0$ and $1$.
Letting $\omega$ approach zero, it follows $\rho(0) = \infty$. This
divergence 
describes the Dyson singularity in the DOS. The above reasoning is
independent of the probability distribution of $\Delta(x)$. However,
it should be noted that $\Delta(x)$ must not be dominated by one
sign. If $\varphi(x) \approx n \pi$ and $(-1)^n \Delta(x)$ is
negative, $\varphi(x)$ will fluctuate
around the stable position near $n \pi +
\omega/\Delta_s$. $\varphi(x) = (n+1)\pi$ can only be reached if $(-1)^n
\Delta(x)$ is positive on average over a finite interval. We therefore
conclude that we expect a Dyson singularity if $\Delta(x)$ is real and
fluctuates around $\Delta_{\rm av} \equiv \langle \Delta(x) \rangle$
with $\Delta_{\rm av}$ sufficiently small.

For complex $\Delta(x)$, fluctuations of the phase of $\Delta(x)$
can be mapped via the gauge transformation (\ref{eq:Gbardef}) onto a
forward scattering potential. Since the amplitude $|\Delta(x)|$ is
always positive and the phase fluctuations lead to an effective local
shift of the frequency $\omega$, there should be no Dyson
singularity. Instead, we expect a suppression of the DOS, i.e.\ a
pseudogap.

\subsection{Numerical algorithm}
\label{sec:numerical_algorithm}

In the following, we present an exact algorithm which for stepwise
constant potentials 
allows for simultaneous numerical calculations of the integrated DOS
and the inverse localization length. 
By choosing the step size sufficiently small, the integrated DOS
and the inverse localization length may be calculated for arbitrary
given potentials. Let us partition the interval $(0,L)$ into $N$
intervals $(x_n,x_{n+1})$ of length $\delta_n = x_{n+1} -x_{n}$ with
$x_0=0 < x_1 < \dots < x_N=L$, 
such that $\Delta_n \equiv \Delta (x)$ and $V_n \equiv V (x)$ for
$x_n<x<x_{n+1}$. Let us also define $\tilde \omega_n$ as 
$\tilde \omega_n \equiv \omega -V_n$.

To find an exact analytic solution for the given stepwise constant
potentials of the equations of motion
(\ref{eq:phiII}) and (\ref{eq:zeta}), let us consider again the
related $S$-matrix $S(L,0;\omega)$, which 
can be written as the finite product
\begin{equation}
  \label{eq:productS}
  S(L,0;\omega) = \prod_{n=0}^{N-1} S_{n} \equiv \prod_{n=0}^{N-1}
  S(x_{n+1},x_{n};\omega)
   \;,
\end{equation} 
where $S_{n}$ is given for $\omega_{n}^2 < |\Delta_{n}|^2$ by 
Eq.\ (\ref{eq:S-matrix_const}) and for $\omega_{n}^2 > |\Delta_{n}|^2$ by
Eq.\ (\ref{eq:S-matrix_constII}).
Eq.\ (\ref{eq:productS}) implies the recurrence relation 
$S(x_{n+1},0;\omega) = S_{n} \, S(x_{n},0;\omega)$,
which can be cast into the following recurrence relations for
$\varphi_{n} \equiv \varphi(x_n)$ and $\zeta_n \equiv \zeta(x_n)$:
\begin{eqnarray}
  \label{eq:recurrence:phi}
  \varphi_{n+1} \hspace{-2.3mm} & = & \hspace{-2.3mm} \varphi_n - 2 \,
  \textrm{Im} \left( \ln z_n \right)
  \;, \qquad \\ 
  \label{eq:recurrence:zeta}
  \zeta_{n+1} \hspace{-2.3mm} & = & \hspace{-2.3mm} \zeta_n + 2 \,
  \textrm{Re} \left(\ln z_n \right)                                 \;, \qquad 
\end{eqnarray}
where $z_n = \left(S_{n}\right)_{22} + \left(S_{n}\right)_{21}\,
  \exp\left(i\varphi_n \right)$.
Note that Eqs.\ (\ref{eq:recurrence:phi}) and
(\ref{eq:recurrence:zeta})  are integrated forms of the equations of
motion (\ref{eq:phiII}) and (\ref{eq:zeta}).
The real and imaginary part of $\ln z_n$ can be determined by writing
$\ln z_n$ as
$\ln z_n \equiv \ln |z_n| +i \arg \left(z_n \right)$.
The argument of $z_n$, $\arg \left(z_n \right)$ can be obtained
up to a multiple of $2\pi$ from
\begin{equation}
  \label{eq:argument:z}
  \arg \left(z_n \right) = 2\pi m_n + \textrm{sgn} \left[ \textrm{Im}
    \, z_n \right] \arccos \left(\frac{\textrm{Re} \, z_n }{|z_n|}
  \right) \;.
\end{equation}
To find the integer
\begin{equation}
m_n = \left[\frac{\arg \left(z_n \right)}{2\pi} + \frac{1}{2}
\right]_{\textrm{\small int}} \;, 
\end{equation}
we define $z_n(x)$ by $z_n$ with $\delta_n$ replaced by
$x-x_n$. $z_n(x)$ is an analytic function of $x$ and at $x=x_{n+1}$
agrees with $z_n$. For $\tilde \omega_n^2 < |\Delta_n|^2$, it follows from
Eq.\ (\ref{eq:S-matrix_const}) that $\textrm{Im} \left[ z_n(x) \right] \propto
\sinh[\sqrt{|\Delta_{n}|^2-\tilde \omega_{n}^2}\, (x-x_n)]$  does not
change its 
sign for any $x>x_n$, such that $|\arg \left[z_n (x) \right]| < \pi$ and
$m_n$ has to be zero. For $\tilde \omega_n^2 > |\Delta_n|^2$, however, 
$\textrm{Im} \left[ z_n(x) \right] \propto
\sin[\sqrt{\tilde \omega_{n}^2 - |\Delta_{n}|^2}\, (x-x_n)]$, such
that $\left|\left[\arg \left(z_n (x) \right)/\pi
  \right]_{\textrm{\small int}}\right| = \left[ 
\sqrt{\tilde \omega_{n}^2 - |\Delta_{n}|^2}\, (x-x_n)/\pi
\right]_{\textrm{\small int}}$. Since the constant of proportionality
is negative for $\tilde \omega_n > |\Delta_n|$ and positive for
$\tilde \omega_n < -|\Delta_n|$, it follows
\begin{equation}
  m_n = \left[\frac{1}{2} - \frac{\textrm{sgn} (\tilde \omega_n)
      \sqrt{\tilde \omega_{n}^2 - |\Delta_{n}|^2} \, \delta_n}{2\pi}  
  \right]_{\textrm{\small int}} \;. 
\end{equation}
To summarize, we can simultaneously calculate the integrated DOS and
the inverse localization length for arbitrary stepwise constant
potentials using 
the following iterative algorithm with the initial values $\varphi_0 =
\zeta_0 = 0$,
\newline
\hfill
\parbox{0mm}
{\begin{eqnarray*}
\\
\end{eqnarray*}}
\parbox{11cm}{
\begin{eqnarray*}
  {\setlength{\fboxsep}{2mm} \fbox {$ \ \displaystyle  
  \begin{array}{rcl}  \varphi_{n+1} \hspace{-2.3mm} & = & \displaystyle
    \hspace{-2.3mm} \varphi_n -2 \left[
  2\pi m_n + \textrm{sgn} \left[ \textrm{Im}
    \, z_n \right] \arccos \left(\frac{\textrm{Re} \, z_n }{|z_n|}
  \right) \right] \;,  \\
  \rule[0mm]{0mm}{5mm} 
  \zeta_{n+1} \hspace{-2.3mm} & = & \hspace{-2.3mm} \zeta_n + 2 
  \ln \left| z_n \right|
  \;, \\ 
\end{array} $}}
\end{eqnarray*}}
\parbox{19mm}
{\begin{eqnarray}
    \label{eq:phi:recurrence} \\
   \rule[-1mm]{0mm}{7mm}
    \label{eq:zeta:recurrence}
\end{eqnarray}}
\newline
where $z_n$ and $m_n$ are given by
\newline
\hfill
\parbox{0mm}
{\begin{eqnarray*}
\\
\end{eqnarray*}}
\parbox{11cm}{
\begin{eqnarray*}
  {\setlength{\fboxsep}{2mm} \fbox {$ \ \displaystyle  
   \begin{array}{rcl}      \rule[-3mm]{0mm}{3mm} 
 z_n \hspace{-2.3mm} & = & \hspace{-2.3mm} \left(S_{n}\right)_{22} +
 \left(S_{n}\right)_{21}\, 
  \exp\left(i\varphi_n \right) \;, \\
  m_n \hspace{-2.3mm} & = & \hspace{-2.3mm} \left\{ \begin{array}{ll}
      \displaystyle 
      \left[\frac{1}{2} - \frac{\textrm{sgn} (\tilde
          \omega_n)\sqrt{\tilde \omega_{n}^2 -  
        |\Delta_{n}|^2} \, \delta_n}{2\pi}
  \right]_{\textrm{\small int}} & ,\quad  \tilde \omega_{n}^2 -
        |\Delta_{n}|^2 > 0 \\ 
        \rule[0mm]{0mm}{5mm} 
        0 \,  & , \quad \tilde \omega_{n}^2 -
        |\Delta_{n}|^2 \le 0 \end{array} 
    \right. , \end{array} $}}\quad
\end{eqnarray*}}
\parbox{19mm}
{\begin{eqnarray}
   \rule[-8mm]{0mm}{12.8mm}
   \label{eq:z_n} \\
   \rule[-9mm]{0mm}{9mm}
    \label{eq:m_n}
\end{eqnarray}}
\newline
and the matrix elements of $S_n$ are determined by Eqs.\
(\ref{eq:S-matrix_const}) and (\ref{eq:S-matrix_constII}).

\subsubsection{Generation of disorder}

In the case of finite correlation lengths, specific extensive physical
quantities (which are obtained by relating
extensive quantities to the length of the system) show a 
self-averaging effect as the length of the chain increases
\cite{Lifshits88}, i.e. they become independent of the concrete
realization of the disorder. 
This self-averaging effect can be understood by partitioning a very
long macroscopic chain into a large number of chains, each one still
being much longer than the correlation length and any other
microscopic length scale. In this case, boundary effects between
adjacent parts of the original chain may be neglected and we are
practically left with an ensemble of a large number of independent
chains. Physical quantities can now be calculated for each chain
individually, assuming independently all possible values with their
respective statistical weight. Specific extensive quantities are now
given by the 
ensemble-average, giving a non-random value in the thermodynamic limit. 
The DOS and the inverse localization length can
therefore be calculated by generating one typical very long chain.

\subsubsection*{Gaussian disorder}

To generate Gaussian disorder at the sample points $x_n$ with the
first two moments satisfying 
\begin{equation}
  \label{eq:Gaussian:1+2}
  \langle \Delta(x) \rangle = \Delta_{\textrm{\scriptsize av}} \quad, \quad
  \langle \tilde \Delta(x) \tilde \Delta(x') \rangle =
  \Delta_s^2 e^{-|x-x'|/\xi} \;, 
\end{equation}
where $\tilde \Delta(x) \equiv \Delta(x) -\Delta_{\textrm{\scriptsize
    av}}$, we use a realization of an Ornstein-Uhlenbeck process
described in more general form in Ref.\ \cite{Bartosch00PhD}
Using the Box-Muller algorithm \cite{numericalrecipes},
we generate independent Gaussian random
numbers $g_n$ with 
$\langle g_n \rangle = 0$ and $\langle g^2_n \rangle = 1$.
For real $\Delta ( x )$,  we generate $\tilde \Delta_n = \Delta_n -
\Delta_{\textrm{\scriptsize av}}$ recursively by
\begin{equation}
\tilde{\Delta}_0  =  \Delta_{s} g_0 \; \; , \;  \;
\tilde{\Delta}_{n+1} = a_n 
\tilde{\Delta}_n
 +  \sqrt{ 1 - a_n^2} \, \Delta_{s} g_{n+1}
 \; ,
 \label{eq:genial}
\end{equation}
where $a_n = e^{- | \delta_n | / \xi}$.
This Markov process  
leads to a Gaussian random process with the desired
correlation functions. The Markov property of the algorithm allows us to
generate the disorder simultaneously with the iteration of the
recurrence relations (\ref{eq:phi:recurrence}) and
(\ref{eq:zeta:recurrence}), so that the algorithm presented above
practically needs no
memory space and, in principle, arbitrary long chains can be
considered. If we choose $| \delta_n | / \xi \ll 1$
(in practical calculations we choose $| \delta_n | / \xi \approx
0.0001$ to $0.05$ depending on $\Delta_s \xi$ and make sure that lessening of
$| \delta_n | / \xi$ does not change the results), the (integrated)
DOS and the inverse localization length may be calculated with
arbitrary accuracy numerically.

Of course, the above algorithm can also be used to
generate $V_n$ and
in the complex case,
$\textrm{Re} \, \Delta_n$ and
$\textrm{Im} \,\Delta_n$ can be generated by replacing
$\Delta_{s}$ by $\Delta_{s}/\sqrt{2}$ (see Ref.\ \cite{Bartosch00PhD}).

\subsubsection{Results}

First numerical calculations of the DOS of the FGM in the regime of
finite correlation lengths were done by myself in collaboration with
Peter Kopietz using an algorithm similar to the one presented here
\cite{Bartosch99d}. Simultaneously, Millis and Monien presented their
data obtained by an exact diagonalization of a lattice regularization of
the FGM \cite{Millis00b}. However, these authors did not make any
attempts to relate their results to the continuous FGM which would
have allowed for a more direct comparison with the solutions given by
Sadovskii \cite{Sadovskii79}.

In contrast to the algorithm described in Ref.\ \cite{Bartosch99d}, the
algorithm presented here does not only allow for a numerical
calculation of the 
(integrated) DOS, it is also capable of a simultaneous evaluation of the
localization length which for finite $\xi$ has never been published
before.

\subsubsection*{Commensurate case}

\begin{figure}[tb]
\begin{center}
\psfrag{omega}{\large\hspace{0.5mm}$\omega / \Delta_s$}
\psfrag{dos}{\large$\hspace{1mm}\rho(\omega)/\rho_{0}$}
\psfrag{ll}{\large\hspace{-5.8mm}$\ell^{-1}(\omega)/\Delta_s$}
\psfrag{0}{\Large $0$}
\psfrag{0.5}{\Large $0.5$}
\psfrag{1}{\Large $1$}
\psfrag{1.5}{\Large $1.5$}
\psfrag{2}{\Large $2$}
\psfrag{3}{\Large $3$}
\psfrag{0.2}{\small$0.2$}
\psfrag{0.50}{\small$0.5$}
\psfrag{1.0}{\small $1.0$}
\psfrag{2.0}{\small $2.0$}
\psfrag{10.0}{\small $10$}
\psfrag{100.0}{\small $100$}
\psfrag{1000.0}{\small $1000$}
\epsfig{file=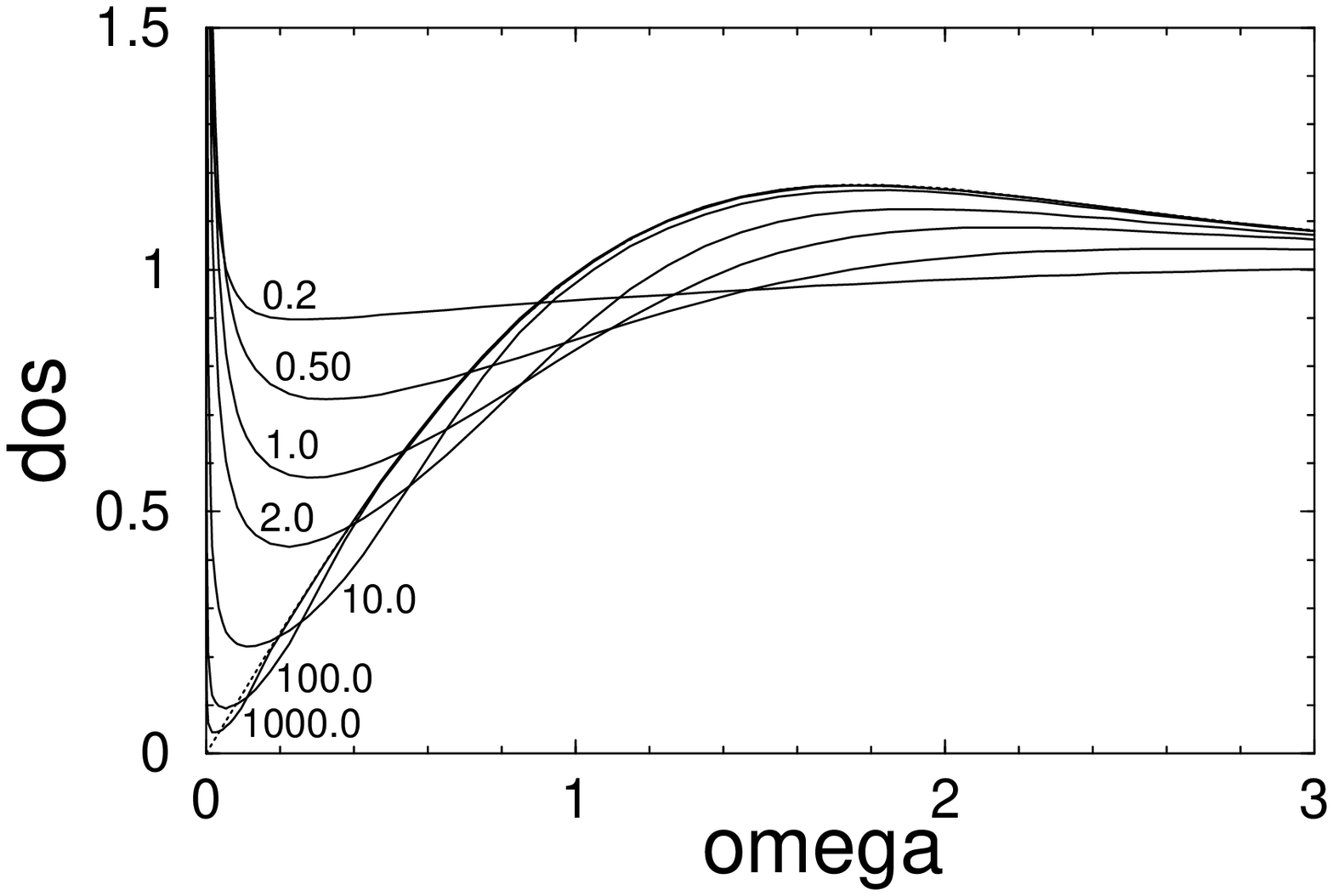,width=11.50cm}
\vspace{-2mm}
\epsfig{file=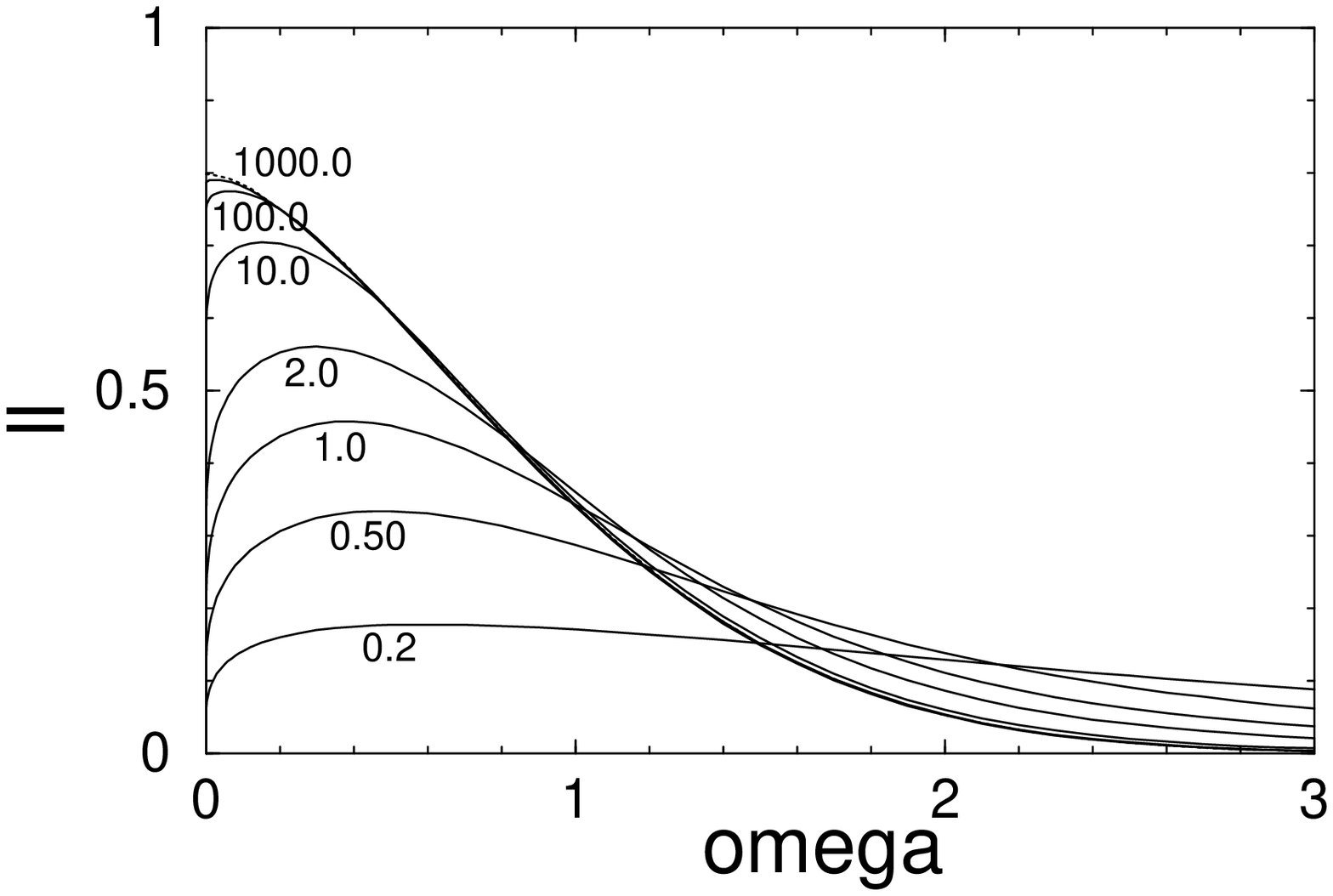,width=11.50cm}
\end{center}
\vspace{-8mm}
\caption{Plot of the DOS $\rho(\omega)$ and the inverse localization length
  $\ell^{-1}(\omega)$ for real $\Delta(x)$ with Gaussian
  statistics, $\Delta_s L = 10^8$, and finite correlation lengths
  $\Delta_s \xi=1000,100,10,2.0,1.0,0.5$, and $0.2$. For any finite
  $\xi$, we find $\rho(0)=\infty$ and $\ell^{-1}(\omega)=0$. The dashed
  line represents the exact result derived in Section
  \ref{chap:whitenoise} and for $\omega \gtrsim 0.2 \Delta_s$ is almost
  indistinguishable from the result for $\Delta_s \xi = 1000$.}
\label{fig:real_finite_xi}
\end{figure}

In Fig.\ \ref{fig:real_finite_xi},
we show our numerical results for the DOS $\rho ( \omega )$ and
inverse localization length $\ell^{-1}(\omega)$ for real
$\Delta(x)$ (with $\Delta_{\rm av}=0$ and $V(x)=0$), which refers to the
symmetric phase of a commensurate system with no forward scattering. 
Except for $\Delta_s \xi = 1000,0.2$ we have chosen
the same values of the dimensionless parameter
$\Delta_s \xi$ as in Fig.\ 7  
of Ref.\ \cite{Sadovskii79}.
One clearly sees the Dyson singularity in the DOS which exists  
for any finite value of $\xi$ and overshadows the pseudogap
at sufficiently small energies. 
One can also see that this Dyson singularity is accompanied by a
singularity in the inverse localization length. The inverse
localization length drops to zero at
$\omega = 0$, in accordance with the exact result $\ell^{-1} (0) =
\langle 
\Delta_{\rm av} \rangle$ [see Eq.\
(\ref{eq:localizationlengthforomega-0})].

\begin{figure}[tb]
\begin{minipage}{0.48\linewidth}
\begin{center}
\psfrag{1/xi}{\hspace{-1.0mm}\small$1/\Delta_{s} \xi$}
\psfrag{alpha}{\small\hspace{0.7cm}\hspace{-5.0mm}$\alpha(\xi)$}
\psfrag{0.001}{\small\hspace{0mm}$0.001$}
\psfrag{0.010}{\small\hspace{0.0mm}$0.01$}
\psfrag{0.100}{\small\hspace{0.0mm}$0.10$}
\psfrag{1.000}{\small\hspace{0.0mm}$1.00$}
\psfrag{10.000}{\small\hspace{0.0mm}$10.00$}
\psfrag{0.0}{\small\hspace{-0.2mm}$0.0$}
\psfrag{0.5}{\small\hspace{-0.2mm}$0.5$}
\psfrag{1.0}{\small\hspace{-0.2mm}$1.0$}
\psfrag{1.5}{\small\hspace{-0.2mm}$1.5$}
\psfrag{2.0}{\small\hspace{-0.2mm}$2.0$}
\epsfig{file=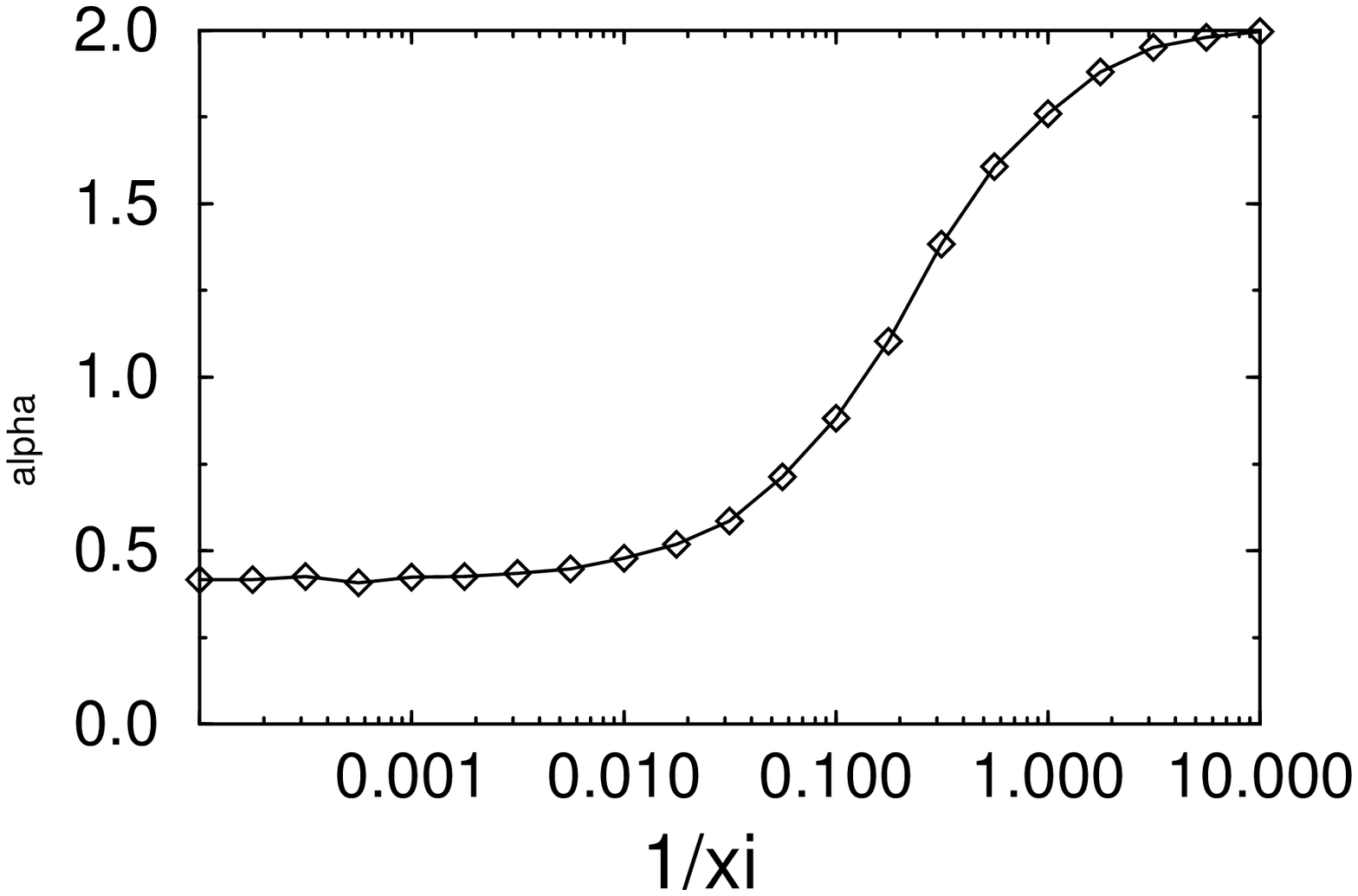,height=4.0cm}

\end{center}

\vspace{-8mm}

\caption{Plot of the exponent $\alpha(\xi)$ defined by Eq.\
  (\ref{eq:Bofxi}) for 
  frequencies between $\omega = 10^{-6} \Delta_s$ and $\omega =
  10^{-11} \Delta_s$. While our results for small $\Delta_s \xi$ are 
  consistent with the white noise result $\alpha = 2$,
  in the opposite limit $\Delta_s \xi \gg 1$ we find $\alpha
  \approx 0.41.$}
\label{fig:exponentofxi}
\end{minipage}
\hfill
\begin{minipage}{0.48\linewidth}
\begin{center}
\psfrag{1/xi}{\small$\hspace{-1.0mm}1/\Delta_{s} \xi$}
\psfrag{A}{\small\hspace{-8.7mm}$A(\xi)\, , \, B(\xi)$}
\psfrag{Axi}{\small$A(\xi)$}
\psfrag{Bxi}{\small$\hspace{-5mm}B(\xi)$}
\psfrag{0.001}{\small\hspace{0mm}$0.001$}
\psfrag{0.010}{\small\hspace{0.0mm}$0.01$}
\psfrag{0.100}{\small\hspace{0.0mm}$0.10$}
\psfrag{1.000}{\small\hspace{0.0mm}$1.00$}
\psfrag{10.000}{\small\hspace{0.0mm}$10.00$}
\epsfig{file=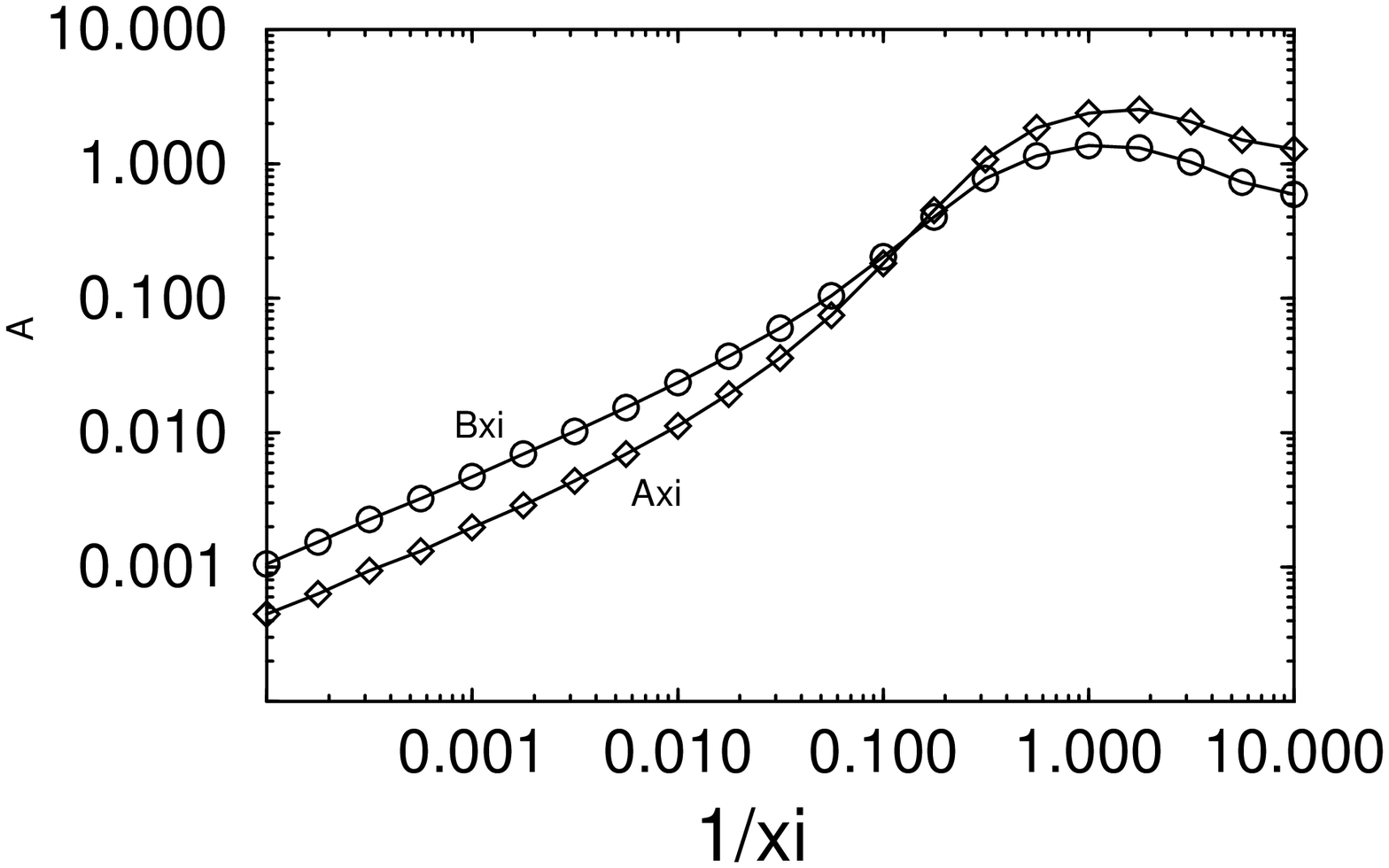,width=6.2cm}
\end{center}
\vspace{-8mm}
\caption{Plot of the weight factors $A(\xi)$ and $B(\xi)$ of the Dyson
  singularity defined by Eqs.\ (\ref{eq:Bofxi}) and (\ref{eq:Aofxi}) for
  frequencies between $\omega = 10^{-6} \Delta_s$ and $\omega =
  10^{-11} \Delta_s$. As the correlation length increases, both weight
  factors scale to zero.} 
\label{fig:weightofsingularity}
\end{minipage}
\end{figure}

\begin{figure}[b]
\begin{minipage}{0.55\linewidth}
\begin{center}
\psfrag{1/xi}{\hspace{-1.0mm}\small$1/\Delta_{s} \xi$}
\psfrag{dos}{\small\hspace{-6.7mm}$\rho(\omega^{\ast})/\rho_{0}$,
  $\omega^{\ast}/\Delta_s$} 
\psfrag{0.01}{\small \hspace{0mm}$0.01$}
\psfrag{0.10}{\small \hspace{0mm}$0.10$}
\psfrag{1.00}{\small \hspace{0mm}$1.00$}
\psfrag{10.00}{\small \hspace{0mm}$10.00$}
\epsfig{file=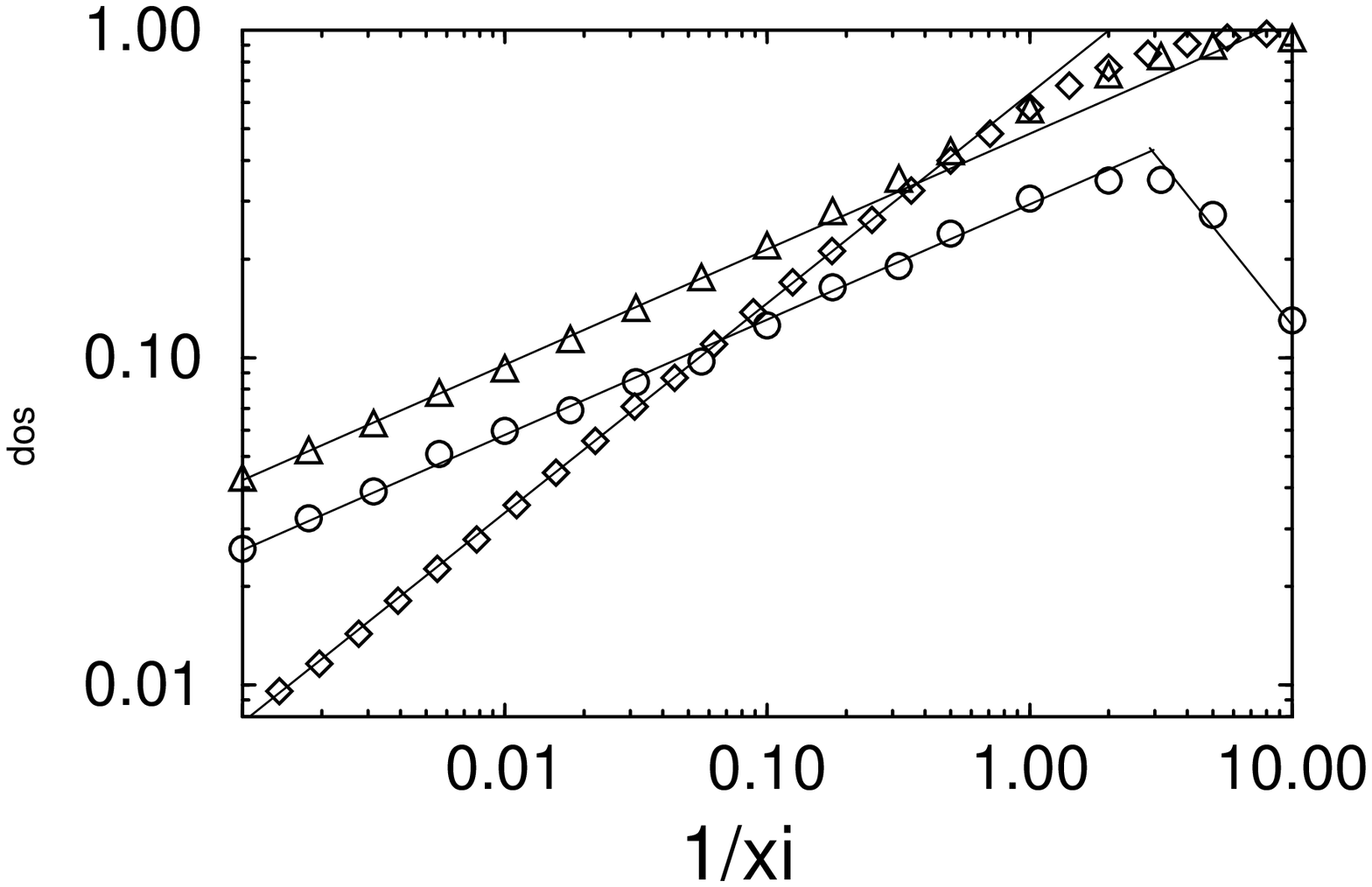,width=6.2cm}
\end{center}
\end{minipage}
\hfill
\begin{minipage}{0.4\linewidth}
\caption{Double-logarithmic plot of $\rho(\omega^{\ast})/\rho_{0}$ as
  a function of 
  $1/\Delta_{s}\xi$ for real $\Delta (x)$ (triangles) and complex
  $\Delta (x)$ (diamonds), where $\omega^{\ast}$ is the energy
  for which the DOS assumes its minimum. While $\omega^{\ast} = 0$ for
  complex $\Delta (x)$, the circles give the double-logarithmic plot
  of $\omega^{\ast}/ \Delta_s$ for real $\Delta (x)$ as a function of
  $1/\Delta_{s}\xi$.}
\label{fig:dosxidependence}
\end{minipage}
\end{figure}

The Dyson singularity in the DOS is missed by
Sadovskii's algorithm \cite{Sadovskii79}. 
For a more quantitative description of the Dyson singularity we have
plotted the logarithm of the integrated DOS $\mathcal{N}/\Delta_s$
versus the logarithm of $-\ln(\omega/\Delta_s)$. For frequencies
between $\omega = 10^{-11} \Delta_s$ and $\omega = 10^{-6} \Delta_s$, 
we find
that the data can be very well fitted by a straight line, such that
\begin{equation}
  \label{eq:Bofxi}
  \mathcal{N}(\omega) = \rho_0\,
  \frac{\Delta_s B(\xi)}{|\ln(\omega/\Delta_s)|^{\alpha(\xi)}} \;, 
\end{equation}
which implies for the DOS
\begin{equation}
  \label{eq:Aofxi}
  \rho(\omega) = \rho_0\,
  \frac{A(\xi)}{(\omega/\Delta_s)|\ln(\omega/\Delta_s)|^{1+\alpha(\xi)}} \;,
\end{equation}
with $A(\xi) = \alpha(\xi)\, B(\xi)$. Plots of the exponent
$\alpha(\xi)$ and the weight factors of the Dyson singularity $A(\xi)$
and $B(\xi)$ are shown in Figs.\ \ref{fig:exponentofxi} and 
\ref{fig:weightofsingularity}.
For $\Delta_s \xi \ll 1$ our data is consistent with the white noise
result $\alpha = 2$.
As $\Delta_s 
\xi$ increases, the exponent 
$\alpha(\xi)$ decreases, assuming for large correlation lengths $\xi$
the finite value
\begin{equation}
  \label{eq:alpha:constant}
  \alpha(\xi) \approx 0.41 \; , \quad \Delta_s \xi \gtrsim 500 \;.
\end{equation}
Fitting the data for $A(\xi)$ in the regime between $\Delta_s \xi = 500$ and
$\Delta_s \xi = 10000$ to a power law shows that the weight of the
singularity of the DOS vanishes as
\begin{equation}
  A(\xi) = 0.175 \, (\Delta_s \xi)^{-0.65} \;. 
\end{equation}
The plot of the DOS given
in Fig.\ \ref{fig:real_finite_xi} shows that for large correlation
lengths $\xi$ the Dyson singularity only overshadows a pseudogap, such
that $\rho(\omega)$ takes a minimal value at a certain frequency
$\omega^{\ast} (\xi)$. A double-logarithmic plot of
$\rho(\omega^{\ast})$ versus $(\Delta_s \xi)^{-1}$ is given by the
triangles in Fig.\ \ref{fig:dosxidependence}.
The straight line gives a fit to a power-law:
\begin{equation}
  \rho ( \omega^{\ast} ) / \rho_0 = C \, ( \Delta_s \xi )^{- \mu }
\end{equation}
We find 
\begin{equation}
  C = 0.482 \pm 0.010 \;,\quad \mu = 0.3526 \pm 0.0043
\end{equation}
The circles in Fig.\ \ref{fig:dosxidependence}
show $\omega^{\ast}$  where $\rho ( \omega )$ is minimal.
The long solid line is a fit
to a power-law 
\begin{equation}
  \omega^{\ast} / \Delta_s = D \, ( \Delta_s \xi )^{- \gamma} \;.
\end{equation}
Here, we find
\begin{equation}
  D = 0.2931 \pm 0.0074 \;,\quad \gamma = 0.3513 \pm 0.0051 \;.
\end{equation}
such that within numerical accuracy $\mu=\gamma$. The proportionality of 
$\rho(\omega^{\ast})$ to the energy scale $\omega^{\ast}$, which
can be 
interpreted as the width of the Dyson singularity,
can also directly be seen in Fig.\ \ref{fig:real_finite_xi}.
Finally we note that for $\Delta_s \xi \lesssim 0.2$ our algorithm
produces results consistent with the white noise limit $\Delta_{s} \xi
\ll 1$. From the exact solution of Ovchinnikov and
Erikhman \cite{Ovchinnikov77} we obtain $\rho(\omega^{\ast})/\rho_{0} \to
0.9636$ and $\omega^{\ast} 
\to 1.2514 \, \Delta_{s}^{2} \xi$ which determines the short solid line in
Fig. \ref{fig:dosxidependence}, describing $\omega^{\ast}(\xi)$
in the white-noise limit.

\subsubsection*{Incommensurate case}

\begin{figure}[tb]
\begin{center}
\psfrag{omega}{\large\hspace{0.5mm}$\omega / \Delta_s$}
\psfrag{dos}{\large$\hspace{1mm}\rho(\omega)/\rho_{0}$}
\psfrag{ll}{\large\hspace{-5.8mm}$\ell^{-1}(\omega)/\Delta_s$}
\psfrag{0}{\Large $0$}
\psfrag{0.5}{\Large $0.5$}
\psfrag{1}{\Large $1$}
\psfrag{1.5}{\Large $1.5$}
\psfrag{2}{\Large $2$}
\psfrag{3}{\Large $3$}
\psfrag{0.2}{\small$0.2$}
\psfrag{0.50}{\small$0.5$}
\psfrag{1.0}{\small $1.0$}
\psfrag{2.0}{\small $2.0$}
\psfrag{10.0}{\small $10$}
\psfrag{100.0}{\small $100$}
\psfrag{1000.0}{\small $1000$}
\epsfig{file=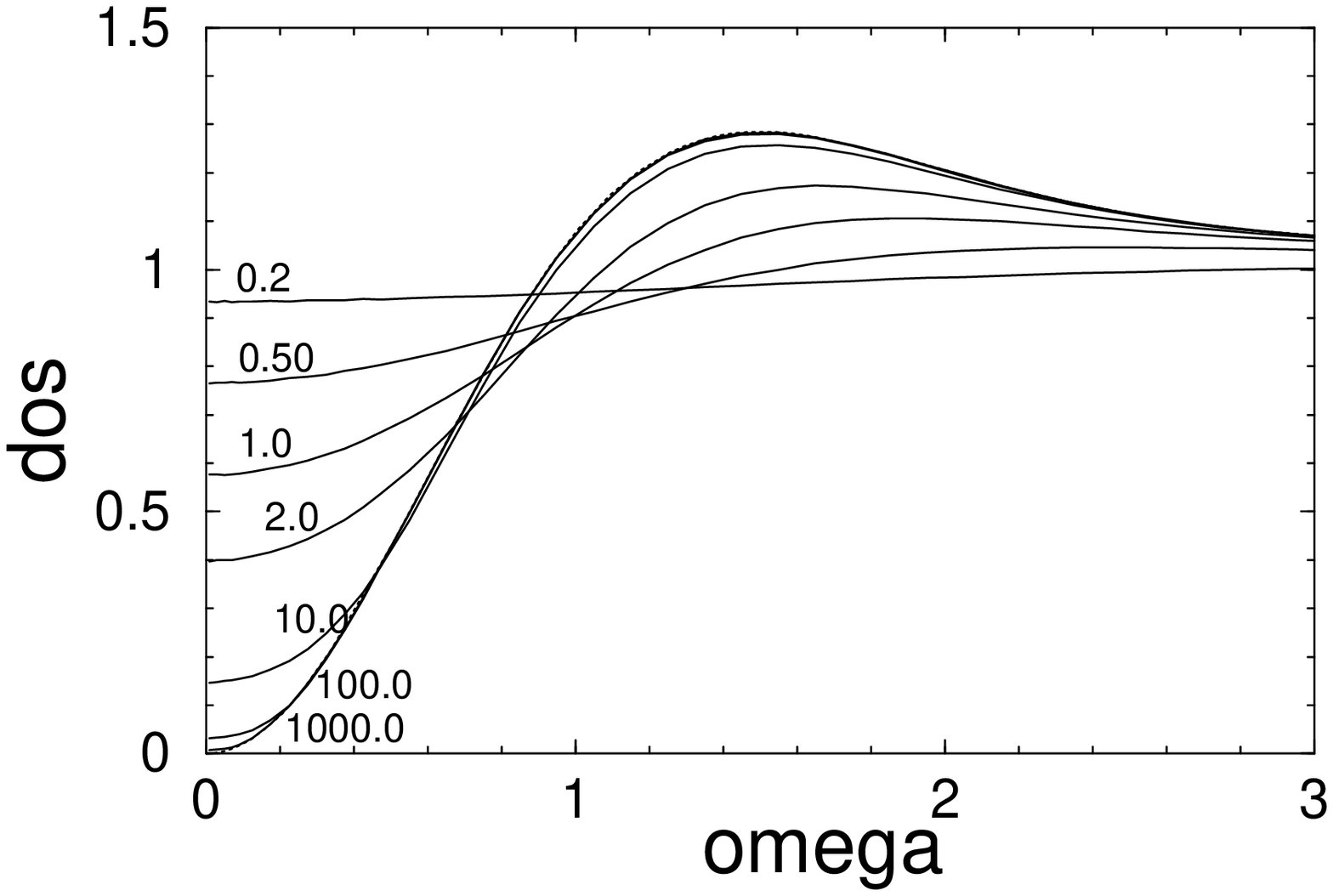,width=11.50cm}
\vspace{-2mm}
\epsfig{file=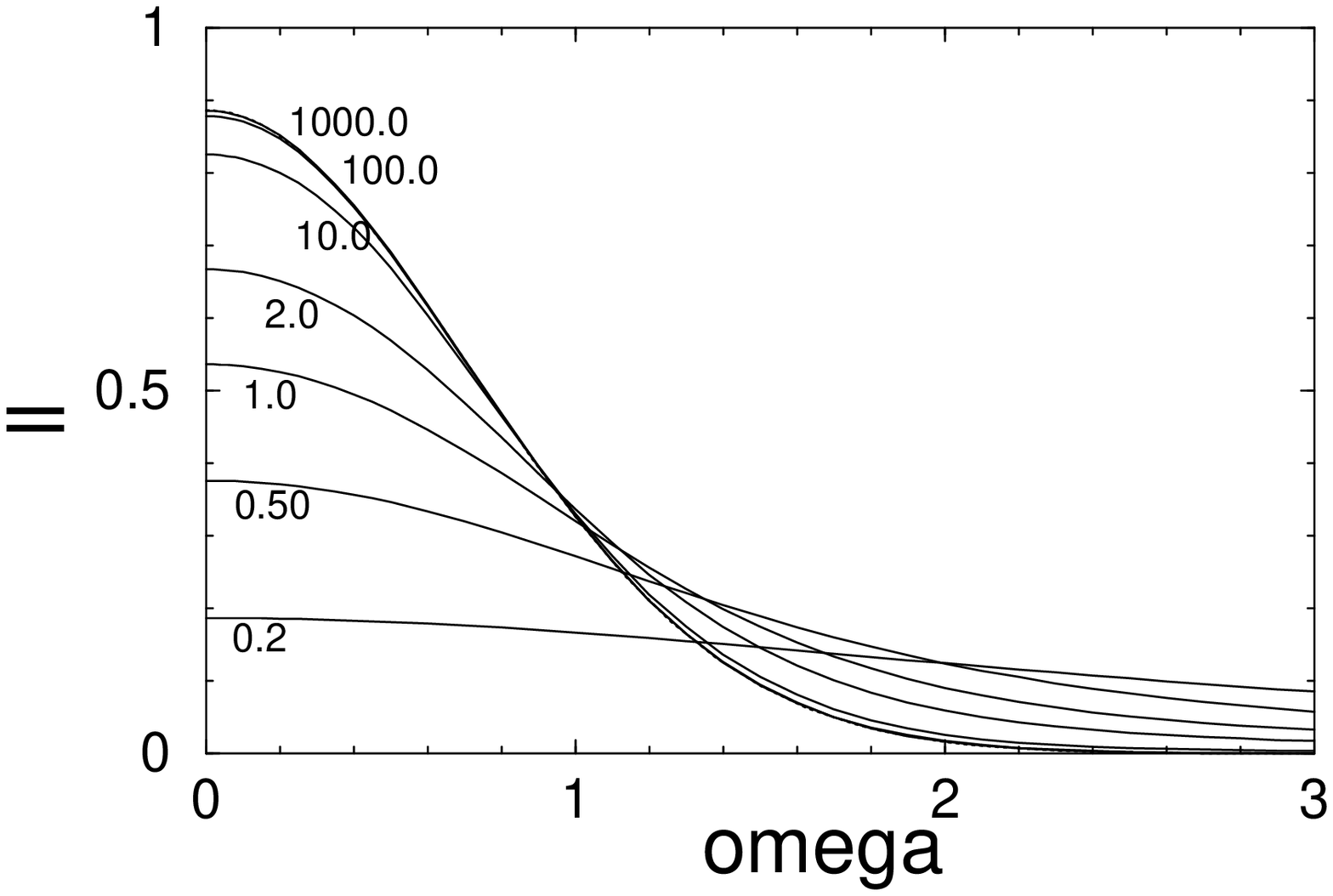,width=11.50cm}
\end{center}
\vspace{-8mm}
\caption{Plot of the DOS $\rho(\omega)$ and the inverse localization length
  $\ell^{-1}(\omega)$ for complex $\Delta(x)$ with Gaussian
  statistics, $\Delta_s L = 10^8$, and finite correlation lengths
  $\Delta_s \xi=1000,100,10,2.0,1.0,0.5$, and $0.2$. The dashed
  line represents the exact result derived in Section
  \ref{chap:whitenoise}. This line is hardly recognizable because it
  is almost indistinguishable from the line for $\Delta_s \xi = 1000$.}
\label{fig:complex_finite_xi}
\end{figure}

The DOS $\rho (\omega)$ and
inverse localization length $\ell^{-1}(\omega)$ for complex
$\Delta(x)$ (and $\Delta_{\rm av}=V(x)=0$), which refers to the
symmetric phase of the commensurate case with no forward scattering are
presented in Fig.\ \ref{fig:complex_finite_xi}. 
Neither the DOS nor
the inverse localization length involve a singularity. In fact, a direct
comparison of our results for the DOS with those obtained from
Sadovskii's algorithm shows a good agreement.

For a more quantitative comparison, the diamonds in Fig.\
\ref{fig:dosxidependence} show 
the DOS $\rho (0)$ at the Fermi energy. A fit to a
power-law gives
\begin{equation}
  \rho (0) / \rho_0 = C \, ( \Delta_s \xi )^{- \mu } \;,
\end{equation}
with
\begin{equation}
 C = 0.6397 \pm 0.0066 \;,\quad \mu = 0.6397 \pm 0.0024 \;.
\end{equation}
Note that within numerical accuracy we find $C = \mu$.
This result should be compared with Sadovskii's approximate result
$C=0.541 \pm 0.013$ and $\mu = 1/2$.

\subsection{Phase fluctuations only}

As already discussed at the end of Section 
\ref{chap:FGM}, far below the mean-field
critical temperature $T_c^{\rm MF}$, amplitude fluctuations of the
complex order parameter $\Delta(x) \equiv |\Delta(x)| e^{i\vartheta(x)}$ are
frozen out and only phase fluctuations survive.
In terms of the ``superfluid velocity'' $V(x)=\partial_x \vartheta(x)/2$,
the process of averaging can be written as
\begin{equation}
  \label{eq:averaging:phase}
  \langle \, \dots \, \rangle = \frac{\int \mathcal{D} \{V\} \, \dots
    \, e^{-\beta
      F\{V\} }}{\int \mathcal{D} \{V\}  e^{-\beta
      F\{V\} }} \;,
\end{equation}
where $ F\{V\}$ is the free energy functional given in Eq.\ 
(\ref{eq:freeenergyphasefluctuations}).
Since $F\{V\}$ is Gaussian and local, the process of averaging is
described by Gaussian white noise. The first two moments of $V(x)$ are
given by $\langle V(x) \rangle = 0$ and
\begin{equation}
  \label{eq:phase:V:secondmoment}
  \langle V(x) V(x') \rangle 
  =
  2\, \frac{1}{4 \xi(T)}\, \delta(x-x') \;, 
\end{equation}
with [see Eq.\ref{eq:xiofT:phase}]
$\xi(T) = {s \rho_s(T)}/{2 T}$.
Well below the mean-field temperature $T_c^{\rm MF}$, we can use the
BCS gap equation
$\Delta_s = 1.764 \,
T_c^{\rm MF}$ and $\rho_s (T) 
\approx \rho_0 = \pi^{-1}$ to get for the dimensionless
parameter\footnote{The prefactor can be expressed in terms of the
  Euler constant $\gamma$, such that
  $\Delta_s \xi(T) = s T_c^{\rm MF}/2 e^{\gamma} T$.}
$\Delta_s \xi(T)$
\begin{equation}
  \label{eq:phase:DeltaxiofT}
  \Delta_s \xi(T) = 0.281 \, s \,T_c^{\rm MF}/T \;. 
\end{equation}
As already shown in Section \ref{chap:FGM}, Eq.\
(\ref{eq:phase:V:secondmoment})
implies the exponentially decaying correlation function
$\langle \Delta(x) \Delta^{\ast}(x')  \rangle 
 = \Delta_s^2 \,
  \exp\left({-|x-x'|/\xi(T)}\right)$.

To calculate physical quantities like the DOS or the inverse
localization length,
we use the gauge invariance of
these quantities under the gauge
transformation (\ref{eq:Gbardef}) and map the phase fluctuations 
of the order parameter $\Delta(x) = \Delta_s
e^{i\vartheta(x)}$ onto the effective forward scattering potential $V(x) =
\partial_x \vartheta(x)/2$.

\subsubsection{Density of states and inverse localization length}

\begin{figure}[tb]
\begin{center}
\psfrag{omega}{\large\hspace{0.5mm}$\omega / \Delta_s$}
\psfrag{dos}{\large$\hspace{1mm}\rho(\omega)/\rho_{0}$}
\psfrag{ll}{\large\hspace{-5.8mm}$\ell^{-1}(\omega)/\Delta_s$}
\psfrag{0}{\Large $0$}
\psfrag{0.5}{\Large $0.5$}
\psfrag{1}{\Large $1$}
\psfrag{1.5}{\Large $1.5$}
\psfrag{2}{\Large $2$}
\psfrag{3}{\Large $3$}
\psfrag{0.1}{\small $0.1$}
\psfrag{0.3}{\small $0.3$}
\psfrag{1.0}{\small $1.0$}
\psfrag{3.0}{\small $3.0$}
\psfrag{10}{\small $10$}
\psfrag{inf}{\small $\infty$}
\epsfig{file=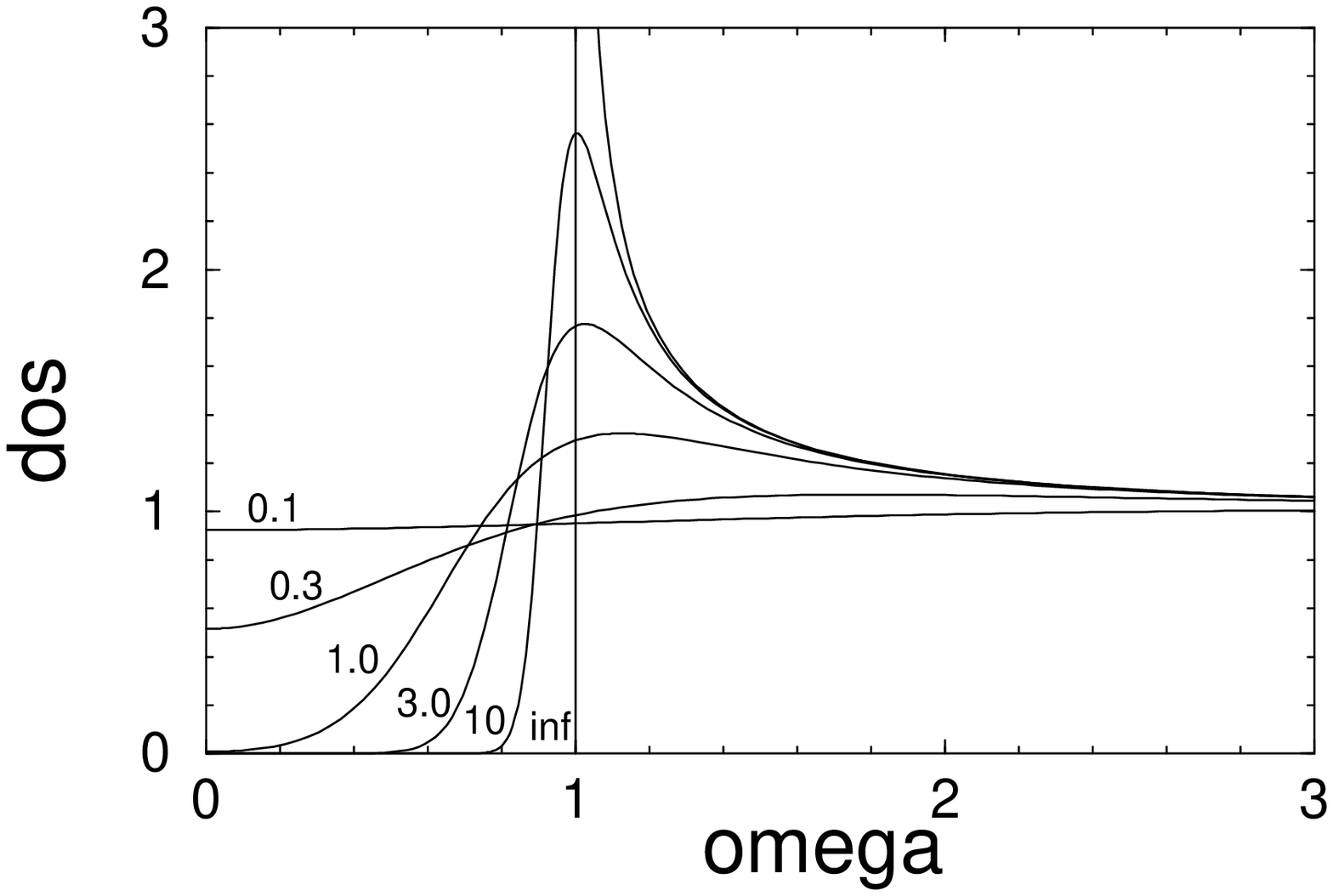,width=11.50cm}
\vspace{-2mm}
\epsfig{file=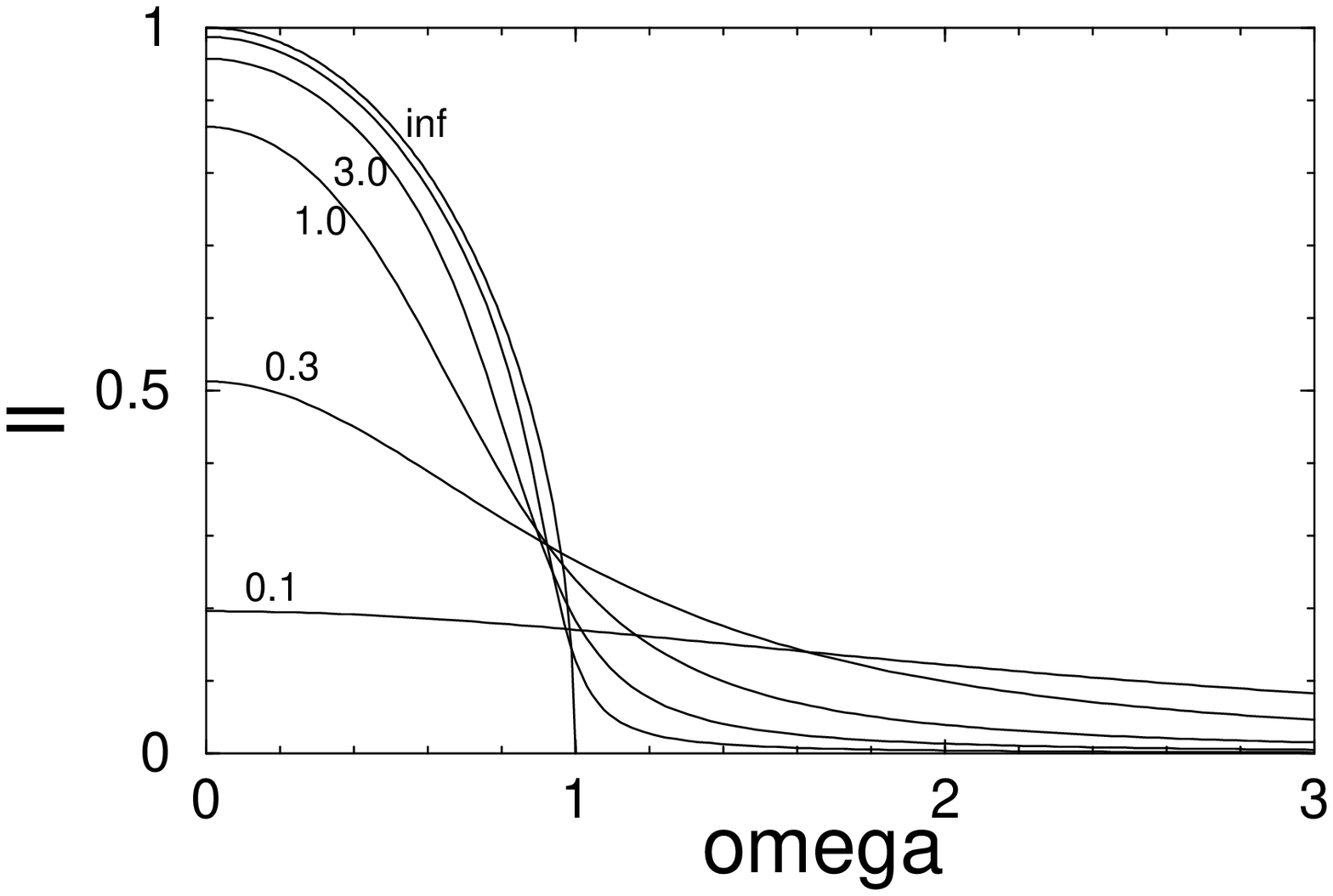,width=11.50cm}
\end{center}
\vspace{-8mm}
\caption{Plot of the DOS $\rho(\omega)$ and the inverse localization length
  $\ell^{-1}(\omega)$ for phase fluctuations only and $\Delta_s \xi =
  0.1,0.3,1.0,3.0,10.0$ and $\infty$.}
\label{fig:DOSandLLphase}
\end{figure}

\begin{figure}[tb]
\begin{minipage}{0.6\linewidth}
\begin{center}
\psfrag{invxi}{\small\hspace{-4.0mm}$1/\Delta_s \xi$}
\psfrag{dos}{\small$\hspace{-1.0mm}\rho(0)/\rho_{0}$}
\psfrag{0}{\small $0$}
\psfrag{0.5}{\small $0.5$}
\psfrag{1}{\small $1$}
\psfrag{5}{\small $5$}
\psfrag{10}{\small $10$}
\epsfig{file=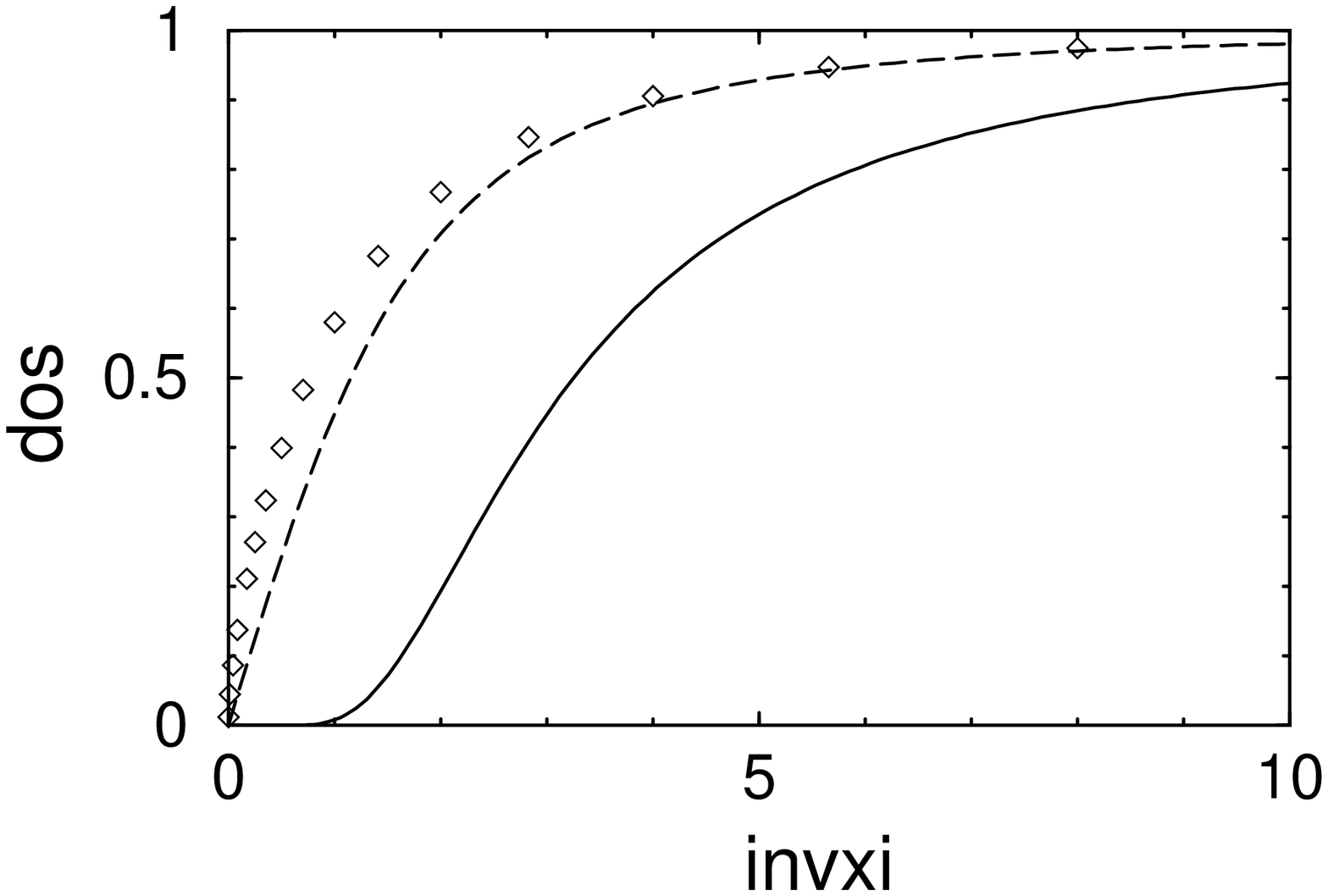,width=6.2cm}
\end{center}
\end{minipage}
\hfill
\begin{minipage}{0.4\linewidth}
\caption{Plot of the DOS $\rho(0)/\rho_0$ as a function of $1/\Delta_s
  \xi$. The solid line gives the DOS for phase fluctuations only while
  the dashed line is the result found in the leading order Born
  approximation [see Eq.\ \ref{eq:DOS_B:0}] and the diamonds give the DOS
  evaluated 
  for Gaussian statistics [see Subsection \ref{sec:numerical_algorithm}]}
\label{fig:DOSat0}
\end{minipage}
\end{figure}

The DOS and the inverse localization length for the remaining
problem
involving only a constant gap parameter and forward scattering 
described by Gaussian white noise were
already calculated in the previous section. With the exception of the
constant shift in the inverse localization length, the results are
identical with those for the incommensurate case without forward
scattering and $\langle \Delta(x) \rangle = \Delta_0 \neq
0$. Substituting in Eqs.\ (\ref{eq:Gamma:incommensurateII}) and 
(\ref{eq:Gamma:incommensurateIII})
$\Delta_0$ by $\Delta_s$, $\tilde D$ by $1/4\xi(T)$ and setting $D=0$
(since there is only forward scattering) which implies a completely
different interpretation of the resulting equations, we get
$\Gamma (\omega) = -i\omega + \Delta_s \, {I_{1-i4\omega \xi}
\left(4 \Delta_s \xi \right)}/{I_{-i4\omega \xi}
\left(4 \Delta_s \xi \right)}$, or, equivalently,
\begin{equation}
  \label{eq:Gamma:phaseIII}
  {\setlength{\fboxsep}{2mm} \fbox {$ \ \displaystyle 
  \Gamma (\omega) = \Delta_s \, \frac{I_{-i4\omega \xi}^{\prime}
\left(4 \Delta_s \xi\right)}{I_{-i4\omega \xi}
\left(4 \Delta_s \xi\right)} \;.$}}
\end{equation}
It follows from Eq.\ (\ref{eq:IDOSic})
that the integrated DOS is given by
\begin{equation}
  \label{eq:IDOS:phase}
  {\setlength{\fboxsep}{2mm} \fbox {$ \ \displaystyle 
  \mathcal{N}(\omega) = \rho_0 \, \frac{\sinh\left(4 \pi \omega \xi
    \right)}{4 \pi \xi} \, \frac{1}{\left|I_{i4\omega \xi}
      \left(4 \Delta_s \xi \right)\right|^2} \; .$}}
\end{equation}
Plots of the DOS and the inverse localization length for
characteristic values of $\Delta_s \xi$ are shown in Fig.\
\ref{fig:DOSandLLphase}. 
At the Fermi energy, the DOS simplifies to
\begin{equation}
  \label{eq:phase:DOSat0}
  \rho(0) = \frac{\rho_0}{[I_0(4\Delta_s\xi)]^2} \;,
\end{equation}
such that, as the temperature is lowered and the correlation length
grows, the DOS at the Fermi energy vanishes exponentially,
\begin{equation}
  \label{eq:phase:DOSat0II}
  \rho(0) \sim 8 \pi \rho_0 \Delta_s \xi \exp(-8 \Delta_s \xi) \;,
  \quad 4 \Delta_s \xi \gg 1 \;.
\end{equation}
This result is in contrast to the power-law behavior of the DOS as
predicted by Gaussian statistics.
A plot of the DOS at the Fermi energy is shown in Fig.\
\ref{fig:DOSat0} as a function of $1/\Delta_s \xi$. 
For a comparison, we have also plotted $\rho(0)$ for Gaussian
statistics and the result found in the Born approximation, which at low
temperatures can only poorly describe the quantitative behavior of the
pseudogap. For $T = T_c^{\rm MF}/4$ which corresponds to $\Delta_s \xi
\approx 2.0$, we find that the DOS $\rho(0)$ for phase fluctuations
only is less than
$10^{-5} \rho_0$ while the Born approximation and the numerical exact
result for 
Gaussian statistics suggest a value of order $\rho_0/4$. 

As can also be seen in Fig.\ \ref{fig:DOSat0}, there is no pseudogap 
for $\Delta_s \xi \lesssim 0.1$. Note, however, that these correlation
lengths correspond to temperatures of order $T_c^{\rm MF}$, where 
amplitude fluctuations are important.
Nevertheless, using only phase fluctuations
gives a good qualitative description of the DOS for all temperatures
$T$.

For temperatures well below the mean-field temperature $T_c^{\rm MF}$
where our theory becomes quite accurate,
we have, according to Eq.\ (\ref{eq:phase:DeltaxiofT}), $\Delta_s \xi
\gg 1$, such that the Bessel function 
$
  I_{i\nu}(\nu z) = e^{-\pi\nu/2} \, J_{i\nu}(i\nu z)
$
may be approximated by an Airy function.
For large correlation lengths we expand the resulting equation
for $\omega$ around
$\Delta_s$ and obtain to leading order in $1/4\Delta_s\xi$ a maximum of
the DOS $\rho(\omega)$ at $\Delta_s$, described by the inverted parabola
\begin{equation}
  \label{eq:maximum_DOS}
  \rho(\omega) \sim \rho_0 \left[a \, (4\Delta_s\xi)^{1/3} - b \,
    (4\Delta_s\xi)^{5/3} \,
    \left(\frac{\omega}{\Delta_s}-1\right)^2 \right] \;,
\end{equation}
with
$a = \frac{1}{2^{4/3}\pi} \, \frac{c_2}{c_1^3} \approx 0.731$ and
$b = \frac{1}{2^{2/3}\pi} \,
  \frac{1}{c_1^2}\left[3\left(\frac{c_2}{c_1}\right)^3-1\right]
  \approx 0.258$,
where $c_1 = {\rm Ai}\, (0) = 3^{-2/3}/\Gamma(2/3) \approx 0.355$ and
$c_2 = -{\rm  Ai}^{\prime}\,  (0) = 3^{-1/3}/\Gamma(1/3) \approx
0.259$. Note that
Eq.\ (\ref{eq:maximum_DOS}) implies that the maximum of the DOS
diverges as $(4\Delta_s\xi)^{1/3} \propto T^{-1/3}$. 

Away from $\omega =
\Delta_s$, 
we find for $\omega > \Delta_s$
\begin{equation}
  \label{eq:DOS:phase:asymptotics:plus}
  \rho(\omega) \sim \rho_0\left(1+\frac{\Delta_s^2}{4\omega^2} \right)
  \;, \quad 4\Delta_s \xi \left(\frac{\omega}{\Delta_s} -
    1\right)^{3/2} \gg 1 \;,
\end{equation}
which is independent of $\Delta_s \xi$ and agrees with the mean-field
result.
And for $\omega < \Delta_s$ we find
\begin{eqnarray}
  \rho(\omega) 
 \hspace{-2.3mm} & \sim \hspace{-2.3mm} & 
 8 \rho_0 \sqrt{\Delta_s^2 -\omega^2}\, \xi \,
  \arccos\left(\omega/\Delta_s\right) \,\left[1+\exp(-8\pi\omega
    \xi)\right] \nonumber \\
  & & \hspace{10mm} \times \,\exp\left(-8 [\sqrt{\Delta_s^2 -\omega^2}\, \xi -\omega \xi \, \arccos
    \left(\omega/\Delta_s\right)]\right) 
  \;,  \nonumber \\
  & & \hspace{55mm} \quad 4\Delta_s \xi \left(1- \frac{\omega}{\Delta_s}
  \right)^{3/2} \gg 1 \;. \qquad \quad
  \label{eq:DOS:phase:asymptotics:minus}
\end{eqnarray}
If $\omega \ll \Delta_s$ and $4\omega^2\xi/\Delta_s \ll 1$, this
result simplifies to
\begin{equation}
  \label{eq:phase:DOSatsmallomega}
  \rho(0) \sim 8 \pi \rho_0 \Delta_s \xi \cosh(4\pi \omega \xi) \exp(-8
  \Delta_s \xi) \;.
\end{equation}

\subsubsection{Pauli paramagnetic susceptibility}

The Pauli paramagnetic susceptibility is defined as the contribution
of the conduction electrons to the susceptibility and can be written
in terms of the DOS. A magnetic field $H$ shifts the energy levels of
the electrons by an amount $\pm \mu_B H$, where $\mu_B$ is the Bohr
magneton and the sign depends on the spin orientation of the electron
with respect to the field. The resulting different occupation of spin-up
and spin-down states leads to a magnetization density $M(T)$ which for
small magnetic fields is linear in $H$. If $\rho(\omega) = \rho(-\omega)$
as in the FGM, the susceptibility $\chi(T) \equiv dM(T)/dH$ can be written
as
\begin{equation}
  \label{eq:susceptibilityIII}
  \chi(T) = \frac{\mu_B^2}{T} \int_{0}^{\infty} d\omega\, \rho(\omega)
  \, \frac{1}{\cosh^2(\omega/2 T)} \;.
\end{equation}
Placing the asymptotic expression
(\ref{eq:DOS:phase:asymptotics:minus}) with $\xi(T)=\rho_s/T$ into
this equation, we find
\begin{eqnarray}
  \chi(T)/\chi_0 
  \hspace{-2.3mm} & \sim  \hspace{-2.3mm} & 
  \frac{16 \rho_s}{T^2} \int_{0}^{\infty}
  d\omega\,
  \sqrt{\Delta_s^2 - \omega^2}
  \arccos\left(\frac{\omega}{\Delta_s}\right) \,\frac{1+\exp[-8\pi
    \rho_s \omega/T]}{(1+\exp[-2\omega/T])^2} \nonumber \\
  & & \hspace{-15mm}
  \times \, \exp\left(-\frac{\Delta_s}{T} \,
    \left[\frac{\omega}{\Delta_s}\left(1 - 8 \rho_s
        \arccos\left(\frac{\omega}{\Delta_s}\right)\right) + 8 
      \rho_s
      \sqrt{1-\left({\omega}/{\Delta_s}\right)^2} 
      \right]\right) \;, \qquad \quad
\end{eqnarray}
where $\chi_0 = 2 \mu_B^2 \rho_0$.
For $\rho_s>\rho_0/4$ (and $T \lll T_c^{\rm MF}$), the integrand is
sharply 
peaked at $\omega = \cos(1/8\rho_s) \Delta_s$, such that the integral
may be evaluated by a saddle point integration, resulting in
\begin{eqnarray}
  {\setlength{\fboxsep}{2mm} \fbox {$ \ \displaystyle  
  \begin{array}{rcl}  \displaystyle
   \frac{\chi(T)}{\chi_0} 
   \hspace{-2.3mm} & \sim  \hspace{-2.3mm} & \displaystyle
   \sqrt{\frac{\pi}{\rho_s}}\,
   \left(\frac{\sin(1/8\rho_s) \, \Delta_s}{T}\right)^{3/2} 
   \exp\left(-\frac{8\rho_s \sin(1/8\rho_s) \, \Delta_s}{T}\right)\;,
   \\
   & & \hspace{35mm} 
   T \ll 4\rho_s \Delta_s
   \left(1-\cos(1/8\rho_s)\right)^{3/2} \;. \end{array} $}}
\qquad
   \label{eq:susceptibilityVI}
\end{eqnarray}
Note, that the temperature restriction is necessary for the asymptotic
expansion of the DOS to be valid. 
Although the susceptibility $\chi(T)$ vanishes exponentially, the
exponent $8\rho_s \sin(1/8\rho_s) \Delta_s/T$ is smaller than the
exponent $8\rho_s \Delta_s/T$, which governs the DOS at the Fermi
energy. However, as $\rho_s$ approaches $\rho_0/4$, the two exponents
become identical, such that for $\rho_s \le \rho_0/4$ the DOS and the
susceptibility have the same exponential dependence on $T$.

\begin{figure}[tb]
\begin{minipage}{0.6\linewidth}
\begin{center}
\psfrag{T}{\small\hspace{-4mm}$T/T_c^{\rm MF}$}
\psfrag{chi}{\small$\hspace{-9.0mm}\chi(T)/\chi_0,\
  \rho(0)/\rho_{0}$}
\psfrag{sus}{\small\hspace{0.5mm}$\chi(T)/\chi_0$}
\psfrag{dos}{\small\hspace{-0.5mm}$\rho(0)/\rho_0$}
\psfrag{0}{\small $0$}
\psfrag{0.5}{\small $0.5$}
\psfrag{1}{\small $1$}
\epsfig{file=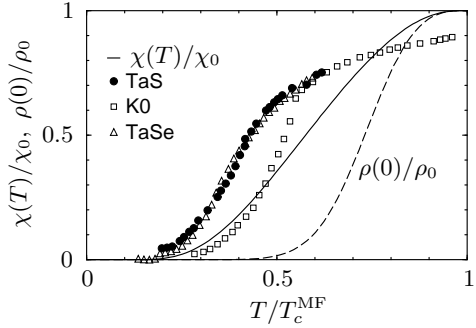,width=6.2cm}
\end{center}
\end{minipage}
\begin{minipage}{0.4\linewidth}
\caption{Plot of the susceptibility $\chi(T)$ calculated for $\xi(T) =
  \rho_s(T)/ T$ [with $\rho_s(T)$ given by Eq.\ (\ref{eq:nT})] and 
  $\Delta_s(T)$ determined by the minimum of the Ginzburg-Landau
  functional. For a  
  comparison, we also show as the dashed line the DOS at the Fermi
  energy, $\rho(0)$.}
\label{fig:susceptibility}
\end{minipage}
\end{figure}

A numerical evaluation of the susceptibility $\chi(T)/\chi_0$ for
$\xi(T) = \rho_s(T)/T$ given by Eq.\ (\ref{eq:nT}) and 
  $\Delta_s(T)$ determined by the BCS gap equation (\ref{eq:BCSgap}) 
is shown in Fig.\ \ref{fig:susceptibility}.
For a comparison, we have also plotted the DOS $\rho(0)/\rho_0$ as a
function of temperature. The two are not identical because for small
temperatures (and $\rho_s = \rho_0$), the major contribution to
the integral in Eq.\ (\ref{eq:susceptibilityIII}) comes from the
frequency region just below $\Delta_s$.
In Fig.\ \ref{fig:susceptibility}, we also show susceptibility data
taken from Ref.\ \cite{Johnston85} for incommensurate quasi
one-dimensional conductors which undergo a Pererls transition.

It is quite surprising that our plot of the susceptibility is very
similar to the plot 
obtained by Lee, Rice and Anderson \cite{Lee73} which
perfectly fits experimental data
\cite{Gruener94,Johnston85,Johnston84}. However, to explain
experimental data, Lee, Rice and Anderson \cite{Lee73} had to base
their calculations on a {\em real} order 
parameter with a correlation length which for low temperatures
increases exponentially as the temperature is lowered. Only the
exponentially increasing correlation length of a real order parameter
could lead to an 
exponentially decreasing susceptibility and the prediction of
$T_c^{\rm 3D} \approx T_c^{\rm MF}/4$. Here, we have shown that these
predictions should also hold for a complex order parameter with a
correlation length which increases as $1/T$.
Since most Peierls chains are incommensurate and the susceptibility of
many incommensurate Peierls chains 
has been compared with
the theory by Lee, Rice and Anderson \cite{Lee73},
our results are of major experimental relevance. For a comparison
between theory and experiment, it  should be
recalled that we have only used a strictly one-dimensional model with
phase fluctuations only. At higher temperatures, one should also
include amplitude fluctuations.

\subsubsection{Thermodynamic quantities}

The DOS encapsulates the whole thermodynamics. Let us first consider
the electronic free energy $F_{\rm el} (T)$ with
respect to the gapped state for which we have
$\rho_{\infty}(\omega) = \rho_0 \,\theta(\omega^2 - \Delta_s^2)
\,|\omega|/(\omega^2 - \Delta_s^2)$: 
\begin{equation}
  \label{eq:F_el}
  F_{\rm el} (T) - F_{\rm el}^{\xi=\infty} (T)= -\frac{s L}{\beta}
  \int_{-\infty}^{\infty} d\omega 
  \left[\rho(\omega) - 
    \rho_{\infty}(\omega) \right] \ln \left(1+e^{-\beta \omega}
  \right) \;. \qquad
\end{equation}
Partial integration leads to
\begin{eqnarray}
  F_{\rm el} (T) - F_{\rm el}^{\xi=\infty} (T)
  \hspace{-2.3mm} & =  \hspace{-2.3mm} & 
  -s L \int d\omega
  \left[\mathcal{N}(\omega) - 
    \mathcal{N}_{\infty} (\omega) \right]
  \frac{1}{e^{\beta \omega} +1}  
  \nonumber \\
  \hspace{-2.3mm} & =  \hspace{-2.3mm} & 
  s \rho_0 L \, \textrm{Im} \int d\omega 
  \left[\Gamma (\omega) - \Gamma_{\infty} (\omega) 
    \right] \frac{1}{e^{\beta \omega} +1} 
\;, \qquad
  \label{eq:F_elII}
\end{eqnarray}
where
$\Gamma(\omega) = \ell^{-1}(\omega) - i\pi \mathcal{N}(\omega)$ 
and $\Gamma_{\infty} (\omega) =
\sqrt{\Delta_s^2 - (\omega+i0)^2}$. The square root has to be
taken such that $\Gamma_{\infty} (\omega) \to -i\omega$ for $\omega
\to \infty$. Since $\Gamma(\omega)$ is analytic
in the upper half plane, the integral may be done by closing the
integral in the upper half plane and using the residue theorem. We
find
\begin{equation}
  F_{\rm el} (T) - F_{\rm el}^{\xi=\infty} (T) = 
  -s \rho_0 L \, \frac{2\pi}{\beta} \sum_{\tilde \omega_n > 0} \left[\textrm{Re} \,\Gamma(i\tilde \omega_n)
    -\tilde \omega_n \right] \;.
  \label{eq:F_elIII}
\end{equation}
For the FGM with phase fluctuations only, we obtain by placing Eq.\
(\ref{eq:Gamma:phaseIII}) into 
this equation 
\begin{equation}
  F_{\rm el} (T) - F_{\rm el}^{\xi=\infty} (T)  =  
  s \rho_0 L \Delta_s \, \frac{2\pi}{\beta} \sum_{\tilde \omega_n > 0}
  \left[\sqrt{1 +\left(\frac{\tilde \omega_n}{\Delta_s}\right)^2}  -
    \frac{I_{4\tilde\omega_n \xi}^{\prime} 
\left(4 \Delta_s \xi \right)}{I_{4\tilde\omega_n \xi}
\left(4 \Delta_s \xi \right)} \right] \;.
  \label{eq:F_el:phase}
\end{equation}
Up to an irrelevant additive constant, $F_{\rm el}^{\xi=\infty} (T)$ is
given by 
\begin{equation}
  F_{\rm el}^{\xi=\infty} (T) = - s \rho_0 L 2 \Delta_s^2 \int_1^{\infty} du\,
  \sqrt{u^2-1} \, \frac{1}{e^{\Delta_s u/T} + 1} \;,
  \label{eq:F_el:inf}
\end{equation}
which for small temperatures is exponentially
small:
\begin{equation}
  F_{\rm el}^{\xi=\infty} (T) \sim - s \rho_0 L \sqrt{2\pi} \Delta_s^2
  \,e^{-\Delta_s/T} \;,\quad T \ll \Delta_s \;.
  \label{eq:F_el:inf:asymp}
\end{equation}
While in the general case we have to add Eqs.\ (\ref{eq:F_el:phase})
and (\ref{eq:F_el:inf}) to get the free energy $F_{\rm el}(T)$, for
low temperatures we can neglect the exponentially small contribution
given by Eq.\ (\ref{eq:F_el:inf:asymp}), such that $F_{\rm el}(T)$ is
determined by the right-hand side of Eq.\ (\ref{eq:F_el:phase}).

For $T \ll
T_c^{\rm MF}$, the dimensionless correlation length $\Delta_s \xi$ is
large, and a uniform asymptotic expansion of $I_{\nu}(\nu z)$ and 
$I_{\nu}^{\prime}(\nu z)$ [see A\&S, Eqs.\ (9.7.7) and (9.7.9)] can be
used to find for the leading terms of the free energy 
\begin{equation}
  \label{freeenergyexpansion}
  {\setlength{\fboxsep}{2mm} \fbox {$ \ \displaystyle 
  F_{\rm el} (T) \sim s\rho_0 L \Delta_s^2 \left[\frac{\pi}{4} \,
    \frac{1}{4\Delta_s\xi} - \frac{1}{12} \,
    \frac{1}{(4\Delta_s\xi)^2}\right]  \;.$}}
\end{equation}

\subsubsection*{Electronic specific heat}

An experimentally accessible thermodynamic quantity is the electronic
specific heat which can be expressed in terms of the free energy as
\begin{equation}
  \label{Cel}
  C_{\rm el} (T) = -T\,\frac{d^2F_{\rm el}(T)}{dT^2} \;.
\end{equation}
The low-temperature behavior of $C_{\rm el} (T)$ can be
obtained from Eq.\ 
(\ref{freeenergyexpansion}):
Using $\xi(T) = s \rho_s(0)/2T$, it directly follows 
\begin{equation}
  \label{eq:ClowT}
  {\setlength{\fboxsep}{2mm} \fbox {$ \ \displaystyle 
  C_{\rm el}(T) \sim \frac{1}{8} \left(\frac{\rho_0}{s
      \rho_s(0)}\right)^2 \, C_{\rm el}^0 (T) \;,$}}
\end{equation}
where the specific heat of free electrons is given by
\begin{equation}
C_{\rm el}^0 (T) = s\,\frac{\pi^2}{3}\, \rho_0  L  T \;.
\end{equation}
Although the DOS exhibits a pseudogap and vanishes exponentially near
the Fermi energy as the temperature is lowered, the electronic
specific heat $C_{\rm el}(T)$ vanishes only linearly in $T$, as for free
electrons. 

We conclude this section with a summary of the central results of the
FGM valid at low temperatures where phase fluctuations dominate in
Table \ref{tab:summary}.
\begin{table}[htb]
    \caption{Asymptotic low temperature results for the FGM
      describing electrons with spin. Note that we have reintroduced
      the Fermi velocity $v_F$ and note also that the mean-field critical
      temperature $T_c^{\rm MF}$ serves as the only energy scale. For
      generalizations of the formulas see the text.}
    \label{tab:summary}
  \begin{center}
    \begin{tabular}{||ll||}
      \hline
      \rule[-5mm]{0mm}{13mm}
      superfluid density & $\displaystyle \rho_s(T) \sim \rho_0 =
      \frac{1}{\pi v_F }$ \\ 
      \rule[-5mm]{0mm}{12mm}
      correlation length & $\displaystyle \xi(T) \sim \frac{v_F}{\pi
        T}$ \\
      \rule[-5mm]{0mm}{12mm}
      density of states & $\displaystyle \frac{\rho(0)}{\rho_0} \sim 
      \frac{14.1\,T_c^{\rm MF}}{T} \exp\left(-\frac{4.49\, T_c^{\rm 
              MF}}{T}\right)$ \\ 
      \rule[-5mm]{0mm}{12mm}
      inverse localization length & $\displaystyle \ell^{-1}(0) \sim
      \frac{1.76\, T_c^{\rm MF}}{v_F}$ \\
       \rule[-5mm]{0mm}{12mm}
      susceptibility & $\displaystyle \frac{\chi(T)}{\chi_0} \sim 1.74
      \left(\frac{T_c^{\rm MF}}{T}\right)^{3/2}
      \exp\left(-\frac{1.72\,T_c^{\rm MF}}{T}\right)$ \\  
      \rule[-6mm]{0mm}{13mm}
      electronic specific heat  & $\displaystyle \frac{C_{\rm el}(T)}{C_{\rm
        el}^0(T)} \sim \frac{1}{32} $ \\
      \hline
    \end{tabular}
  \end{center}
\end{table}

\section{Conclusion}

In this work, we have discussed the density of states (DOS) of the
fluctuating gap model (FGM) and related quantities like the
inverse localization length, the Pauli paramagnetic susceptibility and
the low-temperature specific heat. We introduced the FGM as an
effective low-energy model describing the electronic properties of
Peierls chains and emphasized the fact that the FGM also finds
its applications in other physical contexts: Spin chains can be mapped
by a Jordan-Wigner transformation onto the FGM and in order to explain the
pseudogap-phenomenon in underdoped cuprates above a phase
transition, higher-dimensional generalizations of the FGM have been
used. 

With the rediscovery of the FGM in the context of high-temperature
superconductivity, a previously unnoticed subtle error in
Sadovskii's widely used Green function of the FGM was brought to light.
This error called for a reinvestigation of the FGM.

After setting up a non-perturbative theory which, in principle, allows
to express the one-particle Green function as a functional of an
arbitrary given realization of the disorder, we derived a simple
equation of motion whose solution determines the DOS and the inverse
localization length. Starting from this equation, we could rederive
all known results for the FGM in the 
white noise limit.

Considering the equation of motion governed by the phase which
determines the 
DOS, we argued that the Dyson singularity found in the white noise
limit for commensurate Peierls chains should not be an artifact of the white
noise limit, but should be present for any finite correlation length
in contradiction to Sadovskii's solution.
Our following numerical calculation of the DOS and inverse
localization length confirmed this prediction and showed also that for
large correlation lengths, the Dyson singularity only overshadows a
pseudogap. Although Sadovskii's algorithm misses this singularity, his
solutions for the incommensurate case where there are no singularities in
the DOS give a fairly
good approximation to the exact result. 

In the
pseudogap-regime below the mean-field critical temperature,
fluctuations of the order 
parameter cannot be described by Gaussian statistics. Instead, as the
temperature is lowered, amplitude fluctuations get gradually frozen
out, and the amplitude takes on a value given by the minimum of the
Ginzburg-Landau functional and only
long-wavelength gapless phase fluctuations survive. 
Using a gauge transformation to map the phase fluctuations of the
order parameter onto an effective forward scattering potential, we
could even find an exact solution for the FGM
involving only phase fluctuations which should be valid in
the low temperature regime. We found that the low-temperature specific
heat is linear in $T$ and that both the DOS at the Fermi energy and
the Pauli paramagnetic susceptibility vanish exponentially as the
temperature $T$ is lowered, the ratio of the former to the latter also
vanishing exponentially.
The Pauli paramagnetic susceptibility has been measured in various
experiments and is in good agreement with our results.

Having discussed quantities related to the DOS, one would also like to
calculate quantities like the spectral function.
This has been done for a special non-Gaussian
probability distribution involving amplitude and phase fluctuations in
Ref.\ \cite{Bartosch00b}, but accurate results for realistic
probability distributions (e.g.\ for phase fluctuations only) are not
known yet.

I would like to thank K.\ Sch\"onhammer, R.\ Hayn and especially P.\
Kopietz for helpful discussions. This work was financially supported by the
DFG (Grant No. Ko 1442/4-2).

\end{document}